\documentclass[11pt,a4paper]{article}
\usepackage[utf8]{inputenc}

\usepackage{epic,eepic,epsfig,amsmath,amssymb,amsthm,mathabx,mathrsfs}
\usepackage{t1enc,tabularx,subcaption,graphicx}
\usepackage{array}

\usepackage[francais,english]{babel}

\usepackage{caption}

\usepackage{color}

\usepackage{bbm}

\usepackage{hyperref}

\def\virgp{\raise 2pt\hbox{,}}

\renewcommand{\geq}{\geqslant}
\renewcommand{\leq}{\leqslant}
\def\N{{\mathbb N}}

\def\R{{\mathbb R}}

\def\virgp{\raise 2pt\hbox{,}}
\def\cdotpv{\raise 2pt\hbox{;}}

\def\1{\mathbbm{1}}

\newtheorem{theorem}{Theorem}[section]
\newtheorem{corollary}[theorem]{Corollary}
\newtheorem{proposition}[theorem]{Proposition}

\theoremstyle{remark}
\newtheorem{remark}{Remark}[section]

\theoremstyle{definition}
\newtheorem{definition}{Definition}[section]

\theoremstyle{definition}

\theoremstyle{definition}

\newcolumntype{M}[1]{>{\centering}m{#1}}

\begin{document}

\title{Rethinking Arrow--Debreu: A New Framework for Exchange}

\author{Nizar Riane$^\dag$}

\maketitle

\centerline{$^\dag$ Universit\'e Mohammed V de Rabat, Maroc\footnote{nizar.riane@gmail.com}}

\vskip 0.5cm

\begin{abstract}
This paper revisits the Arrow--Debreu general equilibrium framework through the lens of effective trade, emphasizing the distinction between theoretical and realizable market interactions. We develop the Effective Trade Model (ETM), where transactions arise from bilateral feasibility rather than aggregate supply and demand desires. Within this framework, we establish the main properties of the price-demand correspondence and prove the existence of Nash equilibria, incorporating production, money, and network topology. Our results show that equilibrium is shaped by transaction constraints, subjective pricing, and decentralized negotiation, rather than by universal market-clearing conditions, and thereby call into question the foundations of welfare theory.
\end{abstract}

\vskip 1cm

\noindent \textbf{Keywords}: General equilibrium -- Effective trade -- Bilateral exchange -- Generalized games -- Welfare -- Networks -- Quantity theory of money.

\vskip 1cm

\noindent \textbf{JEL Classification}: C62 -- C72 -- D43 -- D51 -- D61 -- D85 -- L14

\vskip 0.5cm

\vskip 0.5cm

\section{Introduction}

\hskip 0.5cm The Arrow--Debreu Model (ADM) \cite{ArrowDebreu1954} has long been a cornerstone of modern economic theory, providing a rigorous mathematical framework for analyzing microeconomic decisions. It offers a formal resolution to an old question posed by Léon Walras in \cite{Walras1874}, leveraging advancements in the theory of multifunctions and fixed points.\\

While the ADM is meticulously formulated, it has not been universally embraced by economists. Criticism often targets its assumptions, including perfect competition, the existence of forward markets for every commodity and all conceivable contingencies, and the absence of money as a store of value. This has led some scholars, such as Mark Blaug, to express strong reservations. Blaug famously characterized Arrow and Debreu's seminal paper \cite{ArrowDebreu1954} as marking "the beginning of what has since become a cancerous growth in the very center of microeconomics" (see \cite{Blaug1998}). He further argued that "by the time we got to Arrow and Debreu, general equilibrium theory had ceased to make any descriptive claim about actual economic systems and had become a purely formal apparatus about a quasi-economy".\\ 

Beyond Blaug's critiques of its theoretical abstraction, it is important to recognize that general equilibrium theory has undergone significant refinements since the publication of Arrow and Debreu's 1954 paper. Debreu's \textit{The Theory of Value} \cite{Debreu1959} further formalized the framework, particularly in economies facing uncertainty. The role of money within the model has been extensively debated, with insightful contributions from Robert Clower \cite{Clower1967}, Frank Hahn \cite{Hahn1989}, and Joseph M. Ostroy and Ross M. Starr \cite{OstroyStarr1974} and others. The conceptualization of transaction was reformulated by L. Shapley and M. Shubik through the introduction of trading posts in \cite{ShapleyShubik1977}. The question of equilibrium convergence was explored using nontraditional techniques and more realistic approaches (see, for example, \cite{Hahn1962}). Additionally, time was explicitly incorporated into the system \cite{Radner1972}, and market imperfections---such as equilibrium with indivisible goods---were examined to better align the theory with economic realities \cite{Svensson1984}.\\

In his book \emph{The Methodology of Economics: Or, How Economists Explain}~\cite{Blaug1992}, Mark Blaug provocatively asks whether general equilibrium theory should be regarded as an empirical theory or merely as a conceptual framework. To explore this question, one can draw an analogy with classical physics: just as Newtonian mechanics assumes the existence of a Galilean reference frame--a perfectly inertial system in which the laws of motion take their simplest form--general equilibrium theory is grounded in idealized assumptions of complete markets and perfect competition. However, in practice, the Earth is a non-inertial frame of reference: its rotation gives rise to apparent forces, such as the Coriolis effect, which must be accounted for to correctly describe physical phenomena. Likewise, real-world markets are subject to imperfections and incompleteness, forming what one might call a "non-Galilean" economic environment. In such contexts, the neat predictions of general equilibrium theory must be modified to incorporate frictions, constraints, and institutional complexities. Thus, while general equilibrium provides a powerful and elegant framework for understanding economic coordination, its empirical relevance depends on our ability to account for the deviations from its foundational assumptions.\\

Contemplating the ADM, one should recognize that it fundamentally relies on the concepts of theoretical demand and supply—idealized representations of preferences and resource endowments. However, these abstractions often diverge from real-world market transactions, where effective demand and supply, shaped by constraints and actual economic behavior, determine outcomes. A critical reassessment of these distinctions is essential for a deeper understanding of exchange systems and money.\\

In the real economy, effective demand and supply, determined by what individuals can actually transact, take precedence. Transactions reflect tangible constraints, such as budgets, production capabilities, and specifically, the other party capacities and needs, making it a bilateral consent.\\

Utility in real-world exchanges depends on effective allocations---what individuals receive or consume through transactions---rather than theoretical allocations. This distinction underscores the importance of addressing the outcomes of exchanges rather than unattainable aspirations. Markets are not arenas for idealized redistribution but mechanisms for facilitating feasible transactions shaped by constraints, negotiation, and network interactions.\\

Real-world exchanges also separate the roles of buying and selling. Transactions are inherently bilateral: they occur between two parties negotiating terms and quantities, rather than through collective redistribution. The ADM's implicit assumption of a pooled tribal distribution system fails to capture the sequential and decentralized nature of actual market interactions. In practice, individuals bring goods to the market not for pooling but for specific exchanges, governed by mutually acceptable terms.\\

Moreover, the quantity exchanged in a transaction is inherently constrained. It is determined by the minimum of what both parties can support, reflecting their respective limitations. This contrasts with theoretical models that emphasize optimal allocations without considering such practical constraints.\\

Equilibrium is not a pure quantity adjustment process but a dynamic interplay between prices and quantities. Individuals do not merely optimize quantities based on given prices; they actively set prices and negotiate quantities in response to their circumstances and opportunities. This dynamic adjustment process better reflects how markets operate, with equilibrium emerging from these decentralized interactions.\\

In the monetary sphere, the bilateral nature of exchange also provides a lens to trace the circulation of money in the economy from a local perspective. Rather than relying on a global analysis dictated by the quantity theory of money, this approach emphasizes localized flows of currency, capturing the intricacies of individual transactions and their cumulative effects, and also the impact of the topology of the network relationships.\\

Under this vision, monetary policy takes on a more nuanced role, since it may simultaneously affect quantities, prices, velocity, and redistribution. By addressing these interconnected factors, monetary interventions could reshape not only aggregate outcomes but also the microeconomic structures underpinning economic activity.\\

Several strands of the literature have sought to move beyond the frictionless and fully connected exchange environment of the classical Arrow--Debreu model. Dubey \cite{Dubey1982} develops a strategic market game in which agents choose both prices and quantities, and shows that, under appropriate conditions, Nash equilibria coincide with competitive (Walrasian) equilibria even with a finite number of traders. Kranton and Minehart \cite{Kranton2001} introduce an endogenous buyer--seller network, showing how trading relationships can arise from firms' incentives to share suppliers and pool demand uncertainty, and how competitive link formation may generate socially efficient networks under appropriate conditions. Kakade, Kearns, and Ortiz \cite{2004Kakade} incorporate network restrictions directly into the Arrow--Debreu framework: an underlying graph determines which agents can trade, allowing local prices and local market-clearing conditions to emerge. They establish the existence of graphical equilibria under conditions analogous to those of Arrow and Debreu and develop an efficient algorithm for computing approximate equilibria on tree-structured networks. Two other traditions address related questions. Non-Walrasian equilibrium theory \cite{Benassy1975,Dreze1975} studies economies in which prices are rigid and markets clear through quantity rationing. The literature on decentralized trade through bilateral meetings and bargaining \cite{RubinsteinWolinsky1985} studies how the terms of each transaction are determined between the two parties.\\

The present framework differs from each of these approaches in a specific respect. In the strategic market game of Dubey \cite{Dubey1982}, agents submit price--quantity orders to a market for each good, which clears at a single price; this common clearing price is what allows Nash equilibria to coincide with Walrasian equilibria. In our model, there is no such market: each seller sets its own price, trades are bilateral, and the quantity exchanged between two agents is the minimum of their proposals, so that no clearing mechanism equalizes prices across sellers. In non-Walrasian equilibrium theory \cite{Benassy1975,Dreze1975}, prices are assumed to be rigid, and quantity signals arise from a rationing scheme; here, the bilateral constraints act as quantity signals arising from the proposals of the trading partners, and the rigidity of prices is an outcome. In the network models of Kranton and Minehart \cite{Kranton2001} and of Kakade, Kearns, and Ortiz \cite{2004Kakade}, prices result from competitive mechanisms, auctions or local market clearing, whereas they are chosen here by the sellers themselves. In the bargaining literature \cite{RubinsteinWolinsky1985}, the terms of each transaction are negotiated between the two parties; here, they result from the prices and proposals chosen independently by each party. To our knowledge, the combination of individual seller prices, bilateral proposals on a network, and a general equilibrium formulation as a generalized game is new.\\

In this paper, we model the bilateral transaction itself as endogenous. Each consumer simultaneously chooses a price vector and a collection of bilateral supply and demand proposals, and a transaction is realized only to the extent that the proposals of the two parties are compatible. This formulation leads to a generalized game in which an agent's feasible set depends on the prices and proposals of the other agents. Beyond consumption, agents may also have preferences over their trading partners, in separable or hierarchical form, and over how their prices compare with those of the market.\\

We establish the existence of transaction equilibria and describe their structure. Autarky is always a transaction equilibrium, a coordination failure intrinsic to bilateral exchange, but equilibria with trade exist whenever the economy contains a trade cycle with gains from trade. At a transaction equilibrium, every seller sells a single good, at a positive price set at the bound of the normalization, so that the allocation is determined by the bilateral quantities, as in a fixed-price equilibrium. Preferences over trading relationships and prices restore concavity: best responses become unique and continuous, specialization is no longer imposed, and prices respond to market conditions. We characterize equilibria through first-order conditions and a sensitivity analysis, and study a planner benchmark, in which trade is circular and incomes form the stationary distribution of the chain of sales, together with price adjustment dynamics. Finally, both welfare theorems fail: autarky can be an equilibrium, so that equilibria need not be Pareto optimal, and Pareto-optimal allocations need not be reachable through trade from the initial endowments.\\

We extend the framework to production, money, and topological constraints; in the latter case, incomplete bilateral access generates a middleman imperfection. The framework thus provides a unified way of studying how individual pricing decisions, bilateral compatibility, and the structure of exchange jointly shape equilibrium and welfare.\\

Our paper is organized as follows:

\begin{enumerate}
\item The first part introduces the effective trade framework, develops the price--demand correspondence, establishes the existence of transaction equilibria, describes their structure, and studies their welfare properties.

\item The second part characterizes transaction equilibria through first-order conditions, studies a planner benchmark and market convergence, and analyzes market imperfections arising from indivisibilities and the topology of exchange, together with preferences over trading relationships and prices. A numerical illustration concludes this part.

\item The third part extends the framework to incorporate production and money, derives the quantity of money equation, and discusses possible extensions to dynamic settings with time and uncertainty.
\end{enumerate}

\vskip 0.5cm

\section{New way to consider the Consumer Problem\label{New way to consider the Consumer Problem}}

\hskip 0.5cm Consider an economy $\mathcal{E}=\left(w^i,u^i,\mathscr{X}^i\right)_{1\leq i \leq n}$ consisting of $L$ goods and $n$ consumption units. Each unit $i$ is endowed with a real-valued utility function $u^i$ and an initial endowment vector \mbox{$w^i=(w^i_1,\hdots,w^i_L)\in \mathscr{X}^i$}. The consumption possibilities set $\mathscr{X}^i$ is a non-empty, closed, bounded and convex subset of $\R_+^L$.\\

We suppose that each consumption unit (or consumer, for simplicity) selects its own supply price vector (not a price taker)
\begin{align}
\label{eqn:1}
p^i \in P=\left\{ p \mid p \in \mathbb{R}_+^L, \, p \neq 0, \, \sum_{j=1}^L p_j = 1 \right\}
\end{align}

\noindent and its \textbf{potential} bilateral transactions matrix (representing theoretical demands and supplies)
\begin{align}
\label{eqn:2}
X^i=(x^i_{ji},x^i_{ij})_{1\leq j \leq n}\in \mathbb{X}^i \subset (\mathbb{R}_+^L)^{2n}
\end{align}

\noindent with $x^i_{ii} = 0$, where $x^i_{ij}$ denotes the quantity that consumer $i$ wants to sell to consumer $j$, $x^i_{ji}$ denotes the quantity that consumer \( i \) wants to buy from consumer $j$, interpreted in the sense of a directed multigraph, and $\mathbb{X}^i$ is a non-empty, convex, closed, and bounded set of admissible bilateral transactions, with boundedness justified, for instance, by requiring that bilateral transactions cannot exceed the total quantities of goods available in the economy.\\

The \textbf{effective supply (transaction)} from consumer $i$ to consumer $j$ is given by
\begin{align}
\label{eqn:3}
q_{ij}=\min\left(x^i_{ij},x^j_{ij}\right).
\end{align}

Conversely, the \textbf{effective demand} of consumer $i$ from consumer $j$ is defined as
\begin{align}
\label{eqn:4}
q_{ji}=\min\left(x^i_{ji},x^j_{ji}\right).
\end{align}

Define the final holding of consumer $i$ after trade as
\begin{align}
\label{eqn:5}
x^i = w^i + \sum_{j\neq i} q_{ji} - \sum_{j\neq i} q_{ij}.
\end{align}

We introduce the notations $X = (X^1, \dots, X^n)$ and $p = (p^1, \dots, p^n)$. The transactions of consumer $i$ must satisfy the \textbf{transaction balance condition}
\begin{align}
\label{eqn:6}
\sum_{j\neq i} p^j \cdot q_{ji} &\leq \sum_{j\neq i} p^i \cdot q_{ij},
\end{align}

\noindent where purchases from $j$ are valued at the seller's price $p^j$ and sales at the agent's own price $p^i$; a seller thus charges the same price vector to all its buyers. This condition comes together with the requirement that the final holding be feasible, $x^i \in \mathscr{X}^i$.\\

The consumer problem is
\begin{align}
\label{eqn:7}
\max_{(p^i,X^i)\in \mathbf{B}^i(p^{-i},X^{-i})} u^i(x^i)
\end{align}

Under the budget constraint
\begin{align}
\label{eqn:8}
\mathbf{B}^i(p^{-i},X^{-i}) &=\left\{(p^i,X^i)\in P \times \mathbb{X}^i \ \mid \  \sum_{j\neq i} p^j \cdot  q_{ji} \leq  \sum_{j\neq i} p^i \cdot  q_{ij} \ , \   w^i + \sum_{j\neq i} q_{ji}- \sum_{j\neq i} q_{ij} \in \mathscr{X}^i  \right\}
\end{align}

The notation $-i$ refers to the indices $j \in \{1, \dots, n\}$ such that $j \neq i$. An equivalent way to express the maximization problem is to redefine $\displaystyle x^i = w^i + \sum_{j\neq i} x^i_{ji} - \sum_{j\neq i} x^i_{ij}$ and then solve
\begin{align}
\label{eqn:9}
\max_{(p^i,X^i)\in B^i(p^{-i},X^{-i})} u^i(x^i)
\end{align}

\noindent where the budget constraint becomes
\begin{align}
\label{eqn:10}
\begin{split}
B^i(p^{-i},X^{-i}) =\left\{\right. &(p^i,X^i)\in P \times \mathbb{X}^i \ \mid \  \sum_{j\neq i} p^j \cdot  x_{ji}^i \leq \sum_{j\neq i} p^i \cdot  x_{ij}^i \ , \   w^i + \sum_{j\neq i} x_{ji}^i - \sum_{j\neq i} x_{ij}^i \in \mathscr{X}^i \ , \ x_{ij}^i \leq x_{ij}^j \ , \\
& \left. x_{ji}^i \leq x_{ji}^j \right\}
\end{split}
\end{align}

\vskip 0.5cm

\begin{remark}{\label{MaximizationEquivalence}\ }\\
Problems \ref{eqn:7} and \ref{eqn:9} have the same value. Indeed, in \ref{eqn:7} both the constraints and the utility depend on $X^i$ only through the effective trades $(q_{ij},q_{ji})_{j\neq i}$. Replacing each proposal by the corresponding effective trade leaves these trades unchanged, since $\min(\min(a,b),b)=\min(a,b)$, and yields a point of $B^i(p^{-i},X^{-i})$; conversely, on $B^i(p^{-i},X^{-i})$ we have $q_{ij}=x^i_{ij}$ and $q_{ji}=x^i_{ji}$. Problem \ref{eqn:9} thus selects, among the solutions of \ref{eqn:7}, those without excess proposals: the supplier has no interest in exceeding demand, and the demander has no interest in exceeding supply. In particular, at any equilibrium of the game associated with \ref{eqn:9}, the constraints of $i$ and $j$ together give $x^i_{ij}=x^j_{ij}$ for all $i\neq j$: bilateral proposals coincide, making any disequilibrium in the transaction logically inconceivable. In \ref{eqn:7}, by contrast, a disequilibrium may emerge between theoretical demands and supplies.
\end{remark}

\vskip 0.5cm

The feasible set in problem \ref{eqn:9} is not convex, because of the bilinear term $p^i\cdot\sum_{j\neq i}x^i_{ij}$. We therefore transform the consumer problem to a double maximization problem
\begin{align}
\label{eqn:11}
\max_{p^i \in P} \max_{X^i\in \mathcal{B}^i(p,X^{-i})} u^i(x^i)
\end{align}

Under the budget constraint
\begin{align}
\label{eqn:12}
\begin{split}
\mathcal{B}^i(p,X^{-i}) =\left\{\right. & X^i \in \mathbb{X}^i \ \mid \  \sum_{j\neq i} p^j \cdot  x_{ji}^i \leq \sum_{j\neq i} p^i \cdot  x_{ij}^i \ , \   w^i + \sum_{j\neq i} x_{ji}^i - \sum_{j\neq i} x_{ij}^i \in \mathscr{X}^i \ , \ x_{ij}^i \leq x_{ij}^j \ , \\
& \left. x_{ji}^i \leq x_{ji}^j \right\}
\end{split}
\end{align}

Let us assume the following properties of the utility functions:
\begin{enumerate}
\item[\emph{a}] \textbf{Continuity:} $u^i(\cdot)$ is continuous on $\mathscr{X}^i$ .
\item[\emph{b}] \textbf{Strict monotonicity:} $u^i(\cdot)$ is increasing on $\mathscr{X}^i$ in each argument.
\item[\emph{c}] \textbf{Quasi-concavity:} $u^i(\cdot)$ is quasi-concave on $\mathscr{X}^i$.
\end{enumerate}

\noindent and the following properties of the admissible sets and endowments:
\begin{enumerate}
\item[\emph{d}] \textbf{Free reduction of trades:} if $X^i\in\mathbb{X}^i$ and $0\le \widetilde X^i\le X^i$, then $\widetilde X^i\in\mathbb{X}^i$; moreover, $\mathbb{X}^i$ contains $[0,\bar\omega]^{2nL}$ for some $\bar\omega>0$.
\item[\emph{e}] \textbf{Free retention:} if $x\in\mathscr{X}^i$ and $y\in\R_+^L$ satisfy $x\le y\le w^i+\sum_{j\neq i}x^i_{ji}$, then $y\in\mathscr{X}^i$.
\item[\emph{f}] \textbf{Positive endowments:} $w^i_k>0$ for every good $k$, and $[0,w^i]\subseteq\mathscr{X}^i$.
\end{enumerate}

Continuity ensures that consumers do not experience sudden jumps in utility, quasi-concavity expresses a preference for diversified bundles, and strict monotonicity expresses the principle that more is always preferred to less. Assumption \emph{d} states that an agent can always trade less, and in small enough quantities; assumption \emph{e} that keeping goods instead of selling them is always feasible; and assumption \emph{f} that every agent owns a positive quantity of every good, part of which it can give up. We recall Berge's Maximum Theorem from Theorem \ref{Berge's Maximum Theorem} in the appendix.

\vskip 0.5cm

For a fixed price profile $p$, the set-valued correspondence $\mathcal{B}^i(p,X^{-i})$ is non-empty (it contains the autarky situation for any arbitrary vector of prices) and convex-valued, since all its constraints are linear in $X^i$; it is closed and bounded, hence compact-valued. Its continuity properties are the following.
\begin{itemize}
\item \textbf{Upper semicontinuity:} Fix $(p,X^{-i})\in P^n\times\prod_{j\neq i}\mathbb{X}^j$. Let $(p^m,X^{-i,m})\to(p,X^{-i})$ and let $X^{i,m}\in\mathcal{B}^i(p^m,X^{-i,m})$ be a sequence such that $X^{i,m}\to X^i$. We show that $X^i\in\mathcal{B}^i(p,X^{-i})$. Since, for every $m$,
\begin{align*}
\sum_{j\neq i}p^{j,m}\cdot x_{ji}^{i,m} \leq p^{i,m}\cdot\sum_{j\neq i}x_{ij}^{i,m},
\end{align*}
and since the dot product is continuous, taking limits gives
\begin{align*}
\sum_{j\neq i}p^j\cdot x_{ji}^i \leq p^i\cdot\sum_{j\neq i}x_{ij}^i.
\end{align*}
Moreover, since $\mathscr{X}^i$ is closed, $w^i+\sum_{j\neq i}x_{ji}^i-\sum_{j\neq i}x_{ij}^i \in\mathscr{X}^i$. Finally, since $x_{ij}^{i,m}\leq x_{ij}^{j,m}$ and $x_{ji}^{i,m}\leq x_{ji}^{j,m}$ for every $m$, and since $X^{-i,m}\to X^{-i}$, taking limits yields $x_{ij}^i\leq x_{ij}^j$ and $x_{ji}^i\leq x_{ji}^j$. Hence $X^i\in\mathcal{B}^i(p,X^{-i})$. Thus the graph of $\mathcal{B}^i$ is closed. Since $\mathcal{B}^i$ is compact-valued, it follows that $\mathcal{B}^i$ is upper semicontinuous everywhere.

\item \noindent\textbf{Lower semicontinuity:} Lower semicontinuity may fail when an agent can obtain a good offered at a zero price while it has no sale of positive value: if the price of that good becomes slightly positive, the agent can no longer afford it. We therefore establish lower semicontinuity at every $(p,X^{-i})$ satisfying the following condition:
\begin{enumerate}
\item[(C)] every good offered to $i$ is offered at a positive price: $x^j_{ji,k}>0$ implies $p^j_k>0$.
\end{enumerate}
Fix such a point $(p,X^{-i})$ and $X^i\in\mathcal{B}^i(p,X^{-i})$, and let $(p^m,X^{-i,m})\to(p,X^{-i})$. Define the budget gap
\begin{align*}
R^i(p,X^i) &= \sum_{j\neq i}p^j\cdot x_{ji}^i - p^i\cdot\sum_{j\neq i}x_{ij}^i,
\end{align*}
which is continuous, and linear in $X^i$. Let $\varepsilon_m\downarrow0$ be such that $(1-\varepsilon_m)x^j_{ij,k}\le x^{j,m}_{ij,k}$ and $(1-\varepsilon_m)x^j_{ji,k}\le x^{j,m}_{ji,k}$ for all $j\neq i$ and all $k$ with $x^j_{ij,k}>0$ or $x^j_{ji,k}>0$; such a sequence exists since $X^{-i,m}\to X^{-i}$. For every $Z\in\mathbb{X}^i$ satisfying the bilateral constraints at $(p,X^{-i})$, the plan $(1-\varepsilon_m)Z$ then belongs to $\mathbb{X}^i$ by \emph{d}, satisfies the bilateral constraints at $(p^m,X^{-i,m})$, and has final holding $\varepsilon_m w^i+(1-\varepsilon_m)z^i$, which belongs to $\mathscr{X}^i$ by convexity whenever $z^i\in\mathscr{X}^i$.

We distinguish two cases, according to whether $i$ has a sale opportunity of positive value at $(p,X^{-i})$, that is, whether there exist $j\neq i$ and a good $k$ such that $p^i_k>0$ and $x^j_{ij,k}>0$.

\emph{Case 1: $i$ has a sale opportunity of positive value.} Let $\eta=\min(x^j_{ij,k},w^i_k,\bar\omega)>0$, and let $\widehat X^i$ be the plan selling $\eta$ units of good $k$ to $j$ and nothing else. By \emph{d} and \emph{f}, $\widehat X^i\in\mathbb{X}^i$ and its final holding $w^i-\eta e_k$ belongs to $\mathscr{X}^i$; it satisfies the bilateral constraints, and $R^i(p,\widehat X^i)=-\eta\,p^i_k<0$. For $\lambda\in(0,1]$, the plan $X^i_\lambda=(1-\lambda)X^i+\lambda\widehat X^i$ satisfies all the constraints at $(p,X^{-i})$ by convexity, and $R^i(p,X^i_\lambda)\le-\lambda\eta\,p^i_k<0$. Since $R^i(p^m,(1-\varepsilon_m)X^i_\lambda)=(1-\varepsilon_m)R^i(p^m,X^i_\lambda)\to R^i(p,X^i_\lambda)<0$, there exists $M(\lambda)$ such that $(1-\varepsilon_m)X^i_\lambda\in\mathcal{B}^i(p^m,X^{-i,m})$ for all $m\ge M(\lambda)$. A standard diagonal argument yields $\lambda_m\downarrow0$ such that $X^{i,m}=(1-\varepsilon_m)X^i_{\lambda_m}\in\mathcal{B}^i(p^m,X^{-i,m})$ for all $m$ large enough, and $X^{i,m}\to X^i$.

\emph{Case 2: $i$ has no sale opportunity of positive value.} Every sale allowed by the bilateral constraints concerns a good to which $i$ assigns a zero price, so that $p^i\cdot\sum_{j\neq i}x^i_{ij}=0$, and the budget constraint forces $\sum_{j\neq i}p^j\cdot x^i_{ji}=0$. By (C), every good that $i$ can buy has a positive price, hence $X^i$ involves no purchase. The plans $X^{i,m}=(1-\varepsilon_m)X^i$ involve no purchase either, so that $R^i(p^m,X^{i,m})\le0$; they belong to $\mathcal{B}^i(p^m,X^{-i,m})$ and converge to $X^i$.

\noindent Condition (C) is used only in Case 2; Case 1 relies on assumptions \emph{d} and \emph{f}, which allow $i$ to create a strict budget slack by selling part of its endowment. Hence $\mathcal{B}^i$ is lower semicontinuous, and therefore continuous, at every point satisfying (C). When the price profile is held fixed and only $X^{-i}$ varies, condition (C) is not needed: $R^i(p,(1-\varepsilon_m)X^i)=(1-\varepsilon_m)R^i(p,X^i)\le0$, so that $\mathcal{B}^i(p,\cdot)$ is continuous for every $p$.
\end{itemize}

\vskip 0.5cm

\begin{remark}{\ }\\
Even under the assumption of strict concavity, the utility function $u^i$ remains only concave with respect to the elements of $X^i$, since the final holding $x^i=w^i+\sum_{j\neq i}x^i_{ji}-\sum_{j\neq i}x^i_{ij}$ is a non-injective linear function of $X^i$. The optimal final holding is then unique, but the transactions leading to it are not: buying a good from one partner or from another, or buying and reselling the same good, yields the same holding. One potential remedy involves introducing a strictly convex transaction cost on $X^i$, which selects a unique transaction plan and rules out circular trades. However, we will not impose such conditions in the sequel, preserving the generality of the formulation. An alternative and potentially richer approach for the quantities involves incorporating preferences over the connections between agents, capturing human relationships, marketing influences, consumer loyalty, and other social phenomena; preferences over prices can be introduced in the same way. These extensions, developed in Section \ref{Market topology}, reflect how external social dynamics shape individual decisions and, consequently, the equilibrium structure of the economy.
\end{remark}

\vskip 0.5cm

Since $\mathcal{B}^i$ is upper semicontinuous with non-empty compact values and $u^i$ is continuous, the problem
\begin{align*}
\max_{X^i\in \mathcal{B}^i(p,X^{-i})} u^i(x^i)
\end{align*}
admits a solution, the demand correspondence $X^i(p,X^{-i})$ is non-empty, compact-valued and, by assumption \emph{c}, convex-valued, and, by part 1 of Berge's maximum theorem \ref{Berge's Maximum Theorem}, the value function $v^i(p^i,p^{-i},X^{-i})=\max_{X^i\in \mathcal{B}^i(p,X^{-i})} u^i(x^i)$ is upper semicontinuous. Consequently, the problem
\begin{align}
\label{eqn:13}
\max_{p^i \in P} v^i(p^i,p^{-i},X^{-i})
\end{align}
admits a solution, and the price correspondence $p^i(p^{-i},X^{-i})$ is non-empty and compact-valued. At every point where $\mathcal{B}^i$ is continuous, part 2 of Berge's maximum theorem \ref{Berge's Maximum Theorem} further ensures the continuity of $v^i$ and the upper semicontinuity of the demand correspondence; for a fixed price profile, this holds at every $X^{-i}$.

\vskip 0.5cm

\begin{remark}{\label{QuasiConvexValue}\ }\\
The price correspondence need not be convex-valued: $v^i$ is quasi-convex in $p^i$, and in general not quasi-concave. Let $S^i(X^i)=\sum_{j\neq i}x^i_{ij}$ denote the sales and $c^i(X^i)=\sum_{j\neq i}p^j\cdot x^i_{ji}$ the cost of purchases of consumer $i$, and let $F^i$ be the set of $X^i\in\mathbb{X}^i$ satisfying the constraints of $\mathcal{B}^i$ other than the budget inequality, which do not involve $p^i$. Since the maximum defining $v^i$ is attained, $v^i(p^i)<t$ means that no plan reaching the utility level $t$ is affordable at $p^i$, that is, every $X^i\in F^i$ with $u^i(x^i)\ge t$ violates the budget inequality at $p^i$. Hence, for every $t\in\R$,
\begin{align*}
\{p^i\in P \mid v^i(p^i)<t\}=\bigcap_{X^i\in F^i,\ u^i(x^i)\ge t}\{p^i\in P \mid p^i\cdot S^i(X^i)<c^i(X^i)\},
\end{align*}
which is an intersection of convex sets, hence convex. Consequently, writing $p^i=\sum_{k=1}^L p^i_k e_k$ as a convex combination of the vertices of the simplex, quasi-convexity gives $v^i(p^i)\le\max_k v^i(e_k)$: the maximum of $v^i$ over $P$ is attained at a vertex $e_k$ of the simplex, that is, at a price vector putting all its weight on a single good $k$.
\end{remark}

\vskip 0.5cm

One should also note that, in Remark \ref{MaximumCharacterization}, assumption \emph{b} can be weakened to the following condition:

\begin{enumerate}
\item[\emph{b}'] \textbf{Upward local non-satiation:} $\forall x \in \mathscr{X}^i$, $\forall \varepsilon>0$ and $\forall k \in \{1,\hdots, L\}$, there exists $t\in(0,\varepsilon]$ such that $u^i(x)< u^i(x+t\,e_k)$.
\end{enumerate}

Upward local non-satiation ensures that, for any consumption bundle, an arbitrarily small increase in the quantity of any single good strictly improves the agent's utility, without requiring full monotonicity. Theorem \ref{Specialization} and Proposition \ref{PositivePrices}, by contrast, use the fact that holding a given positive additional quantity of a good is strictly better, and therefore require assumption \emph{b}.\\

\vskip 0.5cm

\begin{remark}{\label{MaximumCharacterization}\ }\\
Let $X^{i,\star}$ solve $\max_{X^i\in\mathcal{B}^i(p,X^{-i})}u^i(x^i)$, and suppose that the budget constraint is not binding:
\begin{align*}
\sum_{j\neq i}p^{j}\cdot x_{ji}^{i,\star} &< p^{i}\cdot\sum_{j\neq i}x_{ij}^{i,\star}.
\end{align*}
Since the left-hand side is non-negative, the right-hand side is positive, so that there exist $j\neq i$ and a good $k$ with $x_{ij,k}^{i,\star}>0$ and $p^i_k>0$. Denote by $\sigma>0$ the budget slack, that is, the difference between the right-hand side and the left-hand side, let $\bar\varepsilon=\min\left(x_{ij,k}^{i,\star},\,\sigma/p^i_k\right)>0$, and let $e_k$ be the $k$-th unit vector in $\mathbb{R}^L$. For $\varepsilon\in(0,\bar\varepsilon]$, define
\begin{align*}
\widetilde{x}_{ij,k}^i=x_{ij,k}^{i,\star}-\varepsilon,
\qquad
\widetilde{x}_{ij,\ell}^i=x_{ij,\ell}^{i,\star}
\quad(\ell\neq k),
\end{align*}
all other components of $X^{i,\star}$ being unchanged. Then
\begin{align*}
\widetilde{x}^i=x^{i,\star}+\varepsilon e_k.
\end{align*}
The new plan belongs to $\mathbb{X}^i$ by \emph{d}, the bilateral constraints remain satisfied since a sale has been reduced, $\widetilde{x}^i\in\mathscr{X}^i$ by \emph{e}, and, since $\varepsilon\,p^i_k\le\sigma$,
\begin{align*}
\sum_{j\neq i}p^{j}\cdot x_{ji}^{i,\star} &\leq p^{i}\cdot\sum_{j\neq i}\widetilde{x}_{ij}^i.
\end{align*}
Hence $\widetilde{x}^i$ is feasible for every $\varepsilon\in(0,\bar\varepsilon]$. Under \emph{b}, taking $\varepsilon=\bar\varepsilon$ gives $u^i(\widetilde{x}^i)>u^i(x^{i,\star})$; under \emph{b'}, taking $\varepsilon=t$, where $t\in(0,\bar\varepsilon]$ is given by \emph{b'}, gives the same inequality. In both cases, this contradicts the optimality of $X^{i,\star}$. Therefore,
\begin{align*}
\sum_{j\neq i}p^{j}\cdot x_{ji}^{i,\star} &= p^{i}\cdot\sum_{j\neq i}x_{ij}^{i,\star}.
\end{align*}
\end{remark}

\vskip 0.5cm

\begin{remark}{\label{HomegeneityRemark}\ }\\
The budget constraints are homogeneous of degree zero in the whole price profile $p=(p^1,\dots,p^n)$: multiplying all prices by the same $\lambda>0$ leaves every $\mathcal{B}^i$ unchanged, so that any common normalization of the price profile yields the same allocations. The normalization \ref{eqn:1} is stronger: it requires the prices of each agent to sum to one, which fixes the common unit of value in which every agent expresses its prices and ensures the compactness of the price set $P$.
\end{remark}

\vskip 0.5cm

An interesting feature is to compare our configuration with the Arrow--Debreu vision of the economy. If all agents face a single price vector $p\in P$, common to all of them, as in the Arrow--Debreu model, the consumer's problem would be
\begin{align}
\label{eqn:14}
\tilde{v}^{i}(p) = \max_{X^i \in \tilde{\mathcal{B}}^i(p)} u^i(x^i).
\end{align}

Under the budget constraint
\begin{align}
\label{eqn:15}
\tilde{\mathcal{B}}^i(p) = \left\{ X^i \in \mathbb{X}^i \mid p \cdot \sum_{j \neq i} x_{ji}^i \leq p \cdot \sum_{j \neq i} x_{ij}^i \ , \ w^i + \sum_{j \neq i} x_{ji}^i - \sum_{j \neq i} x_{ij}^i \in \mathscr{X}^i \right\}.
\end{align}

When all agents charge the same price vector $p$, our problem becomes
\begin{align}
\label{eqn:16}
\hat{v}^{i}(p,X^{-i})&=\max_{X^i\in \widehat{\mathcal{B}}^i(p,X^{-i})} u^i(x^i)
\end{align}

where
\begin{align}
\label{eqn:17}
\widehat{\mathcal{B}}^i(p,X^{-i}) &=\left\{X^i\in \mathbb{X}^i \ \mid \  p \cdot \sum_{j\neq i} x_{ji}^i \leq p \cdot \sum_{j\neq i}  x_{ij}^i \ , \   w^i + \sum_{j\neq i} x_{ji}^i - \sum_{j\neq i} x_{ij}^i \in \mathscr{X}^i \ ,  \ x_{ij}^i \leq x_{ij}^j \ , \ x_{ji}^i \leq x_{ji}^j\right\}
\end{align}

One could then remark that $\widehat{\mathcal{B}}^i(p,X^{-i})\subseteq \tilde{\mathcal{B}}^i(p)$, and that
\begin{align}
\label{eqn:18}
\hat{v}^{i}(p,X^{-i}) \leq \tilde{v}^{i}(p).
\end{align}

In the Arrow--Debreu model, this common price vector is set by an additional player, the auctioneer. One can deduce the following result.

\vskip 0.5cm

\begin{proposition}{\textbf{Optimality -- First Comparison\label{First Comparison}}\\}
For every price vector $p\in P$ and every $X^{-i}$, $\hat{v}^i(p,X^{-i})\leq\tilde{v}^i(p)$: at given prices, the bilateral constraints can only reduce the utility attainable by a consumer.
\end{proposition}

\vskip 0.5cm

\begin{definition}{\textbf{Definition of a Generalized Game\cite{Kreps2012}}\\}
An $n$-player generalized game $G=\{A^i,\mathcal{F}^i,u^i\}_{1\leq i \leq n}$ for a finite $n\in \N$ consists of, for $i=1,\hdots,n$,
\begin{enumerate}
\item A set of strategies $A^i$,

\item A constraint correspondence $\mathcal{F}^i \ : \ \prod_{j\neq i} A^j \to 2^{A^i}$,

\item A payoff function $u^i : \prod_{j=1}^n A^j \to \R$.
\end{enumerate}

A \textbf{Nash equilibrium} for this generalized game is a strategy profile $(a^i)_{1\leq i \leq n}\in \prod_{i=1}^n A^i$ such that, for $i=1,\hdots,n$
\begin{enumerate}
\item $a^i\in \mathcal{F}^i(a^{-i})$,

\item $a^i$ maximizes $u^i(\cdot,a^{-i})$ over $\mathcal{F}^i(a^{-i})$.
\end{enumerate}
\end{definition}

\vskip 0.5cm

The following proposition outlines the conditions under which a Nash equilibrium exists in a generalized game.

\vskip 0.5cm

\begin{proposition}{\textbf{Existence of Nash Equilibrium in Generalized Games \cite{Debreu1952}, \cite{Kreps2012}\label{Equilibrium in effective trade economy}}\\}
Suppose that $G=\{A^i,\mathcal{F}^i,u^i\}_{1\leq i \leq n}$ is a generalized game for which

\begin{enumerate}
\item Each $A^i$ is a non-empty, compact, convex subset of a Euclidean space,

\item Each $\mathcal{F}^i$ is a continuous, non-empty valued, and convex-valued correspondence,

\item Each $u^i$ is jointly continuous in the full vector of actions and quasi-concave in $a^i$ for each fixed $a^{-i}$.
\end{enumerate}

\noindent Then $G$ has a Nash equilibrium.
\end{proposition}

\vskip 0.5cm

For a fixed price profile $p$, $\mathcal{B}^i(p,\cdot)$ is continuous, and Proposition \ref{Equilibrium in effective trade economy} applies to the quantity game $\{\mathbb{X}^i, \mathcal{B}^i(p,\cdot), u^i\}_{1\leq i \leq n}$. Prices and quantities are determined jointly through the best-response correspondence
\begin{align}
\label{eqn:19}
\Phi(p,X)=\prod_{i=1}^n p^i(p^{-i},X^{-i})\times X^i(p,X^{-i}).
\end{align}
We call \textbf{transaction equilibrium} any fixed point $(p,X)\in\Phi(p,X)$. If $(p,X)$ is a transaction equilibrium, then $u^i(x^i)=v^i(p^i,p^{-i},X^{-i})=\max_{q\in P}\max_{Y\in\mathcal{B}^i((q,p^{-i}),X^{-i})}u^i(y)$ for every $i$: transaction equilibria are exactly the Nash equilibria of the generalized game associated with problem \ref{eqn:9}, in which each consumer chooses its price and its transactions jointly.

\vskip 0.5cm

One notable transaction equilibrium is the \textbf{autarky situation}: $X = (X^1, \dots, X^n) = 0$, for all $p = (p^1, \dots, p^n) \in P^n$. Let $x \sim (p, X)$ denote an allocation $x = (x^1, \dots, x^n)$ derived from the realizable price-quantity matrix $(p, X)$. Based on Remark \ref{HomegeneityRemark}, the following can be deduced.

\vskip 0.5cm

\begin{corollary}{\textbf{Homogeneity at Equilibrium\label{Homogeneity at equilibrium}}\\}
If $(p,X)$ is a transaction equilibrium of the economy, then, for every $\lambda>0$, $(\lambda p,X)$ is a transaction equilibrium of the same economy in which prices are normalized by $\sum_{k=1}^L p^i_k=\lambda$ instead of \ref{eqn:1}.
\end{corollary}

\vskip 0.5cm

The structure of transaction equilibria is given by the following results.

\vskip 0.5cm

\begin{theorem}{\textbf{Specialization of sellers\label{Specialization}}\\}
Under assumptions \emph{b}, \emph{d} and \emph{e}, at every transaction equilibrium each consumer sells at most one good: for every $i$, there exists $k_i\in\{1,\dots,L\}$ such that $x^i_{ij,\ell}=0$ for all $j\neq i$ and all $\ell\neq k_i$.
\end{theorem}

\vskip 0.5cm

\begin{proof}
We use the notation of Remark \ref{QuasiConvexValue}. Let $(p,X)$ be a transaction equilibrium and suppose that consumer $i$ sells at least two goods. Let $k$ be such that $S^i_k(X^i)=\max_m S^i_m(X^i)$, and let $\ell\neq k$ with $S^i_\ell(X^i)>0$. Define $\widetilde X^i$ by setting $\widetilde x^i_{ij,\ell}=0$ for all $j\neq i$, all other components unchanged, and $\widetilde p^i=e_k$. By \emph{d}, $\widetilde X^i\in\mathbb{X}^i$, and the bilateral constraints hold since trades were only reduced. Since $p^i\in P$,
\begin{align*}
c^i(\widetilde X^i)=c^i(X^i)\le p^i\cdot S^i(X^i)\le S^i_k(X^i)=\widetilde p^i\cdot S^i(\widetilde X^i),
\end{align*}
so the budget constraint holds at $(\widetilde p^i,\widetilde X^i)$. The new final holding $\widetilde x^i=x^i+S^i_\ell(X^i)\,e_\ell$ belongs to $\mathscr{X}^i$ by \emph{e}, and $u^i(\widetilde x^i)>u^i(x^i)$ by \emph{b}. This contradicts the fact that, at a transaction equilibrium, $(p^i,X^i)$ maximizes $u^i$ jointly in price and transactions.
\end{proof}

\vskip 0.5cm

\begin{proposition}{\textbf{Positive prices of traded goods\label{PositivePrices}}\\}
Under assumptions \emph{b}, \emph{d} and \emph{e}, at every transaction equilibrium, every good actually traded has a positive price: if $x^i_{ij,k}>0$, then $p^i_k>0$. Consequently, condition (C) holds for every consumer and, under assumption \emph{f} in addition, $\mathcal{B}^i$ is continuous at every transaction equilibrium.
\end{proposition}

\vskip 0.5cm

\begin{proof}
Suppose that $x^i_{ij,k}>0$ and $p^i_k=0$. Removing this sale, with the same price $p^i$, leaves the value of sales and the budget unchanged, is admissible by \emph{d}, and increases the final holding of good $k$, which remains in $\mathscr{X}^i$ by \emph{e}. By \emph{b}, it strictly increases $u^i$, a contradiction. At a transaction equilibrium, bilateral proposals coincide (Remark \ref{MaximizationEquivalence}), so that every good offered to $i$ is actually traded and therefore offered at a positive price: this is condition (C). The continuity of $\mathcal{B}^i$ follows from the lower semicontinuity result above.
\end{proof}

\vskip 0.5cm

At every transaction equilibrium, part 2 of Berge's theorem \ref{Berge's Maximum Theorem} therefore applies: the value function $v^i$ is continuous and the demand correspondence is upper semicontinuous at the equilibrium point, which validates the local analysis of equilibria.

\vskip 0.5cm

Theorem \ref{Specialization} describes an endogenous division of labor: each consumer sells a single good and acquires the others through exchange. Given a specialization profile $\mathbf{k}=(k_1,\dots,k_n)\in\{1,\dots,L\}^n$, we denote by $\mathcal{E}_{\mathbf{k}}$ the economy $\mathcal{E}$ in which the admissible transactions of consumer $i$ are restricted to
\begin{align}
\label{eqn:20}
\mathbb{X}^i_{\mathbf{k}}=\left\{X^i\in\mathbb{X}^i \ \mid \ x^i_{ij,\ell}=0 \quad \forall j\neq i,\ \forall \ell\neq k_i\right\}.
\end{align}

We will denote by $\mathbf{E}$ the set of transaction equilibria of the economy $\mathcal{E}$, by $\mathbf{E}_{\mathbf{k}}$ that of $\mathcal{E}_{\mathbf{k}}$, and $\mathcal{B}=(\mathcal{B}^1,\hdots,\mathcal{B}^n)$. We have the following result.

\vskip 0.5cm

\begin{proposition}{\textbf{Existence and Structure of $\mathbf{E}$ \label{Compactness of equilibria}}\\}
\begin{enumerate}
\item Under assumption \emph{a}, the set $\mathbf{E}$ of transaction equilibria is non-empty.
\item Under assumptions \emph{a}--\emph{e}, $(p,X)\in\mathbf{E}$ if and only if there exists $\mathbf{k}$ such that $(p,X)\in\mathbf{E}_{\mathbf{k}}$ and no consumer can gain by changing its specialization.
\end{enumerate}
\end{proposition}

\vskip 0.5cm

\begin{proof}
\begin{enumerate}
\item The autarky profile $(p,0)$ is a transaction equilibrium for every $p\in P^n$: if $X^{-i}=0$, the bilateral constraints force $X^i=0$, and $v^i$ is then constant in $p^i$. Hence $\mathbf{E}\neq\emptyset$. No convexity of the graph of $\mathcal{B}^i$ is required.

\item By Theorem \ref{Specialization}, every $(p,X)\in\mathbf{E}$ belongs to some $\mathbf{E}_{\mathbf{k}}$, and no consumer gains by deviating, in particular by changing its specialization. Conversely, let $(p,X)\in\mathbf{E}_{\mathbf{k}}$ be such that no consumer gains by changing its specialization. By the argument of Theorem \ref{Specialization}, any profitable deviation in $\mathcal{E}$ can be replaced by a profitable deviation selling a single good, which is either feasible in $\mathcal{E}_{\mathbf{k}}$ or a change of specialization; neither exists, so $(p,X)\in\mathbf{E}$.
\end{enumerate} 
\end{proof}

\vskip 0.5cm

\begin{remark}{\ }\\
Part 1 guarantees existence, but the equilibrium obtained may be autarky. By part 2, an equilibrium with trade in $\mathcal{E}$ is an equilibrium with trade in some $\mathcal{E}_{\mathbf{k}}$ at which no consumer gains by changing its specialization; Proposition \ref{TradeEquilibrium} below gives a sufficient condition for its existence. In the numerical example of Section \ref{Numerical simulation}, all equilibria with trade in Tables 2 and 3 are specialized. Note also that the set $\mathbf{E}$ need not be closed: along a sequence of equilibria, the price of a traded good may converge to zero, and the limit profile is then not an equilibrium, by Proposition \ref{PositivePrices}.
\end{remark}

\vskip 0.5cm

Equilibria with trade can be constructed along cycles of exchange. A \textbf{trade cycle} is a sequence of $r\ge2$ distinct consumers $i_1,\dots,i_r$, with $i_0=i_r$, together with goods $k_1,\dots,k_r$, consumer $i_s$ selling good $k_s$ to its successor $i_{s+1}$ and buying good $k_{s-1}$ from its predecessor $i_{s-1}$, with $i_{r+1}=i_1$ and $k_0=k_r$. For a trade level $t\ge0$, let $g_s(t)=u^{i_s}\big(w^{i_s}-t\,e_{k_s}+t\,e_{k_{s-1}}\big)$ be the utility of consumer $i_s$ when it sells $t$ units of its good and buys $t$ units of the good of its predecessor, and let $T_s=\min(w^{i_s}_{k_s},\bar\omega)$. We say that the cycle has \textbf{gains from trade} if:
\begin{enumerate}
\item[(G)] for every $s$, there exists $t\in(0,T_s]$ such that $w^{i_s}-t\,e_{k_s}+t\,e_{k_{s-1}}\in\mathscr{X}^{i_s}$ and $g_s(t)>g_s(0)$.
\end{enumerate}

\vskip 0.5cm

\begin{proposition}{\textbf{Existence of an equilibrium with trade\label{TradeEquilibrium}}\\}
Under assumptions \emph{a}--\emph{d}, if the economy has a trade cycle with gains from trade, then it has a transaction equilibrium with positive trade. More precisely, let $\tau_s$ maximize $g_s$ over the feasible levels $t\in[0,T_s]$, and let $t^\star=\min_s\tau_s>0$. The profile in which every consumer $i_s$ of the cycle sets $p^{i_s}=e_{k_s}$, sells $t^\star$ units of good $k_s$ to $i_{s+1}$ and buys $t^\star$ units of good $k_{s-1}$ from $i_{s-1}$, all other proposals being zero, is a transaction equilibrium.
\end{proposition}

\vskip 0.5cm

\begin{proof}
The set of feasible levels of consumer $i_s$ is a compact interval $[0,\bar T_s]$, since $\mathscr{X}^{i_s}$ is convex and closed, and $g_s$ is continuous, so that $\tau_s$ exists; by (G), $g_s(\tau_s)>g_s(0)$, hence $\tau_s>0$ and $t^\star>0$. Since $t^\star\le\tau_s$, the bundle $w^{i_s}-t^\star e_{k_s}+t^\star e_{k_{s-1}}$ is feasible, by convexity of $\mathscr{X}^{i_s}$, and the plan belongs to $\mathbb{X}^{i_s}$ by \emph{d}, since $t^\star\le\bar\omega$. Bilateral proposals coincide, and, since every price vector puts all its weight on the good sold, the budget constraint of $i_s$ reads $t^\star\le t^\star$. The profile is therefore feasible.

Consider a deviation $(p^{i_s},X^{i_s})$ of consumer $i_s$. By the bilateral constraints, it can only sell good $k_s$ to $i_{s+1}$, in a quantity $\sigma\le t^\star$, and buy good $k_{s-1}$ from $i_{s-1}$, in a quantity $q\le t^\star$, all other proposals of its partners being zero. Its budget constraint reads $q\le p^{i_s}_{k_s}\sigma\le\sigma$. For a given $q$, selling more than $q$ only reduces its holding of good $k_s$, so that, by \emph{b}, its utility is at most $g_s(q)$, with $0\le q\le t^\star$. Since $g_s$ is quasi-concave by \emph{c}, and $q\le t^\star\le\tau_s$, we have $g_s(t^\star)\ge\min\big(g_s(q),g_s(\tau_s)\big)=g_s(q)$, since $g_s(\tau_s)\ge g_s(q)$. Hence no deviation is profitable for the consumers of the cycle. The other consumers receive no proposal, so that, as in the autarky situation, they cannot trade, and their utility does not depend on their price. The profile is a transaction equilibrium.
\end{proof}

\vskip 0.5cm

Proposition \ref{TradeEquilibrium} shows that the coordination failure behind autarky can be overcome along any cycle of consumers who gain from exchanging their goods one for one, which is the rate at which goods are exchanged at the prices $e_{k_s}$. The trade level is set by the consumer of the cycle who wishes to trade the least, the others being rationed, in accordance with the bilateral nature of the constraints. Condition (G) holds, for instance, when $\partial_{k_{s-1}}u^{i_s}(w^{i_s})>\partial_{k_s}u^{i_s}(w^{i_s})$ for every $s$, under assumption \emph{a'}, and $w^{i_s}$ lies in the interior of $\mathscr{X}^{i_s}$. Several cycles, or a single cycle with different trade levels, yield several equilibria with trade, and the equilibrium is therefore not unique.

\vskip 0.5cm

We recall next the definition of Pareto optimal allocation.

\vskip 0.5cm

\begin{definition}{\textbf{Pareto Optimal Allocation \cite{Starr2012}}\\}
Let $\mathcal{A}$ be the set of allocations $x\sim(p,X)$ induced by a profile $(p,X)\in P^n\times\prod_i\mathbb{X}^i$ satisfying the budget constraints, that is, $X^i\in\mathcal{B}^i(p,X^{-i})$ for every $i$. An allocation $x^{\star}\in\mathcal{A}$ is \textbf{Pareto optimal} if and only if
\begin{align*}
\not\exists\, x\in\mathcal{A} \text{ such that } u^i(x^i) \geq u^i(x^{i,\star}) \, \forall i \, \text{and } u^j(x^j) > u^j(x^{j,\star}) \text{ for at least one } j.
\end{align*}
\end{definition}

\vskip 0.5cm

Due to the autarky situation, and as illustrated by the numerical example in Section \ref{Numerical simulation}, one can deduce the following result.

\vskip 0.5cm

\begin{proposition}{\ }\\
The analogue of the first welfare theorem does not hold: transaction equilibria need not be Pareto optimal. In particular, whenever autarky is not Pareto optimal, the autarky equilibrium is an inefficient transaction equilibrium.
\end{proposition}

\vskip 0.5cm

\begin{definition}{\label{Nash--Pareto equilibria}\ }\\
Let $\mathfrak{P}(A)$ denote the set of Pareto optimal allocations within an allocation set $A$. We define $\mathfrak{P}(\mathbf{E})$ as the set of \textbf{Nash--Pareto optimal} allocations, also called \textbf{Nash--Pareto equilibria}: equilibrium allocations that are not Pareto dominated by any other equilibrium allocation.
\end{definition}

\vskip 0.5cm

Every equilibrium allocation that is Pareto optimal is Nash--Pareto optimal, that is, $\mathbf{E}\cap\mathfrak{P}(\mathcal{A})\subseteq\mathfrak{P}(\mathbf{E})$. The converse inclusion does not hold in general: an equilibrium allocation may be undominated within $\mathbf{E}$ while being dominated by an allocation of $\mathcal{A}$ that is not an equilibrium allocation. We have the following result:

\vskip 0.5cm

\begin{proposition}{\label{Second welfare theorem}\ }\\
A Pareto optimal allocation need not be reachable through trade from the initial endowments: it need not be a transaction equilibrium allocation.
\end{proposition}

\vskip 0.5cm

A counter-example can be found in Section \ref{Numerical simulation}. Moreover, the set of transaction equilibrium allocations and the set of Pareto optimal allocations may be disjoint, as in the economy with topological constraints of Table 3. In the present setting, redistributing endowments would make exchange unnecessary; the relevant question is therefore whether Pareto optimal allocations can be reached through trade, which the numerical example answers negatively.

\vskip 0.5cm

The two welfare theorems therefore fail in the present setting, in the following precise sense: transaction equilibria need not be Pareto optimal, and Pareto optimal allocations need not be reachable through trade from the initial endowments.

\vskip 0.5cm

\section{KKT characterization and sensitivity analysis}

\hskip 0.5cm We next characterize the optimal transactions of a consumer in a smooth economy, taking into account how the proposals of its partners respond to its own transactions. This sensitivity analysis describes the propagation of a change in one bilateral transaction through the network of transaction constraints; the Nash case, in which the proposals of the partners are taken as given, is obtained as a particular case. We make the following assumptions:
\begin{enumerate}
\item[\emph{a'}] \textbf{Differentiability:} $u^i(\cdot)$ is continuously differentiable on $\mathscr{X}^i$.
\item[\emph{a''}] \textbf{Response of the partners:} for given prices $p$, the proposals of the other consumers to consumer $i$ are continuously differentiable functions of the transactions $X^i$ of consumer $i$, denoted $y_{ij,k}(X^i)=x^j_{ij,k}$ and $y_{ji,k}(X^i)=x^j_{ji,k}$, and a constraint qualification (for instance, linear independence of the gradients of the active constraints) holds at the optimum.
\end{enumerate}

\noindent The functions $y$ may be interpreted as the best responses of the partners of consumer $i$, or as the conjectures of consumer $i$ about these responses. For simplicity, we assume that the bounds defining $\mathscr{X}^i$ and $\mathbb{X}^i$ are not binding, so that the only constraints on the final holding and on the transactions are their non-negativity. We fix the price profile $p$, drop the superscript $i$ on the variables of consumer $i$, writing $x$ for $x^i$, $x_{ij,k}$ for $x^i_{ij,k}$ and $x_{ji,k}$ for $x^i_{ji,k}$, and write $Y$ for the vector of the bounds $y_{ij,k}$ and $y_{ji,k}$. Define the Lagrangian function $\mathcal{L}(X, \lambda, \mu, \eta, \theta)$ as
\begin{align}
\label{eqn:21}
\begin{split}
\mathcal{L}(X, \lambda, \mu, \eta, \theta) &= u^i(x) + \lambda \left( \sum_{k=1}^L p^i_k \sum_{j \neq i} x_{ij,k} - \sum_{k=1}^L \sum_{j \neq i} p^j_k x_{ji,k} \right) + \sum_{k=1}^L \sum_{j \neq i} \mu_{ij,k} x_{ij,k} + \sum_{k=1}^L \sum_{j \neq i} \mu_{ji,k} x_{ji,k} \\
& + \sum_{k=1}^L \sum_{j \neq i} \eta_{ij,k}\big(y_{ij,k}(X) - x_{ij,k}\big) + \sum_{k=1}^L \sum_{j \neq i} \eta_{ji,k}\big(y_{ji,k}(X) - x_{ji,k}\big) \\
&+ \sum_{k=1}^L \theta_{k} \left( w^i_{k}+\sum_{j \neq i} x_{ji,k}-\sum_{j \neq i} x_{ij,k} \right).
\end{split}
\end{align}

Under assumptions \emph{a'} and \emph{a''}, every solution satisfies the following Karush--Kuhn--Tucker conditions. Feasibility and sign conditions read, for all $j\neq i$ and all $k$,
\begin{align}
\label{eqn:22}
\begin{split}
& 0 \leq X \leq Y \ , \ 0 \leq p^i \cdot \sum_{j \neq i} x_{ij} - \sum_{j \neq i} p^j \cdot x_{ji}  \ , \ 0\leq w^i+\sum_{j \neq i} x_{ji}-\sum_{j \neq i} x_{ij} \ , \\
& \ \lambda \ , \ \mu_{ij,k} \ , \ \mu_{ji,k} \ , \ \eta_{ij,k} \ , \ \eta_{ji,k} \ , \ \theta_k \geq 0,
\end{split}
\end{align}
and the complementary slackness conditions read, for all $j\neq i$ and all $k$,
\begin{align}
\label{eqn:23}
\begin{split}
\lambda \left( \sum_{k=1}^L p^i_k \sum_{j \neq i} x_{ij,k} - \sum_{k=1}^L \sum_{j \neq i} p^j_k x_{ji,k} \right) &=0 \ , \ \mu_{ij,k}x_{ij,k}=0 \ , \ \mu_{ji,k}x_{ji,k}=0 \ , \\
\eta_{ij,k} (y_{ij,k}-x_{ij,k})=0 \ , \ \eta_{ji,k} (y_{ji,k}-x_{ji,k})&=0 \ , \ \theta_{k} \left(w^i_{k}+ \sum_{j \neq i} x_{ji,k}-\sum_{j \neq i} x_{ij,k}\right)=0 .
\end{split}
\end{align}

We have the following stationarity conditions, where $\partial_k u^i$ denotes the partial derivative of $u^i$ with respect to good $k$ at $x$:
\begin{align}
\label{eqn:24}
\frac{\partial \mathcal{L}}{\partial x_{ij,k}} &= -\partial_k u^i + \lambda p^i_k + \mu_{ij,k} - \eta_{ij,k} - \theta_{k} + \sum_{l=1}^L \sum_{h \neq i} \eta_{ih,l} \frac{\partial y_{ih,l}}{\partial x_{ij,k}}  + \sum_{l=1}^L \sum_{h \neq i} \eta_{hi,l}\frac{\partial y_{hi,l}}{\partial x_{ij,k}} =0,\\
\label{eqn:25}
\frac{\partial \mathcal{L}}{\partial x_{ji,k}} &= \partial_k u^i - \lambda p^j_k + \mu_{ji,k} - \eta_{ji,k} + \theta_{k} + \sum_{l=1}^L \sum_{h \neq i} \eta_{ih,l} \frac{\partial y_{ih,l}}{\partial x_{ji,k}}  + \sum_{l=1}^L \sum_{h \neq i} \eta_{hi,l}\frac{\partial y_{hi,l}}{\partial x_{ji,k}} =0 \, .
\end{align}

By Remark \ref{MaximumCharacterization}, the budget constraint binds at the optimum. Moreover, under assumption \emph{b}, $\partial_k u^i>0$, so that a positive purchase that is neither rationed nor subject to indirect effects implies $\lambda>0$. Conditions \ref{eqn:24}--\ref{eqn:25} can be written
\begin{align*}
\partial_k u^i &= \lambda p^i_k + \mu_{ij,k} - \eta_{ij,k} - \theta_{k} + \sum_{l=1}^L \sum_{h \neq i} \eta_{ih,l} \frac{\partial y_{ih,l}}{\partial x_{ij,k}}  + \sum_{l=1}^L \sum_{h \neq i} \eta_{hi,l}\frac{\partial y_{hi,l}}{\partial x_{ij,k}}, \\
\partial_k u^i &= \lambda p^j_k - \mu_{ji,k} + \eta_{ji,k} - \theta_{k} - \sum_{l=1}^L \sum_{h \neq i} \eta_{ih,l} \frac{\partial y_{ih,l}}{\partial x_{ji,k}} - \sum_{l=1}^L \sum_{h \neq i} \eta_{hi,l}\frac{\partial y_{hi,l}}{\partial x_{ji,k}}  \, .
\end{align*}

Subtracting these two equalities yields, for $i\neq j$ and every good $k$, the \textbf{price--quantity equation}:
\begin{align}
\label{eqn:26}
\begin{split}
\lambda (p^i_k-p^j_k) &= \eta_{ij,k} + \eta_{ji,k} - (\mu_{ij,k} + \mu_{ji,k}) \\
&- \sum_{l=1}^L \sum_{h \neq i} \eta_{ih,l} \left( \frac{\partial y_{ih,l}}{\partial x_{ij,k}} + \frac{\partial y_{ih,l}}{\partial x_{ji,k}} \right) - \sum_{l=1}^L \sum_{h \neq i} \eta_{hi,l}\left(\frac{\partial y_{hi,l}}{\partial x_{ij,k}} + \frac{\partial y_{hi,l}}{\partial x_{ji,k}} \right).
\end{split}
\end{align}

\vskip 0.5cm

\textbf{Interpretation:}
\begin{enumerate}

\item \textbf{Shadow value of the bilateral price differential:} The difference $p^i_k-p^j_k$ represents the price differential for good $k$ between agents $i$ and $j$. The multiplier $\lambda$ is the shadow value associated with agent $i$'s trade-balance constraint
\begin{align*}
\sum_{k=1}^L p^i_k\sum_{j\neq i}x_{ij,k} - \sum_{k=1}^L\sum_{j\neq i}p^j_kx_{ji,k},
\end{align*}
that is, the gain in utility from one additional unit of value obtained from sales. Consequently, $\lambda(p^i_k-p^j_k)$ represents the marginal shadow value of the price differential between the two sides of the $i$--$j$ transaction for good $k$. It measures the contribution, in objective-function units, of the differential between valuing an outgoing unit at $p^i_k$ and an incoming unit at $p^j_k$, holding the remaining components fixed.

\item \textbf{Direct shadow values of bilateral quantity constraints:} The multipliers $\eta_{ij,k}$ and $\eta_{ji,k}$ are the shadow values associated with the upper-bound constraints $x_{ij,k}\leq y_{ij,k}$ and $x_{ji,k}\leq y_{ji,k}$, respectively. A positive $\eta_{ji,k}$ means that consumer $i$ would like to buy more of good $k$ from $j$ but is rationed by $j$'s supply; a positive $\eta_{ij,k}$ means that $i$ would like to sell more of good $k$ to $j$ but is rationed by $j$'s demand. These multipliers are the bilateral counterparts of the quantity signals of non-Walrasian equilibrium theory. The multipliers $\mu_{ij,k}$ and $\mu_{ji,k}$ are associated with the non-negativity constraints $x_{ij,k}\geq0$ and $x_{ji,k}\geq0$ and are positive only when the corresponding flow is at its lower boundary. Thus,
\begin{align*}
\eta_{ij,k}+\eta_{ji,k}
-
\left(\mu_{ij,k}+\mu_{ji,k}\right)
\end{align*}
represents the net direct contribution of the bilateral quantity bounds to the marginal condition for good $k$. The $\eta$-terms capture the shadow value of relaxing the upper trading limits, whereas the $\mu$-terms capture the shadow effects associated with the non-negativity boundaries.

\item \textbf{Indirect effects through the transaction network:} The quantities $y_{ih,l}$ and $y_{hi,l}$ represent the proposed transactions of agent $h$, as functions of agent $i$'s transaction vector $X^i$. Hence,
\begin{align*}
\frac{\partial y_{ih,l}}{\partial x_{ij,k}}, \qquad \frac{\partial y_{ih,l}}{\partial x_{ji,k}}, \qquad \frac{\partial y_{hi,l}}{\partial x_{ij,k}}, \qquad \frac{\partial y_{hi,l}}{\partial x_{ji,k}}
\end{align*}
measure the marginal response of these proposed transactions to a change in the $i$--$j$ transaction of good $k$. Weighted by the corresponding multipliers $\eta_{ih,l}$ and $\eta_{hi,l}$, these derivatives capture the indirect effects of changing the $i$--$j$ transaction on other bilateral feasibility constraints. Accordingly, the terms
\begin{align*}
&-\sum_{l=1}^L\sum_{h\neq i} \eta_{ih,l} \left( \frac{\partial y_{ih,l}}{\partial x_{ij,k}} + \frac{\partial y_{ih,l}}{\partial x_{ji,k}} \right) - \sum_{l=1}^L\sum_{h\neq i} \eta_{hi,l} \left( \frac{\partial y_{hi,l}}{\partial x_{ij,k}} + \frac{\partial y_{hi,l}}{\partial x_{ji,k}} \right)
\end{align*}
represent the indirect marginal effects generated by the dependence of other agents' proposed transactions on agent $i$'s bilateral flow. These terms capture the propagation of a marginal change in one bilateral transaction through the system of transaction constraints. Only the constraints that are binding, $\eta>0$, transmit such effects: a partner whose proposal does not constrain agent $i$ does not affect its marginal conditions, whatever its response.

\item \textbf{Economic meaning of the price--quantity equation:} Equation \ref{eqn:26} can therefore be written as
\begin{align*}
\underbrace{\lambda(p^i_k-p^j_k)} _{\text{shadow value of the price differential}} &= \underbrace{\eta_{ij,k}+\eta_{ji,k} - (\mu_{ij,k}+\mu_{ji,k})}_{\text{direct shadow effects of bilateral bounds}}\\
&- \underbrace{\sum_{l=1}^L\sum_{h\neq i} \eta_{ih,l} \left( \frac{\partial y_{ih,l}}{\partial x_{ij,k}} + \frac{\partial y_{ih,l}}{\partial x_{ji,k}} \right) - \sum_{l=1}^L\sum_{h\neq i} \eta_{hi,l} \left( \frac{\partial y_{hi,l}}{\partial x_{ij,k}} + \frac{\partial y_{hi,l}}{\partial x_{ji,k}} \right) }_{\text{indirect effects through other transaction constraints}}.
\end{align*}

Thus, at an optimum, the shadow value associated with the price differential between agents $i$ and $j$ is balanced by the direct and indirect marginal effects generated by the transaction constraints. The first component reflects the marginal contribution of the bilateral price differential to the trade-balance constraint; the second captures the direct shadow values of the bilateral quantity bounds; and the third captures the effects transmitted through the dependence of other proposed transactions on the $i$--$j$ transaction.\\

In this sense, the marginal condition states that the shadow-valued bilateral price differential, $\lambda(p^i_k-p^j_k)$, is absorbed through two margins: the first captures the adjustment of the bilateral transaction itself, while the second captures the adjustment of the remaining transactions in the network induced by this change. Hence, a bilateral price differential translates into a marginal reallocation between the adjustment of the individual transaction and the adjustment of other agents' transactions in the market.
\end{enumerate}

\vskip 0.5cm

\subsection*{The Nash case}

At a transaction equilibrium, as defined in Section \ref{New way to consider the Consumer Problem}, each consumer takes the proposals of its partners as given: the derivatives of $y$ with respect to $X^i$ vanish, the constraints are linear in $X^i$, and the linearity constraint qualification holds without further assumption. The indirect terms then disappear, and \ref{eqn:24}--\ref{eqn:25} reduce to
\begin{align}
\label{eqn:27}
\partial_k u^i = \lambda p^i_k + \mu_{ij,k} - \eta_{ij,k} - \theta_{k}, \qquad \partial_k u^i = \lambda p^j_k - \mu_{ji,k} + \eta_{ji,k} - \theta_{k}.
\end{align}
The interdependence between agents is then captured by the fixed point, which determines the bounds $y_{ij,k}=x^j_{ij,k}$ and $y_{ji,k}=x^j_{ji,k}$ at equilibrium, rather than by the marginal conditions of each agent. Conditions \ref{eqn:27} yield the following properties, in which consumer $i$ keeps a positive quantity of the goods considered, so that the corresponding $\theta$ vanish, and $\lambda>0$.
\begin{enumerate}
\item \textbf{Reservation prices.} Let $r^i_k=\partial_k u^i/\lambda$ be the reservation price of consumer $i$ for good $k$. If $i$ buys good $k$ from $j$, then $r^i_k=p^j_k+\eta_{ji,k}/\lambda\ge p^j_k$; if $i$ sells good $k$ to $j$, then $r^i_k=p^i_k-\eta_{ij,k}/\lambda\le p^i_k$, with equality in both cases when the trade is not rationed. Consequently, whenever $j$ sells good $k$ to $i$ at a transaction equilibrium,
\begin{align}
\label{eqn:28}
r^j_k\ \le\ p^j_k\ \le\ r^i_k :
\end{align}
each bilateral exchange takes place at a price lying between the reservation prices of the seller and of the buyer, and the price chosen by the seller determines how the bilateral gains from trade are shared.

\item \textbf{Marginal rates of substitution.} If $i$ buys good $k$ from $j$ and sells good $m$ to $h$, both trades being positive and not rationed, then
\begin{align}
\label{eqn:29}
\frac{\partial_k u^i}{\partial_m u^i}=\frac{p^j_k}{p^i_m}.
\end{align}
When all sellers charge the same price vector and no consumer is rationed, this reduces to the Arrow--Debreu condition. In general, marginal rates of substitution differ across consumers, which is the marginal counterpart of the failure of the first welfare theorem.

\item \textbf{Law of one price up to rationing.} If $i$ buys good $k$ from two sellers $j$ and $h$, both purchases being positive, then
\begin{align}
\label{eqn:30}
\lambda\,(p^j_k-p^h_k)=\eta_{ji,k}-\eta_{hi,k}:
\end{align}
a consumer buys the same good from two sellers at different prices only if it is rationed by the cheaper one.

\item \textbf{Price--quantity equation.} Equation \ref{eqn:26} reduces to $\lambda(p^i_k-p^j_k)=\eta_{ij,k}+\eta_{ji,k}-(\mu_{ij,k}+\mu_{ji,k})$: the price differential is absorbed by the direct shadow effects of the bilateral bounds only. By Theorem \ref{Specialization}, a consumer sells at most one good at equilibrium, so that two-way trades of the same good are exceptional.
\end{enumerate}

\vskip 0.5cm

\begin{remark}{\ }\\
When the functions $y$ are interpreted as conjectures of consumer $i$ about the responses of its partners, the general conditions \ref{eqn:24}--\ref{eqn:26} characterize an equilibrium with conjectural variations, which differs from the transaction equilibrium and coincides with it when the conjectured responses vanish. When they are interpreted as the actual best responses of the partners, their derivatives can be computed from the partners' own KKT conditions by the implicit function theorem, and \ref{eqn:26} measures the sensitivity of the transaction network to a marginal change in one bilateral flow. In both cases, prices are taken as given in this section: they are not characterized by first-order conditions, since $v^i$ is quasi-convex in $p^i$ (Remark \ref{QuasiConvexValue}), and their structure at equilibrium is given by Theorem \ref{Specialization} and Proposition \ref{PositivePrices}.
\end{remark}

\vskip 0.5cm

\section{Convergence\label{Convergence}}

\hskip 0.5cm Next, we propose two adjustment processes, based on two underlying principles:
\begin{enumerate}
\item A planner who chooses both the trades and the prices, in order to maximize the social utility of the community while respecting the budget constraints of all consumers. At given prices, a gradient-based algorithm, representing an exchange dynamic driven by agents' collective needs and desires, converges to the trades that the planner would select; the planner then adjusts the prices in the direction that increases the social utility. The resulting allocation serves as a benchmark for decentralized exchange;
\item A non-t\^atonnement adjustment process in which each agent revises its prices as long as further improvement is attainable, which provides a dynamic counterpart of the specialization of sellers.
\end{enumerate}
We then discuss the joint dynamics, in which all agents adjust simultaneously. Throughout this section, we make the following assumption:
\begin{enumerate}
\item[\emph{c}"] \textbf{Concavity:} $u^i(\cdot)$ is concave on $\mathscr{X}^i$.
\end{enumerate}

\vskip 0.5cm

\subsection{The planner's problem\label{PlannerProblemSection}}

\hskip 0.5cm Suppose the economy starts in the autarky situation $X=0$, or in any other non-equilibrium situation. By Remark \ref{MaximizationEquivalence}, bilateral proposals coincide at equilibrium; we therefore describe the transactions by a matrix of matched trades $q=(q_{ij})_{i\neq j}$, where $q_{ij}\in\R_+^L$ is the quantity sent by consumer $i$ to consumer $j$, so that the final holding of consumer $i$ is $x^i(q)=w^i+\sum_{j\neq i}q_{ji}-\sum_{j\neq i}q_{ij}$, and its sales are $S^i(q)=\sum_{j\neq i}q_{ij}$. The set of matched trades that are feasible at the price profile $p\in P^n$ is
\begin{align}
\label{eqn:31}
\mathcal{K}(p)=\Big\{ q\ge0 \ \mid \ (q_{ij},q_{ji})_{j\neq i}\in\mathbb{X}^i \ , \ \sum_{j\neq i} p^j \cdot  q_{ji} \leq p^i \cdot \sum_{j\neq i} q_{ij} \ , \ x^i(q) \in \mathscr{X}^i \ , \ \forall i \Big\}.
\end{align}
For fixed prices, all the constraints in \ref{eqn:31} are linear in $q$, so that $\mathcal{K}(p)$ is a non-empty (it contains autarky), convex and compact set. The planner evaluates the trades through the social utility function
\begin{align}
\label{eqn:32}
U(q)=\sum_{i=1}^n u^i\big(x^i(q)\big),
\end{align}
which is concave under assumption \emph{c}". The planner's problem is solved in two levels: the choice of the trades at given prices, and the choice of the prices.

\vskip 0.5cm

\subsubsection{Trades at given prices}

\hskip 0.5cm Fix the price profile $p$. A way to improve the situation of the community is to update the matched trades in the direction of increasing social utility (the gradient direction), without breaking the budget constraints (projected gradient). Under assumption \emph{a'}, the components of the gradient are
\begin{align}
\label{eqn:33}
\frac{\partial U}{\partial q_{ij,k}}(q)=\partial_k u^j\big(x^j(q)\big)-\partial_k u^i\big(x^i(q)\big):
\end{align}
the flow of good $k$ from $i$ to $j$ increases when the buyer values the good more than the seller at the margin. Such information can be observed bilaterally, in the magnitude of demand and supply expressed by the two partners of each exchange. The prices do not enter the direction \ref{eqn:33}, since they do not affect the final holdings; they enter the process through the feasible set $\mathcal{K}(p)$, on which the trades are projected. The process reads
\begin{align}
\label{eqn:34}
q_{t+1}=\pi_{\mathcal{K}(p)}\big(q_t+\mu_t\, g_t\big),\qquad g_t\in\partial U(q_t),
\end{align}
where $\pi_{\mathcal{K}(p)}$ is the Euclidean projection on $\mathcal{K}(p)$, $\partial U$ is the superdifferential of $U$, and the step sizes $\mu_t>0$ satisfy
\begin{align}
\label{eqn:35}
\sum_{t\ge0}\mu_t=+\infty,\qquad \sum_{t\ge0}\mu_t^2<+\infty.
\end{align}

\vskip 0.5cm

\begin{proposition}{\textbf{Trades at given prices\label{PlannerBenchmark}}\\}
Under assumptions \emph{a}, \emph{a'} and \emph{c}", and for a fixed price profile $p$, let $\hat U(p)=\max_{q\in\mathcal{K}(p)}U(q)$ and let $q^{\mathrm{best}}_t$ be the best iterate of \ref{eqn:34} up to time $t$. Then $U(q^{\mathrm{best}}_t)\to \hat U(p)$, and every limit point $\hat q$ of $(q^{\mathrm{best}}_t)$ maximizes $U$ over $\mathcal{K}(p)$. The induced allocation $x(\hat q)$ is Pareto optimal among the allocations that are feasible at the prices $p$.
\end{proposition}

\vskip 0.5cm

\begin{proof}
Since $\mathcal{K}(p)$ is compact and convex and $U$ is concave and continuously differentiable on it, its supergradients are bounded on $\mathcal{K}(p)$, and the convergence of the best values of the projected supergradient method under \ref{eqn:35} is classical \cite{Shor1998}. By compactness and continuity of $U$, every limit point of $(q^{\mathrm{best}}_t)$ attains $\hat U(p)$. Finally, if an allocation feasible at the prices $p$ gave every consumer at least the utility of $x(\hat q)$ and one consumer strictly more, it would yield a larger value of $U$, a contradiction.
\end{proof}

\vskip 0.5cm

\subsubsection{Choice of the prices}

\hskip 0.5cm The prices are not given to the planner: they are part of its decision. Two properties of the budget constraints determine how this choice must be formulated. First, the value of the purchases of each consumer is the value of the sales of its partners, so that, for every $p$ and every matched trade $q$,
\begin{align}
\label{eqn:36}
\sum_{i=1}^n\Big(p^i\cdot S^i(q)-\sum_{j\neq i}p^j\cdot q_{ji}\Big)=\sum_{i=1}^n p^i\cdot S^i(q)-\sum_{j=1}^n p^j\cdot \sum_{i\neq j}q_{ji}=0.
\end{align}
This is the analogue of Walras' law in the present setting: since every term of the sum \ref{eqn:36} is non-negative on $\mathcal{K}(p)$, \emph{all the budget constraints are binding} on $\mathcal{K}(p)$, and any one of them is implied by the others. Second, zero prices neutralize the budget constraints.

\vskip 0.5cm

\begin{remark}{\textbf{Zero prices\label{ZeroPricesPlanner}}\\}
Let $q$ be a matched trade, satisfying the constraints of \ref{eqn:31} other than the budget constraints, in which every consumer has a good $k_i$ that it does not sell. For $p^i=e_{k_i}$, every traded good has a zero price, the budget constraints read $0\le0$, and $q\in\mathcal{K}(p)$. A planner free to choose $p\in P^n$ can therefore turn trades into transfers, and the price system loses its role. This is excluded at a transaction equilibrium, where every offered good has a positive price (Proposition \ref{PositivePrices}).
\end{remark}

\vskip 0.5cm

For this reason, the planner chooses the prices in the set
\begin{align}
\label{eqn:37}
P_\pi=\left\{p\in P \ \mid \ p_k\ge\pi, \ k=1,\dots,L\right\},\qquad 0<\pi\le \frac{1}{L},
\end{align}
where the floor $\pi$ measures the extent to which the planner may use prices as an instrument of redistribution. The planner's problem reads
\begin{align}
\label{eqn:38}
V_\pi=\max_{p\in P_\pi^n}\hat U(p)=\max\left\{U(q) \ \mid \ p\in P_\pi^n, \ q\in\mathcal{K}(p)\right\}.
\end{align}

\vskip 0.5cm

\begin{proposition}{\textbf{Existence of a planner's optimum\label{PlannerExistence}}\\}
Under assumptions \emph{a} and \emph{c}":
\begin{enumerate}
\item For every $\pi\in(0,1/L]$, the problem \ref{eqn:38} admits a solution $(\hat p,\hat q)$, and $\pi\mapsto V_\pi$ is non-increasing.
\item Assume $L\ge2$ and, in addition, assumptions \emph{b}, \emph{d} and \emph{e}. Let $(p^\star,X^\star)$ be a transaction equilibrium, with matched trades $q^\star$, and let $\pi^\star=\min\{p^{i\star}_{k}\mid S^i_k(q^\star)>0\}$, with the convention $\min\emptyset=+\infty$. Then $V_\pi\ge U(q^\star)$ for every $\pi\le\bar\pi=\min\{\pi^\star/2,\,1/(2(L-1))\}$.
\end{enumerate}
\end{proposition}

\vskip 0.5cm

\begin{proof}
\begin{enumerate}
\item The set $\mathcal{G}_\pi=\{(p,q)\mid p\in P_\pi^n,\ q\in\mathcal{K}(p)\}$ is non-empty, since it contains autarky; it is closed, since all its constraints are continuous in $(p,q)$ and the sets $\mathbb{X}^i$ and $\mathscr{X}^i$ are closed; and it is bounded, since $P_\pi$ and the sets $\mathbb{X}^i$ are bounded. The continuous function $U$ therefore attains its maximum on the compact set $\mathcal{G}_\pi$; equivalently, by part 1 of Theorem \ref{Berge's Maximum Theorem}, $\hat U$ is upper semicontinuous and attains its maximum on $P^n_\pi$. Since $P_{\pi'}\subseteq P_\pi$ for $\pi\le\pi'$, $V_\pi$ is non-increasing in $\pi$.

\item By Remark \ref{MaximizationEquivalence}, bilateral proposals coincide at the equilibrium, so that $q^\star\in\mathcal{K}(p^\star)$. By Theorem \ref{Specialization}, every seller $i$ sells a single good $k_i$, and, by Proposition \ref{PositivePrices}, $p^{i\star}_{k_i}>0$, so that $\pi^\star>0$. A consumer that does not sell has no revenue, and therefore buys nothing, since every traded good has a positive price; its price vector enters no budget constraint, and we set it to $(1/L,\dots,1/L)$. For a seller, define
\begin{align*}
p^i_{k_i}=\frac{p^{i\star}_{k_i}}{2},\qquad p^i_k=\frac{1-p^i_{k_i}}{L-1}\quad (k\neq k_i),
\end{align*}
so that $p^i\in P$. Since consumer $i$ buys from $j$ only the good $k_j$, the budget constraint of $i$ involves only the prices $p^j_{k_j}$ of the goods actually sold; dividing all these prices by $2$ divides both sides of every budget constraint by $2$, and leaves them satisfied. The other constraints of \ref{eqn:31} do not depend on $p$, hence $q^\star\in\mathcal{K}(p)$. Finally, $p^i_{k_i}\ge\pi^\star/2$ and, since $p^i_{k_i}\le 1/2$, $p^i_k\ge 1/(2(L-1))$ for $k\neq k_i$, while $1/L\ge 1/(2(L-1))$ for $L\ge2$. Hence $p\in P^n_\pi$ for every $\pi\le\bar\pi$, and $V_\pi\ge U(q^\star)$.
\end{enumerate}
\end{proof}

\vskip 0.5cm

\noindent\textbf{Prices determined by trades.} Since all prices in $P_\pi$ are positive, a positive trade carries a positive value, and the binding budget constraints state that the value flows circulate. Let $G(q)$ be the trade graph, with an edge $i\to j$ whenever $q_{ij}\neq0$. In the specialized case, the planner's prices are then determined by the trades, through the stationary distribution of a Markov chain. If consumer $i$ sells the single good $k_i$ in quantity $s^i=S^i_{k_i}(q)>0$, let
\begin{align}
\label{eqn:39}
M_{ij}=\frac{q_{ij,k_i}}{s^i},\qquad \nu^i=p^i_{k_i}\,s^i,\qquad \nu^\top M=\nu^\top,
\end{align}
where $M_{ij}$ is the share of the sales of $i$ bought by $j$ and $\nu^i$ is the income of $i$.

\vskip 0.5cm

\begin{proposition}{\textbf{Circular trade and income\label{CircularTrade}}\\}
\begin{enumerate}
\item If $q\in\mathcal{K}(p)$ for some $p\in P^n_\pi$, every edge of $G(q)$ lies on a directed cycle: each connected component of $G(q)$ is strongly connected.
\item Conversely, let $L\ge2$ and let $q$ satisfy the constraints of \ref{eqn:31} other than the budget constraints, with every trading consumer selling a single good, and every connected component $C$ of $G(q)$ strongly connected. Then the prices $p$ with $p^i_{k_i}>0$ that satisfy the budget constraints are exactly those for which the incomes $\nu$ satisfy \ref{eqn:39}, that is,
\begin{align*}
p^i_{k_i}=\theta_C\,\frac{\bar\nu^i_C}{s^i},\qquad i\in C,
\end{align*}
where $\bar\nu_C>0$ is the unique stationary distribution of the irreducible stochastic matrix $M$ restricted to $C$, and $\theta_C>0$. In particular, $q\in\mathcal{K}(p)$ for some $p\in P^n_\pi$ when $\pi$ is small enough.
\end{enumerate}
\end{proposition}

\vskip 0.5cm

\begin{proof}
\begin{enumerate}
\item Let $a_{ij}=p^i\cdot q_{ij}$ be the value of the goods sent by $i$ to $j$; since $p\gg0$, $a_{ij}>0$ if and only if $i\to j$ is an edge of $G(q)$. By \ref{eqn:36}, the budget constraints bind: $\sum_j a_{ij}=\sum_j a_{ji}$ for all $i$. Suppose that the edge $i\to j$ lies on no cycle, and let $R$ be the set of consumers reachable from $j$, so that $i\notin R$. Summing the equalities over $R$, the flows inside $R$ cancel, no edge leaves $R$, and the edge $i\to j$ enters it: $0=\sum_{l\in R}\big(\sum_m a_{lm}-\sum_m a_{ml}\big)\le -a_{ij}<0$, a contradiction.

\item Consumer $i$ buys from $j$ only the good $k_j$, so that its binding budget constraint reads $p^i_{k_i}s^i=\sum_j p^j_{k_j}q_{ji,k_j}=\sum_j \nu^j M_{ji}$, which is \ref{eqn:39}. On a strongly connected component, $M$ is stochastic and irreducible, so that its stationary distribution is unique and positive \cite{Kemeny1983,Horn2013}, which gives the expression of the prices. Choosing $\theta_C$ small enough and distributing the remaining weight of each $p^i$ on the goods that $i$ does not sell yields $p\in P_\pi^n$ for $\pi$ small enough.
\end{enumerate}
\end{proof}

\vskip 0.5cm

\noindent Proposition \ref{CircularTrade} gives the economic content of the prices chosen by the planner: trade must be circular, since a consumer who gives without receiving would need a zero price; and, within each trading community, the incomes are the stationary distribution of the chain of sales, the price of a good being the income of its seller divided by the quantity sold. Only the price levels $\theta_C$ of the communities remain free, within the bounds imposed by $P_\pi$.

\vskip 0.5cm

\noindent\textbf{Price adjustment by the planner.} The value function $\hat U$ need not be concave in $p$, since $\mathcal{K}(p)$ depends on $p$ through bilinear terms. Its gradient is nevertheless explicit. Let $\lambda^i$ be the multiplier of the budget constraint of consumer $i$ in the problem $\max_{q\in\mathcal{K}(p)}U(q)$; by \ref{eqn:36}, the multipliers are defined up to a common additive constant. Assume that, for $p\in P_\pi^n$, the solution $\hat q(p)$ and the multipliers (up to this constant) are unique and continuous, and that the resulting gradient is Lipschitz continuous. By the envelope theorem \cite{MilgromSegal2002},
\begin{align}
\label{eqn:40}
\frac{\partial \hat U}{\partial p^i_k}(p)=\lambda^i S^i_k-\sum_{l\neq i}\lambda^l q_{il,k}=\sum_{l\neq i}\big(\lambda^i-\lambda^l\big)\,q_{il,k},
\end{align}
evaluated at $\hat q(p)$, which does not depend on the additive constant. Raising the price of good $k$ sold by $i$ transfers purchasing power from its buyers $l$ to $i$; it increases the social utility if and only if purchasing power is more valuable, at the margin, to the seller than to its buyers. Prices are thus the planner's instrument of redistribution, and they are locally neutral when the multipliers of trading partners coincide. The planner adjusts the prices by the projected gradient-ascent dynamics
\begin{align}
\label{eqn:41}
\dot p=\Pi_{T_{P_\pi^n}(p)}\big(\nabla \hat U(p)\big),
\end{align}
where, at each instant, $\hat q(p)$ and the multipliers are computed by the process \ref{eqn:34}. By the argument of \ref{eqn:46}--\ref{eqn:49} below, $\hat U$ is non-decreasing along the solutions of \ref{eqn:41}, and its limit points satisfy, for every consumer $i$, the stationarity condition: there exists $c^i\in\R$ such that
\begin{align}
\label{eqn:42}
\sum_{l\neq i}\big(\lambda^i-\lambda^l\big)q_{il,k}=c^i\quad\text{if } p^i_k>\pi,\qquad \sum_{l\neq i}\big(\lambda^i-\lambda^l\big)q_{il,k}\le c^i\quad\text{if } p^i_k=\pi.
\end{align}

\vskip 0.5cm

\begin{proposition}{\textbf{Solution of the planner's problem\label{PlannerSolution}}\\}
Under the assumptions of this subsection, $\hat U$ is non-decreasing along every solution of \ref{eqn:41}, every limit point of a solution satisfies \ref{eqn:42}, and every solution $\hat p$ of \ref{eqn:38} satisfies \ref{eqn:42}. The planner's optimum $V_\pi$ is therefore attained at the stationary point of \ref{eqn:41} with the highest value of $\hat U$, and its trades $\hat q(\hat p)$ are given by Proposition \ref{PlannerBenchmark}.
\end{proposition}

\vskip 0.5cm

\begin{proof}
The first two statements follow from the Lyapunov argument of \ref{eqn:46}--\ref{eqn:49}, applied to $\hat U$ on the compact convex set $P_\pi^n$, whose normal cone at $p$ gives \ref{eqn:42}. A maximizer of the continuously differentiable function $\hat U$ on the convex set $P_\pi^n$ satisfies the first-order condition $\Pi_{T_{P_\pi^n}(\hat p)}\big(\nabla \hat U(\hat p)\big)=0$, which is \ref{eqn:42}.
\end{proof}

\vskip 0.5cm

The planner's optimum is a social optimum for the community. It is, however, not a transaction equilibrium in general: the planner may impose on a consumer trades that this consumer would reduce, or prices under which it would prefer another price vector. By part 2 of Proposition \ref{PlannerExistence}, the gap $V_\pi-U(q^\star)\ge0$ between this benchmark and the value of a transaction equilibrium $q^\star$ measures the inefficiency of decentralized exchange, which is the dynamic counterpart of the failure of the first welfare theorem. Since $\hat U$ is not concave, the dynamics \ref{eqn:41} converges to stationary points, which need not be global maximizers: in practice, the global optimum is selected by comparing the stationary points reached from several initial prices.

\vskip 0.5cm

\subsection{Non-t\^atonnement process\label{Non-tâtonnement process}}

\hskip 0.5cm We now consider a continuous-time adjustment process in which consumer $i$ revises its price vector, the strategies $(p^{-i},X^{-i})$ of the other consumers being fixed. Prices remain in the simplex $P$ of \ref{eqn:1}; as noted in Remark \ref{HomegeneityRemark}, only a normalization common to all agents would be neutral, and the individual normalization is part of the model. For a given price vector $p^i$, let
\begin{align}
\label{eqn:43}
X^i(p^i) \in \text{Arg}\max_{X^i\in\mathcal{B}^i(p,X^{-i})}u^i(x^i)
\end{align}
denote the optimal quantity response, and $v^i(p^i)=u^i\big(x^i(p^i)\big)$ the associated indirect utility. The process is a dynamics of prices: the price vector $p^i$ is its only state variable, and the quantities are not adjusted gradually but given, at every instant, by the optimal response \ref{eqn:43} to the current prices. Trade therefore takes place at the current prices, which are not equilibrium prices, and this is what makes the process a non-t\^atonnement process \cite{Hahn1962}: no auctioneer suspends exchange until equilibrium prices are found. By construction, $X^i(p^i)\in\mathcal{B}^i(p,X^{-i})$, so that the budget, feasibility and bilateral transaction constraints are satisfied at every instant of the adjustment process. We assume, in addition to \emph{a'} and \emph{c}", that the optimal quantity response and the associated multiplier $\lambda^i$ of the budget constraint are unique and continuous in $p^i$, and that the resulting gradient below is Lipschitz continuous.

\vskip 0.5cm

\noindent\textbf{Gradient of the indirect utility.} The price $p^i$ enters the inner problem only through the value of sales $p^i\cdot S^i(X^i)$, where $S^i(X^i)=\sum_{j\neq i}x^i_{ij}$. By the envelope theorem for constrained problems \cite{MilgromSegal2002}, applied to the Lagrangian \ref{eqn:21} at the Nash values of the proposals of the partners,
\begin{align}
\label{eqn:44}
\frac{\partial v^i}{\partial p^i_k}(p^i)=\lambda^i(p^i)\,S^i_k\big(X^i(p^i)\big),\qquad k=1,\dots,L:
\end{align}
raising the price of good $k$ increases the indirect utility in proportion to the quantity of good $k$ sold, valued at the shadow price $\lambda^i$ of purchasing power. In particular, $\nabla v^i\ge0$: in the absence of the normalization, prices would increase without bound, which is why the dynamics must be studied on the simplex.

\vskip 0.5cm

\noindent\textbf{Price dynamics.} We define the price adjustment process by the projected gradient-ascent dynamics
\begin{align}
\label{eqn:45}
\dot p^i=\Pi_{T_P(p^i)}\big(\nabla v^i(p^i)\big),
\end{align}
where $T_P(p^i)$ is the tangent cone of the simplex $P$ at $p^i$ and $\Pi_{T_P(p^i)}$ the Euclidean projection on it. Prices thus adjust in the direction that increases the indirect utility generated by the optimal quantity response, while remaining normalized. Under the assumption that the gradient \ref{eqn:44} is Lipschitz continuous in $p^i$, \ref{eqn:45} admits a solution that remains in $P$ \cite{DupuisNagurney1993}. Along a solution of \ref{eqn:45}, the indirect utility satisfies
\begin{align}
\label{eqn:46}
\dot v^i = \nabla v^i(p^i)\cdot\dot p^i = \left\|\Pi_{T_P(p^i)}\big(\nabla v^i(p^i)\big)\right\|^2 \geq0,
\end{align}
since the projection on a closed convex cone satisfies $g\cdot\Pi_T(g)=\|\Pi_T(g)\|^2$. Hence $-v^i$ is a Lyapunov function for the price adjustment process. Moreover,
\begin{align}
\label{eqn:47}
\dot v^i =0 \quad\Longleftrightarrow \quad \Pi_{T_P(p^i)}\big(\nabla v^i(p^i)\big)=0,
\end{align}
so that the set on which the Lyapunov function ceases to decrease coincides with the stationary set of the projected dynamics. Since the simplex $P$ is compact, the price trajectory remains bounded, and $v^i$, which is continuous and non-decreasing along the trajectory, converges to some value $\bar v^i$. Consequently,
\begin{align}
\label{eqn:48}
\int_0^\infty \left\| \Pi_{T_P(p^i(t))}\big(\nabla v^i(p^i(t))\big)\right\|^2dt &<\infty,
\end{align}
and, by the invariance principle, every limit point of the price trajectory belongs to the stationary set
\begin{align}
\label{eqn:49}
\mathcal{S}^i = \left\{ p^i\in P \ \mid\  \Pi_{T_P(p^i)}\big(\nabla v^i(p^i)\big)=0 \right\}.
\end{align}

\vskip 0.5cm

\noindent\textbf{Characterization of the stationary set.} The projection of a vector $g$ on the tangent cone of the simplex at $p^i$ vanishes if and only if $g$ belongs to the normal cone of $P$ at $p^i$, that is, if and only if there exists $c\in\R$ such that $g_k=c$ whenever $p^i_k>0$ and $g_k\le c$ whenever $p^i_k=0$. By \ref{eqn:44}, $p^i\in\mathcal{S}^i$ if and only if there exists $c\in\R$ such that
\begin{align}
\label{eqn:50}
\lambda^i S^i_k = c \quad\text{if } p^i_k>0, \qquad \lambda^i S^i_k \le c \quad\text{if } p^i_k=0,
\end{align}
where $S^i$ is evaluated at $X^i(p^i)$. We obtain the following result.

\vskip 0.5cm

\begin{proposition}{\textbf{Price dynamics and specialization\label{PriceDynamics}}\\}
Under the assumptions of this subsection, $v^i$ is non-decreasing along every solution of \ref{eqn:45}, and every limit point $p^{i,\star}$ of the price trajectory such that $\lambda^i(p^{i,\star})>0$ puts all its weight on the goods sold in maximal quantity:
\begin{align}
\label{eqn:51}
\left\{k \mid p^{i,\star}_k>0\right\}\subseteq \arg\max_{k}S^i_k\big(X^i(p^{i,\star})\big).
\end{align}
In particular, if a single good $k^\star$ is sold in maximal quantity, then $p^{i,\star}=e_{k^\star}$.
\end{proposition}

\vskip 0.5cm

\begin{proof}
The monotonicity of $v^i$ is \ref{eqn:46}. Every limit point belongs to $\mathcal{S}^i$, and, when $\lambda^i>0$, \ref{eqn:50} states that $S^i_k=c/\lambda^i$ on the support of $p^{i,\star}$ and $S^i_k\le c/\lambda^i$ elsewhere, which is \ref{eqn:51}. If $\arg\max_k S^i_k$ is a singleton $\{k^\star\}$, the support of $p^{i,\star}$ is $\{k^\star\}$, hence $p^{i,\star}=e_{k^\star}$.
\end{proof}

\vskip 0.5cm

The price adjustment therefore transfers the weight of the price vector toward the goods that the consumer sells most, and stops when no transfer increases the value of its sales. This is the dynamic counterpart of Remark \ref{QuasiConvexValue}, according to which the maximum of $v^i$ is attained at a vertex of the simplex, and of Theorem \ref{Specialization}. Contrary to an interior adjustment, the limit prices typically lie on the boundary of the simplex: the prices of the goods that the consumer does not sell carry no weight. When $\lambda^i=0$, the budget constraint is slack, the gradient \ref{eqn:44} vanishes, and every price vector is stationary: prices are then indeterminate, as in the autarky situation.

\vskip 0.5cm

\subsection{Joint dynamics}

\hskip 0.5cm When all consumers adjust simultaneously, according to \ref{eqn:45} for their prices and to their optimal quantity responses for their transactions, the rest points of the joint dynamics are related to the transaction equilibria as follows.

\vskip 0.5cm

\begin{proposition}{\textbf{Rest points of the joint dynamics\label{RestPoints}}\\}
Let $(p,X)$ be a rest point of the joint dynamics at which every consumer $i$ with $\lambda^i>0$ sells a single good $k_i$, in a positive quantity, and let $\mathbf{k}=(k_1,\dots,k_n)$. Then $(p,X)$ is a transaction equilibrium of the specialized economy $\mathcal{E}_{\mathbf{k}}$. It is a transaction equilibrium of $\mathcal{E}$ if and only if no consumer can gain by changing its specialization.
\end{proposition}

\vskip 0.5cm

\begin{proof}
At a rest point, $X^i$ is an optimal quantity response for every $i$. If $\lambda^i>0$, Proposition \ref{PriceDynamics} gives $p^i=e_{k_i}$. In $\mathcal{E}_{\mathbf{k}}$, the budget of consumer $i$ depends on $p^i$ only through $p^i_{k_i}$, and $v^i$ is non-decreasing in it, so that $e_{k_i}$ maximizes $v^i$ over $P$. If $\lambda^i=0$, the Karush--Kuhn--Tucker conditions of the inner problem hold without the budget constraint; by concavity of $u^i$, $X^i$ is then optimal even when the budget constraint is removed, so that no price vector can yield a higher utility, and $p^i$ is optimal. Hence $(p,X)$ is a fixed point of the best-response correspondence of $\mathcal{E}_{\mathbf{k}}$. The last statement is part 2 of Proposition \ref{Compactness of equilibria}.
\end{proof}

\vskip 0.5cm

The rest points of the joint dynamics are thus local equilibria, in which each consumer has optimized its price and its transactions given its specialization; whether they are equilibria of the whole economy depends on the incentives to change specialization, in accordance with the structure of transaction equilibria described in Section \ref{New way to consider the Consumer Problem}. The convergence of the joint dynamics itself is not guaranteed by the previous arguments: each indirect utility $v^i$ depends on the strategies of the other consumers, which move simultaneously, so that the individual Lyapunov functions $-v^i$ do not yield a Lyapunov function for the joint system. Establishing such a function, for instance under a potential structure, or under conditions ensuring that the joint best-response map is a contraction, is left for future research; the numerical example of Section \ref{Numerical simulation} provides an illustration of the rest points.

\vskip 0.5cm

\section{Discrete transactions\label{Discrete transactions}}

\hskip 0.5cm For reasons of realism, we may consider an economy with discrete transactions, in which goods are indivisible and prices are not fully continuous. In such a setting, the continuity and convexity arguments of Section \ref{New way to consider the Consumer Problem} no longer apply. The autarky situation remains a transaction equilibrium. The construction of Proposition \ref{TradeEquilibrium} carries over to the discrete setting when the set $P$ contains the vertices of the simplex and the trade levels along the cycle belong to the grid, since its proof only uses the quasi-concavity of $u^i$ along a segment; outside this case, the existence of an equilibrium with trade is not guaranteed. Suppose that in our original problem the maximum is
\begin{align}
\label{eqn:52}
\max_{(p^i,X^i)\in \mathbb{B}^i(p^{-i},X^{-i})} u^i(x^i).
\end{align}

Under the budget constraint
\begin{align}
\label{eqn:53}
\begin{split}
\mathbb{B}^i(p^{-i},X^{-i}) =\left\{\right. & (p^i,X^i)\in P \times \mathbb{X}^i \ \mid \  \sum_{j\neq i} p^j \cdot  x_{ji}^i \leq \sum_{j\neq i} p^i \cdot  x_{ij}^i \ , \   w^i + \sum_{j\neq i} x_{ji}^i - \sum_{j\neq i} x_{ij}^i \in \mathscr{X}^i \ , \ x_{ij}^i \leq x_{ij}^j \ , \\
& \left. x_{ji}^i \leq x_{ji}^j \right\}
\end{split}
\end{align}

Suppose further that the sets $\mathbb{X}^i$, $\mathscr{X}^i$ and $P$ are finite, for instance the intersections of the sets of the previous sections with a finite grid, and that $0\in\mathbb{X}^i$ and $w^i\in\mathscr{X}^i$. Then $\mathbb{B}^i(p^{-i},X^{-i})$ is finite and non-empty, since it contains the autarky plan for every price vector. Therefore, the maximization problem \ref{eqn:52} admits a solution, and the optimal price-demand correspondence $\big(p^i(p^{-i},X^{-i}),X^i(p^{-i},X^{-i})\big)$ is non-empty-valued. As in Section \ref{New way to consider the Consumer Problem}, the autarky profile $(p,0)$ is a transaction equilibrium for every $p\in P^n$: if $X^{-i}=0$, the bilateral constraints force $X^i=0$, and the utility of consumer $i$ no longer depends on $p^i$.\\

Define $\Phi : P^n \times \prod_{i=1}^n \mathbb{X}^i \to P^n \times \prod_{i=1}^n \mathbb{X}^i$, the discrete counterpart of the best-response correspondence \ref{eqn:19}, as the Cartesian product $\Phi=\prod_{i=1}^n \phi^i$, where
\begin{align}
\label{eqn:54}
\phi^i(p^{-i},X^{-i}) &= \left(p^i(p^{-i},X^{-i}),X^i(p^{-i},X^{-i})\right).
\end{align}
The fixed points of $\Phi$ are the transaction equilibria of the discrete economy. To study the dynamics induced by $\Phi$, let $(p^m,X^m)_{m\geq 0}$ be a sequence generated by $\Phi$, that is,
\begin{align}
\label{eqn:55}
(p^{m+1},X^{m+1})\in\Phi(p^m,X^m).
\end{align}

Equivalently, for every $i$,
\begin{align}
\label{eqn:56}
(p^{i,m+1},X^{i,m+1}) \in \phi^i(p^{-i,m},X^{-i,m}).
\end{align}

Since $X^{i,m+1}$ satisfies the bilateral constraints in \ref{eqn:53}, for every $i\neq j$ we have
\begin{align*}
x_{ij}^{i,m+1} \leq x_{ij}^{j,m}, \qquad x_{ji}^{i,m+1} \leq x_{ji}^{j,m}.
\end{align*}

Similarly,
\begin{align*}
x_{ij}^{j,m+1} \leq x_{ij}^{i,m}, \qquad x_{ji}^{j,m+1} \leq x_{ji}^{i,m}.
\end{align*}

A proposal can therefore never exceed the proposal of the partner at the previous iteration: the process is a \textbf{descending} one, in which bilateral proposals are reduced until they become compatible. In particular, if the process starts in autarky, $X^0=0$, it remains in autarky forever. The relevant initial condition is the one in which every consumer declares itself ready to trade the maximal admissible quantities of $\mathbb{X}^i$, and the process then selects, by successive reductions, the transactions that the partners are willing to carry out, in the manner of a descending auction.

\vskip 0.5cm

\begin{proposition}{\textbf{Simultaneous updating\label{SimultaneousUpdating}}\\}
Let $(p^m,X^m)_{m\ge0}$ be generated by \ref{eqn:55}.
\begin{enumerate}
\item For every $m\ge0$ and every $i\neq j$,
\begin{align}
\label{eqn:57}
x_{ij}^{i,m+2}\leq x_{ij}^{i,m}, \qquad x_{ji}^{i,m+2}\leq x_{ji}^{i,m}.
\end{align}
\item There exists $M\in\mathbb{N}$ such that
\begin{align}
\label{eqn:58}
X^{m+2} = X^{m} \qquad\text{for all }m\geq M:
\end{align}
the transactions eventually become constant or periodic of period two.
\item If $\Phi$ is single-valued, for instance under a deterministic tie-breaking rule, the sequence $(p^m,X^m)$ is eventually periodic. If it is eventually constant, its limit is a transaction equilibrium.
\end{enumerate}
\end{proposition}

\vskip 0.5cm

\begin{proof}
\begin{enumerate}
\item Combining the two pairs of inequalities above at two consecutive iterations gives $x_{ij}^{i,m+2}\le x_{ij}^{j,m+1}\le x_{ij}^{i,m}$, and similarly for $x_{ji}^{i}$.

\item By \ref{eqn:57}, each component of the subsequences $(X^{2m})_{m\geq 0}$ and $(X^{2m+1})_{m\geq 0}$ is non-increasing. Since $\mathbb{X}^i$ is finite, each of these subsequences takes finitely many values, and is therefore eventually constant, which gives \ref{eqn:58}.

\item If $\Phi$ is single-valued, the sequence is generated by a map on the finite set $P^n\times\prod_i\mathbb{X}^i$; some state is visited twice, after which the sequence repeats itself. If $(p^{m+1},X^{m+1})=(p^m,X^m)$, then $(p^m,X^m)$ is a fixed point of $\Phi$, that is, a transaction equilibrium.
\end{enumerate}
\end{proof}

\vskip 0.5cm

Two consumers may thus alternate in their proposals: at every iteration, each one adopts the previous proposal of the other, and the transactions oscillate with period two. The prices are not monotone: once the transactions have become constant or periodic, the prices evolve in the finite set $P^n$, and their cycles may have any period.\\

The conclusion changes if the agents update sequentially rather than simultaneously. Suppose that, at every iteration $m\ge1$, agents update in the order $1,2,\ldots,n$, with each agent observing the choices already updated at iteration $m$ and the choices of the remaining agents at iteration $m-1$. When agent $i$ updates at iteration $m$, its bilateral constraints with each partner $j$ therefore read
\begin{align}
\label{eqn:59}
x^{i,m}_{ij}\le x^{j,m}_{ij},\quad x^{i,m}_{ji}\le x^{j,m}_{ji}\quad (j<i),\qquad x^{i,m}_{ij}\le x^{j,m-1}_{ij},\quad x^{i,m}_{ji}\le x^{j,m-1}_{ji}\quad (j>i).
\end{align}

\vskip 0.5cm

\begin{proposition}{\textbf{Sequential updating\label{SequentialUpdating}}\\}
Under the sequential updating rule \ref{eqn:59}:
\begin{enumerate}
\item For every $m\ge1$ and every $i$, componentwise,
\begin{align}
\label{eqn:60}
X^{i,m+1}\leq X^{i,m}.
\end{align}
\item There exists $M\in\mathbb{N}$ such that $X^m=X^M$ for all $m\ge M$: the transactions become constant after finitely many iterations, and no cycle of the transactions can occur.
\item Under a deterministic tie-breaking rule, the sequence $(p^m,X^m)$ is eventually periodic, with constant transactions. If it is eventually constant, its limit is a transaction equilibrium.
\end{enumerate}
\end{proposition}

\vskip 0.5cm

\begin{proof}
\begin{enumerate}
\item Fix $i$, a partner $j$ and $m\ge1$, and consider the component $x^{i}_{ij}$; the argument for $x^i_{ji}$ is identical. If $j<i$, consumer $j$ updates before $i$ at iteration $m+1$ and observes the proposal of $i$ at iteration $m$, so that, by \ref{eqn:59},
\begin{align*}
x^{i,m+1}_{ij}\le x^{j,m+1}_{ij}\le x^{i,m}_{ij}.
\end{align*}
If $j>i$, consumer $j$ updated after $i$ at iteration $m$, and observed the proposal of $i$ at iteration $m$, so that
\begin{align*}
x^{i,m+1}_{ij}\le x^{j,m}_{ij}\le x^{i,m}_{ij}.
\end{align*}

\item By \ref{eqn:60}, each component of $(X^m)_{m\ge1}$ is non-increasing and takes finitely many values, hence is eventually constant.

\item Once the transactions are constant, the prices are generated by a map on the finite set $P^n$, and are eventually periodic. If $(p^{m+1},X^{m+1})=(p^m,X^m)$, every consumer has chosen a best response to the current choices of all the others, so that $(p^m,X^m)$ is a fixed point of $\Phi$, that is, a transaction equilibrium.
\end{enumerate}
\end{proof}

\vskip 0.5cm

Thus, unlike the simultaneous updating case, in which monotonicity arises only along the even and odd subsequences and a two-cycle of the transactions may occur, sequential updating makes the transaction variables monotone after every complete sweep, and the transactions converge in finitely many iterations. The descending nature of the process has two consequences. First, the limit depends on the initial proposals: starting from autarky, nothing happens, and starting from maximal proposals, the process reduces them until they become compatible. Second, when the process becomes stationary, it reaches a transaction equilibrium, which may be autarky if the proposals are reduced to zero; whether an equilibrium with trade is reached depends on the initial proposals and on the preferences of the consumers.

\vskip 0.5cm

\section{Market topology}\label{Market topology}

\hskip 0.5cm What happens when there are barriers to exchange, and the bilateral exchange graph is incomplete? This occurs when some consumers do not have access to all suppliers and vice versa, due to factors such as wholesale-retail inadequacy, geographic barriers, or information asymmetry.\\

In this situation, the economy $\mathcal{E}=\left(w^i,u^i,\mathscr{X}^i,C^i\right)_{1\leq i \leq n}$ will be characterized with exchange capacities $C^i=(c^i_{ji},c^i_{ij})_{1\leq j \leq n}\in (\R_+^L)^{2n}$ such that $X^i \leq C^i$, a zero capacity $c^i_{ij,k}=0$ meaning that consumer $i$ cannot sell good $k$ to consumer $j$. Define the set of topological constraints \mbox{$\mathfrak{T}^i=\left\{X^i \in \mathbb{X}^i \ \mid \ X^i \leq C^i \right\}$}. The new consumer problem is
\begin{align}
\label{eqn:61}
\max_{p^i \in P} \max_{X^i\in \mathcal{B}^i(p,X^{-i}) \cap \mathfrak{T}^i} u^i(x^i).
\end{align}

The set $\mathfrak{T}^i$ is convex, compact, contains autarky, does not depend on $(p,X^{-i})$, and satisfies assumption \emph{d}: reducing a trade preserves the capacity constraints. The analysis of Section \ref{New way to consider the Consumer Problem} therefore applies with $\mathbb{X}^i$ replaced by $\mathfrak{T}^i$, provided that a sale opportunity of positive value, in the proof of lower semicontinuity, is understood as a sale allowed by a positive capacity $c^i_{ij,k}>0$. In particular, the autarky profile remains a transaction equilibrium. Proposition \ref{TradeEquilibrium} also carries over, for trade cycles whose links have positive capacities: the feasible levels $t\in[0,T_s]$ are further bounded by the capacities $c^{i_s}_{i_si_{s+1},k_s}$ and $c^{i_s}_{i_{s-1}i_s,k_{s-1}}$ of the cycle, and the gains from trade must be attainable within these bounds; the proof is unchanged, since the capacity constraints are box constraints and do not affect the quasi-concavity argument along the segment. The introduction of these new constraints alters the problem, and a direct comparison with the initial formulation leads to the following result (the middleman imperfection).

\vskip 0.5cm

\begin{proposition}{\textbf{Optimality -- Second Comparison\label{Second Comparison}}\\}
For every $(p,X^{-i})$, the value of the topologically constrained problem \ref{eqn:61} cannot exceed the value of the problem \ref{eqn:11} of the initial effective trade model.
\end{proposition}

\vskip 0.5cm

\begin{proof}
For every $(p,X^{-i})$, the feasible set $\mathcal{B}^i(p,X^{-i})\cap\mathfrak{T}^i$ is contained in $\mathcal{B}^i(p,X^{-i})$.
\end{proof}

\vskip 0.5cm

The comparison holds consumer by consumer, at given strategies of the others. It does not extend to equilibria: the equilibria of the constrained and unconstrained economies are reached at different strategies of the others, and need not be ranked. The capacities, that is, the web of relations, shape the price structure and the exchanged quantities.

\vskip 0.5cm

\subsection{Topological Influences on Consumer Choice and Market Coordination\label{Topological Influences on Consumer Choice and Market Coordination}}

\hskip 0.5cm In the following, we address the problem of demand indeterminacy by analyzing consumer preferences within an exchange network. Our focus is on exchange processes structured by network interactions: each agent $i$ may hold intrinsic preferences over goods, as captured by classical demand theory (see, for instance, \cite{Mas-Colell95}), while also developing affinities toward specific trading partners. These affinities—shaped by factors such as marketing, consumer loyalty, market positioning, or interpersonal relationships—can lead agents to favor transactions with certain peers over others. We proceed to formalize and examine this layered preference structure and explore its implications for decentralized exchange.\\

Given a binary relation $R$ on a set $S$, we recall the following properties:

\begin{enumerate}
\item \textbf{Reflexivity:} $\forall x \in S,\ x R x$.

\item \textbf{Transitivity:} $\forall x, y, z \in S,\ (x R y \ \wedge \ y R z) \Rightarrow x R z$.

\item \textbf{Completeness:} $\forall x, y \in S,\ x \neq y \Rightarrow (x R y \ \vee \ y R x)$.
\end{enumerate}

Now, define the binary relation $\succeq$ on a set of alternatives $Z$, where $x \succeq y$ is interpreted as "$x$ is at least as good as $y$". From this, we derive two related notions:

\begin{enumerate}
\item \textbf{Strict preference:} $x \succ y$ if and only if $(x \succeq y) \wedge \neg (y \succeq x)$, meaning $x$ is strictly preferred to $y$.

\item \textbf{Indifference:} $x \sim y$ if and only if $(x \succeq y) \wedge (y \succeq x)$, meaning $x$ and $y$ are considered equally preferable.
\end{enumerate}

\vskip 0.5cm

\begin{definition}{\textbf{Rational Preferences}\\}
A preference relation $\succeq$ on a set $S$ is said to be \textbf{rational} if it satisfies the following two properties:
\begin{itemize}
\item \textbf{Completeness:} For all $x, y \in S$, either $x \succeq y$ or $y \succeq x$ (or both).
\item \textbf{Transitivity:} For all $x, y, z \in S$, if $x \succeq y$ and $y \succeq z$, then $x \succeq z$.
\end{itemize}
\end{definition}

\vskip 0.5cm

We now introduce the concept of a utility function that represents a preference relation:

\vskip 0.5cm

\begin{definition}{\textbf{Utility Function}\\}
A function $u : S \to \mathbb{R}$ is called a \textbf{utility function} representing the preference relation $\succeq$ if, for all $x, y \in S$,
\begin{align*}
x \succeq y \ \Longleftrightarrow \ u(x) \geq u(y).
\end{align*}
\end{definition}

\vskip 0.5cm

Only rational preferences can be represented by a utility function:

\vskip 0.5cm

\begin{proposition}{\textbf{\cite{Mas-Colell95}}\\}
A preference relation $\succeq$ can be represented by a utility function only if it is rational.
\end{proposition}

\vskip 0.5cm

The converse does not hold in general: lexicographic preferences on $\R^2_+$ are rational but cannot be represented by any utility function \cite{Mas-Colell95}. This case will be relevant in Section \ref{Equilibrium Existence under Hierarchical Preferences}.

\vskip 0.5cm

\begin{definition}{\textbf{Continuous Preferences}\\}
A preference relation $\succeq$ on $S$ is said to be \textbf{continuous} if for every pair of sequences $(x_m)$ and $(y_m)$ in $S$ such that $x_m \succeq y_m$ for all $m$, and $x_m \to x$, $y_m \to y$, it follows that $x \succeq y$.
\end{definition}

\vskip 0.5cm

The following fundamental result is due to Debreu:

\vskip 0.5cm

\begin{proposition}{\textbf{Debreu's Theorem \cite{Kreps2012}}\label{Debreu's theorem}\\}
Let $\succeq$ be a rational preference relation on $S\subseteq\R^d$, for instance $S=\R^L_+$. If a continuous function $u$ represents $\succeq$, then $\succeq$ is continuous. Conversely, if $\succeq$ is continuous, then there exists a continuous utility function $u$ that represents it.
\end{proposition}

\vskip 0.5cm

\begin{definition}{\textbf{Convex Preferences}\\}
Let $S$ be a convex subset of a vector space. A preference relation $\succeq$ on $S$ is said to be:
\begin{enumerate}
\item \textbf{Convex} if for all $x, y \in S$ such that $x \succeq y$, and for all $\alpha \in [0,1]$, we have
\begin{align*}
\alpha x + (1-\alpha)y \succeq y.
\end{align*}

\item \textbf{Strictly convex} if for all $x, y \in S$, with $x \neq y$ and $x \succeq y$, and for all $\alpha \in ]0,1[$, we have
\begin{align*}
\alpha x + (1-\alpha)y \succ y.
\end{align*}

\item \textbf{Semi-strictly convex} if they are convex and for all $x, y \in S$ such that $x \succ y$, and for all $\alpha \in ]0,1[$, we have
\begin{align*}
\alpha x + (1-\alpha)y \succ y.
\end{align*}
\end{enumerate}
\end{definition}

\vskip 0.5cm

Convexity is similarly reflected in the properties of utility representations:

\vskip 0.5cm

\begin{proposition}{\textbf{\cite{Kreps2012}}\label{ConvexUtility}\\}
\begin{enumerate}
\item If preferences $\succeq$ are represented by a concave utility function $u$, then $\succeq$ is convex. If $u$ is strictly concave, then $\succeq$ is strictly convex.

\item Suppose that $u$ represents preferences $\succeq$. Then:
\begin{itemize}
\item $u$ is quasi-concave if and only if $\succeq$ is convex.
\item $u$ is strictly quasi-concave if and only if $\succeq$ is strictly convex.
\item $u$ is semi-strictly quasi-concave if and only if $\succeq$ is semi-strictly convex.
\end{itemize}
\end{enumerate}
\end{proposition}

\vskip 0.5cm

Let us now consider the two sets $\{1, \dots, n\}$ and $\{1, \dots, L\}$, and define the corresponding preference relations $\succeq_n$ and $\succeq_L$. Natural questions arise when these relations are considered simultaneously, i.e., when a consumer is faced with a decision to choose between goods owned by different agents. A first question that comes to mind is: which of these relations takes priority? Is it always the case? Can an agent prefer a good solely because it was marketed by a particular agent? What happens when considering the dimensions of preferences? Do the usual smooth properties such as continuity and convexity still hold?\\

Let us now approach this by discussing separable preferences:

\vskip 0.5cm

\begin{definition}{\textbf{Separable Preferences}\\}
Let $J_1, \dots, J_N$ partition $\{1, \dots, L\}$. That is, $J_a \cap J_b = \emptyset$ for $a \neq b$, and $J_1 \cup \dots \cup J_N = \{1, \dots, L\}$. Preferences $\succeq$ are \textbf{strongly separable} in $J_1, \dots, J_N$ if for every $K = J_{a_1}\cup \dots\cup J_{a_k}$, for some set of indices $\{a_1, \dots, a_k\}$ drawn from $\{1, \dots, N\}$, $(x_K, x_{K^c}) \succeq (x'_K, x_{K^c})$ for some $x_{K^c}$ implies that $(x_K, x'_{K^c}) \succeq (x'_K, x'_{K^c})$ for all $x'_{K^c}$.
\end{definition}

\vskip 0.5cm

We have the following result:

\vskip 0.5cm

\begin{proposition}{\textbf{\cite{Kreps2012}}\label{AdditiveSeparability}\\}
Suppose preferences $\succeq$ are continuous and strongly separable in $J_1$ through $J_N$. Suppose further that $\succeq$ is nontrivial on at least three of the commodity index sets $J_1$ through $J_N$ (therefore, $N \geq 3$ is certainly required). Then we can find continuous functions $u_a : \mathbb{R}_+^{J_a} \to \mathbb{R}$ such that
\begin{align*}
u(x) &= \sum_{a=1}^N u_a(x_{J_a})
\end{align*}
is a utility representation of $\succeq$. Conversely, if preferences are represented by a utility function $u$ taking the previous form, then preferences $\succeq$ are strongly separable.
\end{proposition}

\vskip 0.5cm

To incorporate preferences over trading partners within the separability framework, we augment the commodity space by treating trading relationships as additional attributes of an alternative. For agent $i$, let $z^i=(x^i,r^i_1,\dots,r^i_n)$, where $x^i\in\mathscr{X}^i$ denotes the consumption bundle and $r^i_j=(x^i_{ij},x^i_{ji})$, $j\neq i$, describes the trading relationship of $i$ with $j$; preferences are defined on this augmented space, before the consistency between $x^i$ and the relationships is imposed. If preferences on this augmented space are continuous and strongly separable in the consumption bundle and in each relationship, and nontrivial on at least three of these groups, which requires $n\geq3$, Proposition \ref{AdditiveSeparability} yields the additive representation
\begin{align*}
\tilde{u}^i(x^i,r^i)=u^i(x^i)+\sum_{j\neq i}\vartheta^i_j(r^i_j).
\end{align*}

Since both $x^i$ and $r^i$ are induced by the transaction configuration $X^i$, through $x^i=T^i(X^i)$ and $r^i=X^i$, the resulting preferences over transactions take the form
\begin{align}
\label{eqn:62}
\tilde{u}^i(X^i) &= u^i\left(T^i(X^i)\right) + \vartheta^i\left(X^i\right),\qquad \vartheta^i(X^i)=\sum_{j\neq i}\vartheta^i_j(x^i_{ij},x^i_{ji}).
\end{align}

\vskip 0.5cm

\begin{definition}{\textbf{Equivalent Classes of Utility}\\}
For consumer $i$, write $T^i(X^i)=w^i+A^i(X^i)$, where $A^i(X^i)=\sum_{j \neq i} x^i_{ji} - \sum_{j \neq i} x^i_{ij}$ is linear, and let $\ker(A^i) = \left\{ Y : A^i(Y) = 0 \right\}$. Two plans $X^i,Y^i\in\mathbb{X}^i$ are equivalent if $X^i-Y^i\in\ker(A^i)$, that is, if they lead to the same final holding. The equivalence class $\dot{X}^i=(X^i+\ker(A^i))\cap\mathbb{X}^i$ gathers all the transaction plans that correspond to the same utility level of good consumption $u^i$.
\end{definition}

\vskip 0.5cm

The function $u^i\circ T^i$ is constant on each equivalence class: on its own, it cannot select a transaction plan within a class. This is the source of the non-uniqueness of transactions noted in Section \ref{New way to consider the Consumer Problem}. The term $\vartheta^i$ resolves this indeterminacy. We assume that $u^i : \mathscr{X}^i \to \mathbb{R}$ is continuous and concave (assumption \emph{c}"), and that $\vartheta^i : \mathbb{X}^i \to \mathbb{R}$ is continuous and strictly concave. Strict concavity of $\vartheta^i$ can be interpreted as a desire for diversification by the partners. Since $T^i$ is affine, $u^i\circ T^i$ is concave, and $\tilde u^i$ is strictly concave; it is continuous, so that, by Debreu's Theorem \ref{Debreu's theorem} and Proposition \ref{ConvexUtility}, $\tilde{u}^i$ represents continuous and strictly convex preferences $\tilde{\succeq}$ on $\mathbb{X}^i$. Note that concavity of $u^i$ is required here: the sum of a quasi-concave function and of a strictly concave function need not be quasi-concave.\\

Given this construction, what does $X \,\tilde{\succeq}\, Y$ mean for $X, Y \in \mathbb{X}^i$? By definition,
\begin{align}
\label{eqn:63}
X \,\tilde{\succeq}\, Y \quad\Longleftrightarrow\quad u^i(T^i(X))-u^i(T^i(Y))\ \geq\ \vartheta^i(Y)-\vartheta^i(X),
\end{align}
which covers the following situations:

\begin{enumerate}
    \item \textbf{Better net consumption bundle:} $u^i(T^i(X)) \geq u^i(T^i(Y))$ and $\vartheta^i(X) = \vartheta^i(Y)$.
    \item \textbf{Better flow pattern:} $u^i(T^i(X)) = u^i(T^i(Y))$ and $\vartheta^i(X) \geq \vartheta^i(Y)$; in particular, within an equivalence class, the plan is selected by $\vartheta^i$ alone.
    \item \textbf{Better net bundle, but worse flow pattern:} $u^i(T^i(X)) \geq u^i(T^i(Y))$ and $\vartheta^i(X) \leq \vartheta^i(Y)$, the gain in consumption compensating the loss in the flow pattern.
    \item \textbf{Worse net bundle, but better flow pattern:} $u^i(T^i(X)) \leq u^i(T^i(Y))$ and $\vartheta^i(X) \geq \vartheta^i(Y)$, the gain in the flow pattern compensating the loss in consumption.
\end{enumerate}

In practice, the function $\vartheta^i$ can be made more explicit. For example, one could consider $\vartheta^i$ to be some measure of centrality in the network, as there are many possibilities. From a trade perspective, one could imagine that the seller, depending on their position in the market, could have different preferences. For instance, a large seller might seek to gain more influence in the market and thus maximize their spectral influence (see \cite{Riane2024}), while another agent might aim to maximize their prestige by forming relationships with larger sellers and thus maximize their eigenvector centrality. For more details on centrality measures, consult \cite{Bloch2023}. Marketing and social determinants can also influence the form of $\vartheta^i$. Since centrality measures depend on the whole network, $\vartheta^i$ may also depend on the proposals $X^{-i}$ of the other consumers; the results below remain valid provided that $\vartheta^i$ is jointly continuous and strictly concave in $X^i$, a property that must be checked for the chosen measure.\\

Dealing with pricing, one should also consider the agent's perception of reality. An agent may choose a price vector $p^i \in P$, where $P$ is the $L$-dimensional unit simplex. However, the multitude of possible price choices can lead to a loss of decision-making uniqueness: by Remark \ref{QuasiConvexValue}, the indirect utility is quasi-convex in $p^i$. To address this, one could imagine a strictly concave component $\varrho^i$ to the pricing decision, so that the utility of the consumer is given by
\begin{align}
\label{eqn:64}
\tilde{u}^i(p, X^i) &= u^i(T^i(X^i)) + \vartheta^i(X^i) + \varrho^i(p^i,p^{-i}) \, .
\end{align}

A possible choice for $\varrho^i$ is:
\begin{align}
\label{eqn:65}
\varrho^i(p^i,p^{-i}) = -\frac{\kappa}{2} \sum_{j \neq i} \sum_{k=1}^L (p_k^j - p^i_k)^2 \, ,\qquad \kappa>0.
\end{align}

This represents a natural tendency to avoid prices too far from those of the market (the crowd), with intensity $\kappa$. Since $\varrho^i$ does not depend on $X^i$, the sequential structure of the consumer problem is preserved: the inner problem selects the transactions, with value $\tilde v^i(p^i)=\max_{X^i\in\mathcal{B}^i(p,X^{-i})\cap\mathfrak{T}^i}\big(u^i(T^i(X^i))+\vartheta^i(X^i)\big)$, and the outer problem selects the price by maximizing $\tilde v^i+\varrho^i$ over the convex set $P$, the bilinear budget term being absorbed by the inner problem. Adding $\varrho^i$ to the quasi-convex function $\tilde v^i$ does not by itself yield a concave function; it does so when the curvature of $\varrho^i$ dominates that of $\tilde v^i$. Finally, as in Section \ref{PlannerProblemSection}, we restrict the prices to the set $P_\pi$ of \ref{eqn:37}, so that every good has a positive price, condition (C) holds at every point, and $\mathcal{B}^i\cap\mathfrak{T}^i$ is continuous everywhere under assumption \emph{f}.

\vskip 0.5cm

\begin{proposition}{\textbf{Equilibrium with strictly concave preferences\label{ConcaveEquilibrium}}\\}
Assume \emph{a}, \emph{c}", \emph{d}, \emph{e} and \emph{f}, that $\vartheta^i$ is continuous and strictly concave in $X^i$, that the prices are restricted to $P_\pi$, and that $\tilde v^i$ is continuously differentiable in $p^i$, with a gradient that is Lipschitz continuous with constant $\ell^i$, uniformly in $(p^{-i},X^{-i})$. If $\kappa(n-1)>\ell^i$ for every $i$:
\begin{enumerate}
\item For every $(p^{-i},X^{-i})$, consumer $i$ has a unique best response $\big(\tilde p^i,\tilde X^i\big)(p^{-i},X^{-i})$, which depends continuously on $(p^{-i},X^{-i})$.
\item A transaction equilibrium exists, and the set of transaction equilibria is compact.
\end{enumerate}
\end{proposition}

\vskip 0.5cm

\begin{proof}
\begin{enumerate}
\item The feasible set $\mathcal{B}^i(p,X^{-i})\cap\mathfrak{T}^i$ is non-empty, convex and compact, and the objective $u^i\circ T^i+\vartheta^i$ is continuous and strictly concave, so that the inner problem has a unique solution $\tilde X^i(p,X^{-i})$. Since the correspondence $\mathcal{B}^i\cap\mathfrak{T}^i$ is continuous everywhere, part 2 of Theorem \ref{Berge's Maximum Theorem} shows that $\tilde v^i$ is continuous and that the single-valued map $\tilde X^i$ is continuous. Since the gradient of $\tilde v^i$ in $p^i$ is $\ell^i$-Lipschitz, $\tilde v^i-\frac{\ell^i}{2}\|p^i\|^2$ is concave; since $\varrho^i=-\frac{\kappa(n-1)}{2}\|p^i\|^2$ plus terms that are affine in $p^i$,
\begin{align*}
\tilde v^i+\varrho^i=\Big(\tilde v^i-\frac{\ell^i}{2}\|p^i\|^2\Big)-\frac{\kappa(n-1)-\ell^i}{2}\|p^i\|^2+\text{affine terms}
\end{align*}
is strictly concave on the convex compact set $P_\pi$, and has a unique maximizer $\tilde p^i(p^{-i},X^{-i})$, which is continuous by part 2 of Theorem \ref{Berge's Maximum Theorem}, the constraint set $P_\pi$ being constant. The best response $\big(\tilde p^i,\tilde X^i\big((\tilde p^i,p^{-i}),X^{-i}\big)\big)$ is a composition of continuous maps.

\item By part 1, the best-response map $\Phi$ of \ref{eqn:19} is a continuous function from the non-empty, convex and compact set $P_\pi^n\times\prod_i\mathfrak{T}^i$ into itself. By Brouwer's fixed point theorem, it has a fixed point, which is a transaction equilibrium. The set of fixed points of a continuous map is closed, hence compact.
\end{enumerate}
\end{proof}

\vskip 0.5cm

\begin{remark}{\textbf{Specialization is no longer necessary\label{NoSpecialization}}\\}
Theorem \ref{Specialization} rests on a costless deviation: a seller of two goods $k$ and $\ell$ stops selling $\ell$, puts all the weight of its price on $k$, keeps $\ell$, and strictly increases its utility. This deviation is costless for two reasons: the price has no value in itself, and only the final bundle matters. Each of the new terms removes one of these reasons. With $\varrho^i$, moving to the vertex $e_k$ moves the price of $i$ away from those of the market, at a cost that exceeds the gain of the deviation when $\kappa$ is large enough; moreover, $\tilde v^i+\varrho^i$ being strictly concave, its maximum is no longer attained at a vertex of the simplex. With $\vartheta^i$, stopping the sale of $\ell$ weakens or cuts trading relationships, so that the gain of the deviation is $\Delta u^i+\Delta\vartheta^i$, which may be negative; strict concavity of $\vartheta^i$, a taste for the diversity of partners and flows, favors the sale of several goods to several buyers. Specialization is thus no longer imposed, although specialized equilibria remain possible, and the model of Section \ref{New way to consider the Consumer Problem} is recovered as $\kappa\to0$ and $\vartheta^i\to0$. In exchange, Proposition \ref{PositivePrices}, which relies on the same deviation, no longer holds: selling at a zero price may be worthwhile when the relationship has value. This is why the prices are restricted to $P_\pi$. The equilibrium of Proposition \ref{ConcaveEquilibrium} may be autarkic. If $X^{-i}=0$, the bilateral constraints force $X^i=0$, $\tilde v^i$ no longer depends on $p^i$, and the best price of consumer $i$ maximizes $\varrho^i$, that is, it is the average of the prices of the others; the only fixed points of this map are the common price profiles, so that autarky at any common price $p\in P_\pi$ is a transaction equilibrium of the augmented economy. Moreover, the cycle construction of Proposition \ref{TradeEquilibrium} does not carry over: it uses the vertices $e_{k_s}$, which do not belong to $P_\pi$, and, at a common price, a seller whose budget binds with multiplier $\lambda^i>0$ gains $\lambda^i q$ at the margin by raising the price of the good it sells, an incentive that $\varrho^i$ moderates but does not cancel. Along a cycle, the budgets then require the traded flows to have equal values, so that an equilibrium with trade has heterogeneous prices and quantities determined jointly, and cannot be written explicitly. The existence of an equilibrium with trade in the augmented economy therefore remains an open question, although such equilibria exist in the example of Section \ref{Numerical simulation}.
\end{remark}

\vskip 0.5cm

\begin{remark}{\textbf{Partial specialization\label{PartialSpecialization}}\\}
Specialization need not be uniform across agents. In the basic model, the deviation used in the proof of Theorem \ref{Specialization} is strictly profitable only if consumer $i$ values the good it stops selling; if $u^i$ does not depend on that good, a seller may sell several goods at equilibrium, and specialization holds only in the weak sense that a specialized plan is equally good. In the augmented model, the terms $\vartheta^i$ and $\varrho^i$ may differ across agents, the intensity $\kappa$ being replaced by $\kappa^i$: agents that value their trading relationships tend to diversify their sales, while agents with $\vartheta^i=0$ and a small $\kappa^i$ behave as in the basic model. Specialized and non-specialized sellers may therefore coexist, as in the equilibria of Table \ref{tab:4}, in which consumer $b$ sells both goods while $a$ and $c$ sell a single good.
\end{remark}

\vskip 0.5cm

\begin{remark}{\textbf{Uniqueness\label{Uniqueness}}\\}
Part 1 of Proposition \ref{ConcaveEquilibrium} establishes the uniqueness of the best responses, not that of the equilibrium. Autarky at any common price is an equilibrium (Remark \ref{NoSpecialization}), so that the equilibrium is never unique, and, as soon as an equilibrium with trade exists, autarky and trade coexist. This indeterminacy of the prices is specific to autarky, where the budget constraints impose no restriction on the prices; in an equilibrium with trade, the binding budget constraints determine the prices, as the example of Section \ref{Numerical simulation} illustrates. Multiplicity therefore concerns the transactions rather than the prices that support them, and uniqueness can only be sought for a selection of equilibria, such as the limit points of Proposition \ref{LexicographicLimit} or the rest points of the adjustment processes of Section \ref{Convergence}.
\end{remark}

\vskip 0.5cm

\subsection{Equilibrium Existence under Hierarchical Preferences\label{Equilibrium Existence under Hierarchical Preferences}}

\hskip 0.5cm Suppose that each consumer has hierarchical (lexicographic) preferences over goods and trading partners. That is, consumer $i$ first maximizes the utility of its consumption of goods, and in case of ties, selects preferred trading partners. Preferences are described by a pair of continuous utility functions $(u^i, \vartheta^i)$; as recalled above, such preferences are rational but not continuous, and cannot be represented by a single utility function. The consumer's problem is thus a sequential maximization problem.\\

We define consumer $i$'s total outgoing and incoming flows by
\begin{align}
\label{eqn:66}
x^i_{i\rightarrow} &= \sum_{j\neq i} x^i_{ij} \qquad , \qquad x^i_{i\leftarrow} = \sum_{j\neq i} x^i_{ji}.
\end{align}

\medskip

In the first stage, consumer $i$ solves, under assumptions \emph{a}, \emph{b} and \emph{c},
\begin{align}
\label{eqn:67}
\max_{p^i \in P} \max_{X^i\in \mathcal{B}^i(p,X^{-i}) \cap \mathfrak{T}^i} u^i(x^i).
\end{align}
The set $\mathcal{A}^i(p^{-i},X^{-i})$ of its solutions $(p^i,X^i)$ is non-empty and compact, as in Section \ref{New way to consider the Consumer Problem}. In the second stage, consumer $i$ selects among first-stage maximizers by solving
\begin{align}
\label{eqn:68}
\max_{(p^i,X^i)\in \mathcal{A}^i(p^{-i},X^{-i})}  \vartheta^i(x^i_{i\leftarrow}, x^i_{i\rightarrow}),
\end{align}
where $\vartheta^i$ is continuous, strictly increasing and quasi-concave. Since $\vartheta^i$ is continuous and $\mathcal{A}^i(p^{-i},X^{-i})$ is compact, the second-stage problem admits a solution.

\vskip 0.5cm

Berge's theorem and Kakutani's fixed point theorem cannot, however, be applied to the lexicographic best response. First, the correspondence $\mathcal{A}^i$ is not lower semicontinuous in general, so that the second-stage problem is a maximization over a discontinuous constraint correspondence, and its solution need not be upper semicontinuous. Second, $\mathcal{A}^i$ need not be convex-valued, since the indirect utility is quasi-convex in $p^i$ (Remark \ref{QuasiConvexValue}). The existence of an equilibrium nevertheless holds, by the autarky argument.

\vskip 0.5cm

\begin{proposition}{\textbf{Equilibrium in an Effective Trade Economy with Hierarchical Preferences\label{Equilibrium in an Effective Trade Economy with Hierarchical Preferences}}\\}
The game $G=\{(p^i,X^i), \mathcal{B}^i\cap\mathfrak{T}^i, (u^i,\vartheta^i)\}_{i=1}^n$ with lexicographic preferences admits a Nash equilibrium.
\end{proposition}

\vskip 0.5cm

\begin{proof}
At the autarky profile $(p,0)$, the bilateral constraints force $X^i=0$ for every consumer, whatever its price, so that $(p^i,0)$ solves both stages for every $i$ and every $p\in P^n$.
\end{proof}

\vskip 0.5cm

A rigorous way to give content to the hierarchy beyond autarky is to approximate it by continuous preferences. For $\varepsilon>0$, consider the utility
\begin{align}
\label{eqn:69}
\tilde u^i_\varepsilon(X^i)=u^i(x^i)+\varepsilon\,\vartheta^i(x^i_{i\leftarrow}, x^i_{i\rightarrow}),
\end{align}
which falls within the framework of Section \ref{Topological Influences on Consumer Choice and Market Coordination}.

\vskip 0.5cm

\begin{proposition}{\textbf{Lexicographic limit\label{LexicographicLimit}}\\}
Fix $(p^{-i},X^{-i})$, let $F^i$ be the compact set of pairs $(p^i,X^i)$ satisfying the constraints of \ref{eqn:67}, and let $(p^i_\varepsilon,X^i_\varepsilon)$ maximize $\tilde u^i_\varepsilon$ over $F^i$. Then every limit point of $(p^i_\varepsilon,X^i_\varepsilon)$ as $\varepsilon\to0$ solves the lexicographic problem \ref{eqn:67}--\ref{eqn:68}.
\end{proposition}

\vskip 0.5cm

\begin{proof}
Write $z_\varepsilon=(p^i_\varepsilon,X^i_\varepsilon)$, and let $z_\varepsilon\to\bar z$ along a sequence $\varepsilon\to0$. For every $z\in F^i$, $u^i(z_\varepsilon)+\varepsilon\vartheta^i(z_\varepsilon)\ge u^i(z)+\varepsilon\vartheta^i(z)$; letting $\varepsilon\to0$ gives $u^i(\bar z)\ge u^i(z)$, so that $\bar z\in\mathcal{A}^i$. For every $z\in\mathcal{A}^i$, $u^i(z)\ge u^i(z_\varepsilon)$, hence $\varepsilon\vartheta^i(z_\varepsilon)\ge\varepsilon\vartheta^i(z)+u^i(z)-u^i(z_\varepsilon)\ge\varepsilon\vartheta^i(z)$, so that $\vartheta^i(z_\varepsilon)\ge\vartheta^i(z)$ and, letting $\varepsilon\to0$, $\vartheta^i(\bar z)\ge\vartheta^i(z)$.
\end{proof}

\vskip 0.5cm

The lexicographic consumer is thus the limit of consumers who give a vanishing weight to their trading relationships. For $\varepsilon>0$, the equilibria of the economy with utilities \ref{eqn:69} enjoy, under the assumptions of Proposition \ref{ConcaveEquilibrium}, the properties established above; their limit points as $\varepsilon\to0$ provide a selection among the possible outcomes of the hierarchical economy. Since the lexicographic best response is discontinuous, such limit points need not be equilibria of the hierarchical economy, and their characterization is left for future research.

\vskip 0.5cm

\section{Numerical simulation: Exchange with indivisible goods\label{Numerical simulation}}

\hskip 0.5cm In the following, we construct an economy consisting of two markets for indivisible goods, labeled $l = 1,2$, and three consumers, denoted by $i = a, b, c$. Each consumer $i$ is endowed with an initial wealth $w^i = (w^i_1, w^i_2)$ and has a constant elasticity of substitution (CES) utility function of the form $\displaystyle u^i(x^i_1, x^i_2) = \left(\alpha_i (x^i_1)^r + (1 - \alpha_i) (x^i_2)^r \right)^{\frac{1}{r}}$. The allocation of consumer $i$ is defined as $x^i = (x^i_1, x^i_2) = \left(w^i_1 + \sum_{j\neq i} x^i_{ji,1} - \sum_{j\neq i} x^i_{ij,1}, \quad w^i_2 + \sum_{j\neq i} x^i_{ji,2} - \sum_{j\neq i} x^i_{ij,2} \right)$. For numerical analysis, we set the parameters as follows:  
\begin{align*}
w^a = (3,1), \quad w^b = (2,2), \quad w^c = (1,3), \quad r = 0.3, \quad \alpha_a = 0.2, \quad \alpha_b = 0.4, \quad \alpha_c = 0.8.
\end{align*}
In the tables below, consumers $a$, $b$ and $c$ are indexed by $1$, $2$ and $3$, so that $x_{ij,k}$ is the quantity of good $k$ sent by consumer $i$ to consumer $j$, and, since $L=2$, only the first component $p^i_1$ of each price vector is reported, the second being $p^i_2=1-p^i_1$.\\

We begin with a counter-example to prove Proposition \ref{Second welfare theorem}. In the case of a continuous good, we can compute a Pareto-optimal allocation of $\mathcal{A}$, that is, among the allocations that satisfy the budget constraints of all consumers, which is given by the following table:  
\begin{table}[h!]
\centering
\resizebox{\textwidth}{!}{%
\begin{tabular}{|c|c|c|c|}
\hline
$(u^a,u^b,u^c)$ & $(p^a_1,p^b_1,p^c_1)$ & $(x_{12,1}, x_{13,1}, x_{23,1}, x_{21,1}, x_{31,1}, x_{32,1})$ & $(x_{12,2}, x_{13,2}, x_{23,2}, x_{21,2}, x_{31,2}, x_{32,2})$ \\
\hline
$(4.03, 0, 3.68)$  & $(0.49, 0.73, 0)$ & $(1.03, 1.12, 3.03, 0, 0, 0)$ & $(0, 0.66, 0, 5.09, 0, 3.09)$ \\
\hline
\end{tabular}}
\caption{Pareto-optimal allocation in the continuous good case}
\label{tab:1}
\end{table}

In this allocation, consumer $b$ acts as a pure intermediary and ends with $x^b=(0,0)$, so that $u^b=0$; all budget constraints are binding, up to rounding, in accordance with \ref{eqn:36}. Since autarky is always feasible, every consumer obtains at a transaction equilibrium at least its autarky utility, here $u^b(w^b)=2$. Hence no transaction equilibrium realizes this Pareto optimal allocation.\\

Now, consider the discrete case. Transactions are integer, a consumer does not sell more of a good than its endowment of this good (no resale), and two consumers do not exchange the same good in both directions. A transaction profile is feasible if there exists a price profile $p\in P^3$ at which the budget constraints of all consumers are satisfied, that is, if it induces an allocation of $\mathcal{A}$; several price profiles may support the same transaction profile, for instance any price profile in the case of the null solution. The total number of feasible solutions is $3334$, of which $771$ are Pareto optimal.\\

At a transaction equilibrium, bilateral proposals coincide (Remark \ref{MaximizationEquivalence}), so that a deviation of consumer $i$ consists in reducing some of its transactions and choosing a new price vector; by Remark \ref{QuasiConvexValue}, such a deviation is affordable if and only if the largest quantity of a good sold covers the value of the purchases at the prices of the sellers. A feasible transaction profile is a transaction equilibrium if there exists a supporting price profile at which no consumer has an affordable deviation that strictly increases its utility. We present the set of Nash equilibria, including a supporting price vector $(p^a_1, p^b_1, p^c_1)$, the allocation of good 1 $(x_{12,1}, x_{13,1}, x_{23,1}, x_{21,1}, x_{31,1}, x_{32,1})$, the allocation of good 2 $(x_{12,2}, x_{13,2}, x_{23,2}, x_{21,2}, x_{31,2}, x_{32,2})$, and the corresponding utilities $(u^a, u^b, u^c)$:

\begin{table}[htb!]
\centering
\resizebox{\textwidth}{!}{%
\begin{tabular}{|l|c|c|c|}
\hline
$(u^a,u^b,u^c)$ & $(p^a_1,p^b_1,p^c_1)$ & $(x_{12,1}, x_{13,1}, x_{23,1}, x_{21,1}, x_{31,1}, x_{32,1})$ & $(x_{12,2}, x_{13,2}, x_{23,2}, x_{21,2}, x_{31,2}, x_{32,2})$ \\
\hline
$(1.28, 2, 1.28)$ & $(0.27, 0, 1)$ & $(0, 0, 0, 0, 0, 0)$ & $(0, 0, 0, 0, 0, 0)$ \\
\hline
$(1.28, 2, 1.28)$ & $(0.23, 0.23, 0.23)$ & $(0, 0, 0, 0, 0, 0)$ & $(0, 1, 0, 1, 0, 1)$ \\
\hline
$(1.28, 2, 1.28)$ & $(0.65, 0.65, 0.65)$ & $(0, 0, 0, 0, 0, 0)$ & $(1, 0, 1, 0, 1, 0)$ \\
\hline
$(1.28, 2, 1.28)$ & $(0.22, 0.22, 0.22)$ & $(0, 1, 0, 1, 0, 1)$ & $(0, 0, 0, 0, 0, 0)$ \\
\hline
$(1.28, 2, 1.28)$ & $(0.22, 0.22, 0.22)$ & $(1, 0, 1, 0, 1, 0)$ & $(0, 0, 0, 0, 0, 0)$ \\
\hline
$(2, 2, 2)$ & $(1, 0, 0)$ & $(0, 1, 0, 0, 0, 0)$ & $(0, 0, 0, 0, 1, 0)$ \\
\hline
$(2, 2, 2)$ & $(0.34, 0.34, 0.66)$ & $(1, 0, 1, 0, 0, 0)$ & $(0, 0, 0, 0, 1, 0)$ \\
\hline
$(2, 2, 2)$ & $(0.59, 0.41, 0.41)$ & $(0, 1, 0, 0, 0, 0)$ & $(0, 0, 0, 1, 0, 1)$ \\
\hline
$(1.28, 2.02, 2)$ & $(0, 1, 0)$ & $(0, 0, 1, 0, 0, 0)$ & $(0, 0, 0, 0, 0, 1)$ \\
\hline
$(1.28, 2.02, 2)$ & $(0.26, 0.74, 0.26)$ & $(0, 0, 1, 0, 0, 0)$ & $(1, 0, 0, 0, 1, 0)$ \\
\hline
$(1.28, 2.02, 2)$ & $(0.94, 0.94, 0.06)$ & $(0, 1, 0, 1, 0, 0)$ & $(0, 0, 0, 0, 0, 1)$ \\
\hline
\textcolor{red}{$(2.47, 2, 2.47)$}* & \textcolor{red}{$(1, 0, 0)$}* & \textcolor{red}{$(0, 2, 0, 0, 0, 0)$}* & \textcolor{red}{$(0, 0, 0, 0, 2, 0)$}* \\
\hline
\textcolor{red}{$(2.47, 2, 2.47)$}* & \textcolor{red}{$(0.61, 0.39, 0.39)$}* & \textcolor{red}{$(0, 2, 0, 0, 0, 0)$}* & \textcolor{red}{$(0, 0, 0, 2, 0, 2)$}* \\
\hline
\textcolor{red}{$(2.47, 2, 2.47)$}* & \textcolor{red}{$(0.57, 0.43, 0.43)$}* & \textcolor{red}{$(0, 2, 0, 0, 0, 0)$}* & \textcolor{red}{$(0, 0, 0, 1, 1, 1)$}* \\
\hline
\textcolor{red}{$(2.47, 2, 2.47)$}* & \textcolor{red}{$(0.51, 0.51, 0.49)$}* & \textcolor{red}{$(2, 0, 2, 0, 0, 0)$}* & \textcolor{red}{$(0, 0, 0, 0, 2, 0)$}* \\
\hline
\textcolor{red}{$(2.47, 2, 2.47)$}* & \textcolor{red}{$(0.6, 0.6, 0.4)$}* & \textcolor{red}{$(1, 1, 1, 0, 0, 0)$}* & \textcolor{red}{$(0, 0, 0, 0, 2, 0)$}* \\
\hline
\textcolor{red}{$(2, 2.02, 2.47)$}* & \textcolor{red}{$(0.97, 0.97, 0.03)$}* & \textcolor{red}{$(0, 1, 1, 0, 0, 0)$}* & \textcolor{red}{$(0, 0, 0, 0, 1, 1)$}* \\
\hline
\textcolor{red}{$(2, 2.02, 2.47)$}* & \textcolor{red}{$(0.53, 0.53, 0.47)$}* & \textcolor{red}{$(0, 2, 0, 1, 0, 0)$}* & \textcolor{red}{$(0, 0, 0, 0, 1, 1)$}* \\
\hline
\textcolor{red}{$(2, 2.02, 2.47)$}* & \textcolor{red}{$(0.58, 0.58, 0.42)$}* & \textcolor{red}{$(1, 0, 2, 0, 0, 0)$}* & \textcolor{red}{$(0, 0, 0, 0, 1, 1)$}* \\
\hline
\end{tabular}}
\caption{Nash equilibria in the discrete goods case: Pareto-optimal equilibria are highlighted in red, while asterisk is used to mark Nash--Pareto equilibria.}
\label{tab:2}
\end{table}

There are nineteen Nash equilibria corresponding to five utility profiles, eight of which are Nash--Pareto equilibria (and Pareto optimal). In every equilibrium, each seller sells a single good, at a positive price, in accordance with Theorem \ref{Specialization} and Proposition \ref{PositivePrices}, and all budget constraints are binding. Several equilibria are circular exchanges of a single good, which leave the allocation unchanged with respect to autarky. The equilibrium with utilities $(2.47,2,2.47)$ and prices $(1,0,0)$ is the cycle equilibrium of Proposition \ref{TradeEquilibrium}: consumers $a$ and $c$, who sell goods 1 and 2 at the vertices of the simplex, exchange them one for one, at the level $t^\star=2$ preferred by both of them.\\  

Now, suppose we introduce topological constraints—for example, consumers $a$ and $c$ cannot trade. Under these restrictions, and with the same rules, the number of feasible solutions is reduced to $225$, of which $56$ are Pareto optimal. However, the set of Nash equilibria is reduced to two situations, none of them is Pareto optimal.\\

\begin{table}[h!]
\centering
\resizebox{\textwidth}{!}{%
\begin{tabular}{|l|c|c|c|}
\hline
$(u^a,u^b,u^c)$ & $(p^a_1,p^b_1,p^c_1)$ & $(x_{12,1}, x_{13,1}, x_{23,1}, x_{21,1}, x_{31,1}, x_{32,1})$ & $(x_{12,2}, x_{13,2}, x_{23,2}, x_{21,2}, x_{31,2}, x_{32,2})$ \\
\hline
$(1.28, 2, 1.28)$  &$(0.27, 0, 1)$ & $(0, 0, 0, 0, 0, 0)$ & $(0, 0, 0, 0, 0, 0)$ \\
\hline
$(1.28, 2.02, 2)$* & $(0.87, 1, 0)$* & $(0, 0, 1, 0, 0, 0)$* & $(0, 0, 0, 0, 0, 1)$*\\
\hline
\end{tabular}}
\caption{Nash equilibria in the discrete goods case under topological constraints: Nash--Pareto equilibria are marked with an asterisk.}
\label{tab:3}
\end{table}

This numerical example is crucial for understanding the price vector $p$. In our model, price is subjective and is a seller's price: by Proposition \ref{PositivePrices}, a null price $p^i_k$ indicates that agent $i$ does not sell good $k$, although it may buy it at the price of its sellers, as consumer $c$ does in the sixth equilibrium of Table \ref{tab:2}. The price of a good that an agent does not sell plays no role in any budget constraint and is therefore indeterminate; at the vertex solutions of Remark \ref{QuasiConvexValue}, it is zero. This is natural, as price is a market-related concept determined by supply and demand forces. It should not be confused with intrinsic value.\footnote{The distinction between use value and exchange value can be traced back to Adam Smith \cite{Smith1776}.} Furthermore, the topology plays a crucial role in shaping equilibria, influencing welfare losses, and highlighting imperfections caused by intermediaries.

\vskip 0.5cm

\subsection{Relational preferences and price conformity}

\hskip 0.5cm We now compute the equilibria of the same economy when the consumers have the utility \ref{eqn:64} of Section \ref{Market topology}, with the relational term and the price term
\begin{align}
\label{eqn:70}
\vartheta^i(X^i)=\beta\sum_{j\neq i}\Big(\sum_{k=1}^{2}\big(x^i_{ij,k}+x^i_{ji,k}\big)\Big)^{1/2},\qquad \varrho^i(p^i,p^{-i})=-\frac{\kappa}{2}\sum_{j\neq i}\big\|p^j-p^i\big\|^2,
\end{align}
and $\beta=0.1$, $\kappa=1$. The term $\vartheta^i$ values each trading relationship, with decreasing returns in its volume, and therefore rewards the diversification of partners; the term $\varrho^i$ expresses the tendency to stay close to the prices of the market. The feasible transaction profiles are the $3334$ profiles of Table \ref{tab:2}. Given a transaction profile, the price that maximizes $\varrho^i$ among the affordable ones is the projection of the average price of the others on the set of affordable prices; we compute the fixed points of these price responses, and keep a profile $(p,X)$ when no consumer can increase $\tilde u^i$ by reducing some of its transactions and choosing its price accordingly. With $\beta=\kappa=0$, this procedure recovers exactly the nineteen equilibria of Table \ref{tab:2}.\\

The two additional terms change the set of equilibria substantially. The number of equilibria rises from $19$ to $34$, and the number of utility profiles $(u^a,u^b,u^c)$ from $5$ to $9$; $13$ equilibria induce a Pareto optimal allocation of $\mathcal{A}$, against $8$ in Table \ref{tab:2}, Pareto optimality being evaluated, as before, with the utilities $u^i$ of the goods. Above all, the specialization of sellers is no longer imposed: in four equilibria, reported in Table \ref{tab:4}, consumer $b$ sells both goods, as anticipated in Remark \ref{NoSpecialization}. In all of them, $b$ acts as an intermediary that values its relationships with both partners, and the prices are those of a consensus, $p^a=p^b=p^c$. Price conformity also sustains new terms of trade: in the equilibria with utilities $(2.78,2,1.76)$, of which two are reported in Table \ref{tab:4}, consumer $a$ obtains two units of good 2 for one unit of good 1, a terms of trade that consumer $c$ could not refuse without moving its price away from those of the market.

\begin{table}[htb!]
\centering
\resizebox{\textwidth}{!}{%
\begin{tabular}{|l|c|c|c|}
\hline
$(u^a,u^b,u^c)$ & $(p^a_1,p^b_1,p^c_1)$ & $(x_{12,1}, x_{13,1}, x_{23,1}, x_{21,1}, x_{31,1}, x_{32,1})$ & $(x_{12,2}, x_{13,2}, x_{23,2}, x_{21,2}, x_{31,2}, x_{32,2})$ \\
\hline
$(1.28, 2.02, 2)$ & $(0.5, 0.5, 0.5)$ & $(0, 0, 1, 0, 0, 0)$ & $(0, 1, 0, 1, 0, 2)$ \\
\hline
$(2, 2, 2)$ & $(0.5, 0.5, 0.5)$ & $(1, 0, 1, 0, 0, 0)$ & $(0, 0, 0, 1, 0, 1)$ \\
\hline
\textcolor{red}{$(2, 2.02, 2.47)$} & \textcolor{red}{$(0.5, 0.5, 0.5)$} & \textcolor{red}{$(0, 1, 1, 0, 0, 0)$} & \textcolor{red}{$(0, 0, 0, 1, 0, 2)$} \\
\hline
\textcolor{red}{$(2.47, 2, 2.47)$} & \textcolor{red}{$(0.5, 0.5, 0.5)$} & \textcolor{red}{$(1, 1, 1, 0, 0, 0)$} & \textcolor{red}{$(0, 0, 0, 1, 1, 1)$} \\
\hline
\textcolor{red}{$(2.78, 2, 1.76)$} & \textcolor{red}{$(0.75, 0.6875, 0.625)$} & \textcolor{red}{$(0, 1, 0, 0, 0, 0)$} & \textcolor{red}{$(0, 0, 0, 0, 2, 0)$} \\
\hline
\textcolor{red}{$(2.78, 2, 1.76)$} & \textcolor{red}{$(0.75, 0.75, 0.625)$} & \textcolor{red}{$(1, 0, 1, 0, 0, 0)$} & \textcolor{red}{$(0, 0, 0, 0, 2, 0)$} \\
\hline
\end{tabular}}
\caption{Equilibria with relational preferences and price conformity ($\beta=0.1$, $\kappa=1$): the first four equilibria are not specialized, consumer $b$ selling both goods; the last two exhibit new terms of trade. Allocations that are Pareto optimal in $\mathcal{A}$ are highlighted in red.}
\label{tab:4}
\end{table}

The terms $\vartheta^i$ and $\varrho^i$ therefore play distinct roles. In this example, price conformity is the main source of non-specialized equilibria: with $\kappa=0$, non-specialized equilibria appear only for larger values of $\beta$, whereas for $\kappa=1$ and $\kappa=5$ their number grows with both parameters. Since the fixed points of the price responses are computed from a finite set of initial prices, the number $34$ is a lower bound on the number of equilibria.\\

Finally, the example illustrates Remark \ref{Uniqueness}: the uniqueness of the best responses, established in Proposition \ref{ConcaveEquilibrium}, does not imply the uniqueness of the equilibrium. The set of equilibria is larger than in Table \ref{tab:2}, and autarky remains an equilibrium for every common price vector, $p^a=p^b=p^c$, since the budget constraints then do not determine the prices. In the equilibria with trade, by contrast, the binding budget constraints determine the prices: on a grid of step $0.01$, the non-specialized equilibria of Table \ref{tab:4} are supported by the single common price $p^a_1=p^b_1=p^c_1=0.5$, and the last two equilibria by no common shift of the reported prices. Multiplicity thus concerns the transactions, not the prices that support them: the economy remains far from the determinacy of the Arrow--Debreu model, and uniqueness must be sought for a selection of equilibria.

\vskip 0.5cm

\section{Production\label{Production}}

\hskip 0.5cm In this section, we consider an economy $\mathcal{E} = \left(w^{i_d}, u^{i_d}, \mathscr{X}^{i_d}, \mathscr{Y}^{i_s} \right)_{1\leq {i_d} \leq n_d , 1\leq {i_s} \leq n_s}$ consisting of $L$ goods. The economy includes $n_d$ consumption units, each characterized by an endowment $w^{i_d}$, a utility function $u^{i_d}$, and a consumption set $\mathscr{X}^{i_d}$. The consumption units solve the problem \ref{eqn:11} under the assumptions of Section \ref{New way to consider the Consumer Problem}, with the budget constraint \ref{eqn:75} below, which includes the distributed profits.\\

Additionally, there are $n_s$ production units, each defined by its production-possibility set $\mathscr{Y}^{i_s}$; production units have no endowment, and sell what they produce. The total number of agents in the economy is $n = n_d + n_s$. For a transaction plan $X^{i_s}$, the net production of producer $i_s$ is $y^{i_s}(X^{i_s})=\sum_{j\neq {i_s}} x^{i_s}_{{i_s}j} - \sum_{j\neq {i_s}} x^{i_s}_{j{i_s}}$, its sales minus its purchases of inputs, and its feasible set is
\begin{align}
\label{eqn:71}
\mathcal{P}^{i_s}(X^{-{i_s}}) &=\left\{X^{i_s}\in \mathbb{X}^{i_s} \ \mid \ y^{i_s}(X^{i_s}) \in \mathscr{Y}^{i_s} \ , \  x_{{i_s}j}^{i_s} \leq x_{{i_s}j}^j \ , \ x_{j{i_s}}^{i_s} \leq x_{j{i_s}}^j\right\} \,,
\end{align}
which does not depend on the prices. The producer problem is to maximize the profit, in the same sequential form as the consumer problem,
\begin{align}
\label{eqn:72}
\max_{p^{i_s}\in P}\ \max_{X^{i_s}\in \mathcal{P}^{i_s}(X^{-{i_s}})} \pi^{i_s}(p,X^{i_s}),
\end{align}
where
\begin{align}
\label{eqn:73}
\pi^{i_s}(p,X^{i_s})=\sum_{j\neq {i_s}} p^{i_s} \cdot  x_{{i_s}j}^{i_s} - \sum_{j\neq {i_s}} p^j \cdot  x_{j{i_s}}^{i_s}.
\end{align}

Here, $\mathbb{X}^{i_s}$ is non-empty, closed, bounded and convex, and satisfies assumption \emph{d}, and $\mathscr{Y}^{i_s}$ satisfies the classical assumptions:
\begin{enumerate}
\item[\emph{g}] \textbf{No free lunch:} $\mathscr{Y}^{i_s} \cap \R_+^L\subseteq \{0\}$.
\item[\emph{h}] \textbf{Free disposal:} if $y^{i_s}\in \mathscr{Y}^{i_s}$ and $y'^{i_s} \leq y^{i_s}$ then $y'^{i_s}\in\mathscr{Y}^{i_s}$.
\item[\emph{i}] \textbf{Possibility of inaction:} $0\in\mathscr{Y}^{i_s}$.
\item[\emph{j}] \textbf{Convexity:} $\mathscr{Y}^{i_s}$ is a convex set of $\R^L$.
\item[\emph{k}] \textbf{Closedness:} $\mathscr{Y}^{i_s}$ is a closed set of $\R^L$.
\end{enumerate}

We allow each consumption unit ${i_d}$ to receive a share $\theta_{{i_s}{i_d}}\ge0$ of the profit of the ${i_s}$th production unit, where
\begin{align}
\label{eqn:74}
\sum_{i_d=1}^{n_d} \theta_{{i_s}{i_d}} = 1.
\end{align}
Only non-negative profits are distributed, losses being borne by the producer, so that the dividend received by consumer $i_d$ is $D^{i_d}(p,X)=\sum_{i_s=1}^{n_s}\theta_{{i_s}{i_d}}\,\pi^{i_s}(p,X^{i_s})^+$, where $a^+=\max(a,0)$. Thus, the new transaction condition of each consumer ${i_d}$ is
\begin{align}
\label{eqn:75}
\sum_{j\neq {i_d}} p^j \cdot x^{i_d}_{j{i_d}} \leq \sum_{j\neq {i_d}} p^{i_d} \cdot x^{i_d}_{{i_d}j} + D^{i_d}(p,X).
\end{align}
The dividend depends on the strategies of the producers, so that the budget set of a consumer depends, as in a generalized game, on the strategies of the other agents. Since $D^{i_d}\ge0$, the budget set contains autarky, and, by Remark \ref{MaximumCharacterization}, the budget constraint binds at the optimum.

\vskip 0.5cm

\begin{remark}{\textbf{Walras' law with production\label{ProductionWalras}}\\}
Summing the value of sales minus the value of purchases over all agents gives zero, as in \ref{eqn:36}. For a producer, this difference is its profit; for the consumers, it is the opposite of the total profit. Hence, whenever all profits are non-negative, the sum over the consumers of the budget slacks in \ref{eqn:75} is $-\sum_{i_s}\pi^{i_s}+\sum_{i_d}D^{i_d}=0$, and all the budget constraints of the consumers that are satisfied are binding. Since inaction is always feasible, the profit of a producer is non-negative at its optimum, and in particular at every transaction equilibrium.
\end{remark}

\vskip 0.5cm

\begin{proposition}{\textbf{Feasible set of a producer\label{ProducerFeasibleSet}}\\}
The correspondence $\mathcal{P}^{i_s}$ has non-empty, convex and compact values, and is continuous.
\end{proposition}

\vskip 0.5cm

\begin{proof}
The set $\mathcal{P}^{i_s}(X^{-i_s})$ contains $0$, by the possibility of inaction; it is convex, since $y^{i_s}$ is linear and $\mathscr{Y}^{i_s}$ is convex, and compact, since $\mathbb{X}^{i_s}$ is compact and $\mathscr{Y}^{i_s}$ is closed. Its graph is closed, since all its constraints are continuous, so that $\mathcal{P}^{i_s}$ is upper semicontinuous. For lower semicontinuity, let $X^{i_s}\in\mathcal{P}^{i_s}(X^{-i_s})$, let $X^{-i_s,m}\to X^{-i_s}$, and let $\varepsilon_m\downarrow0$ be chosen as in Section \ref{New way to consider the Consumer Problem}. The plan $(1-\varepsilon_m)X^{i_s}$ belongs to $\mathbb{X}^{i_s}$ by assumption \emph{d}, satisfies the bilateral constraints at $X^{-i_s,m}$, and its net production $(1-\varepsilon_m)y^{i_s}(X^{i_s})$ belongs to $\mathscr{Y}^{i_s}$, by convexity and the possibility of inaction. It converges to $X^{i_s}$. No budget constraint is involved, so that no condition on the prices is needed.
\end{proof}

\vskip 0.5cm

\begin{proposition}{\textbf{Existence of an equilibrium with production\label{ProductionAutarky}}\\}
The autarky profile $(p,0)$ is a transaction equilibrium of the economy with production, for every $p\in P^n$.
\end{proposition}

\vskip 0.5cm

\begin{proof}
If the other agents propose no trade, the bilateral constraints force $X^i=0$ for every agent $i$. The profit of every producer is then zero, whatever its price, no dividend is distributed, and every consumer keeps its endowment, whatever its price.
\end{proof}

\vskip 0.5cm

As in Section \ref{New way to consider the Consumer Problem}, the fixed point theorems cannot be applied directly. For a fixed plan $X^{i_s}$, the profit \ref{eqn:73} is linear in $p^{i_s}$, and is maximized at a vertex of the simplex, $p^{i_s}=e_k$, where $k$ is the good sold in maximal quantity. The optimal value of the inner problem, $\max_k S^{i_s}_k(X^{i_s})-\sum_{j\neq i_s}p^j\cdot x^{i_s}_{ji_s}$, where $S^{i_s}(X^{i_s})=\sum_{j\neq i_s}x^{i_s}_{i_sj}$, is a maximum of linear functions, hence convex in $X^{i_s}$, and the set of optimal price-supply pairs is not convex in general. The existence of an equilibrium with production and trade is therefore not guaranteed by Proposition \ref{ProductionAutarky}. Proposition \ref{TradeEquilibrium} still yields an equilibrium with trade among the consumers when they form a trade cycle with gains from trade: if no agent proposes to trade with the producers, the bilateral constraints keep the producers inactive, their profits and the dividends are zero, and the budgets of the consumers are those of the exchange economy. This equilibrium, however, involves no production. We now present two ways to recover the classical tools: the specialized economies, in which the prices are fixed by dominance, and a modification of the objectives, which restores concavity.

\vskip 0.5cm

\subsection{Specialized economies with production\label{Specialized economies with production}}

\hskip 0.5cm Given a specialization profile $\mathbf{k}=(k_1,\dots,k_n)$, which assigns a good to each consumer and to each producer, we denote by $\mathcal{E}_{\mathbf{k}}$ the economy in which the admissible transactions of every agent $i$ are restricted to $\mathbb{X}^i_{\mathbf{k}}$, as in \ref{eqn:20}: agent $i$ sells only good $k_i$. We write $p_{\mathbf{k}}=(e_{k_1},\dots,e_{k_n})$ for the profile of price vectors that put all their weight on the goods sold.

\vskip 0.5cm

\begin{proposition}{\textbf{Dominant prices\label{DominantPrice}}\\}
In $\mathcal{E}_{\mathbf{k}}$, for every agent $i$ and every strategy profile $(p^{-i},X^{-i})$ of the other agents, the price vector $e_{k_i}$ is optimal: for every $p^i\in P$, the value of the problem of agent $i$ at $p^i$ does not exceed its value at $e_{k_i}$.
\end{proposition}

\vskip 0.5cm

\begin{proof}
In $\mathcal{E}_{\mathbf{k}}$, agent $i$ sells only good $k_i$, so that its price enters its problem only through the value of its sales $p^i_{k_i}S^i_{k_i}$, and $p^i_{k_i}\le1$. For a consumer, a plan that satisfies the budget constraint \ref{eqn:75} at $p^i$ satisfies it at $e_{k_i}$, the dividend being unchanged. For a producer, $\pi^{i}\big((e_{k_i},p^{-i}),X^i\big)\ge\pi^i(p,X^i)$ for every plan $X^i$.
\end{proof}

\vskip 0.5cm

\begin{proposition}{\textbf{Equilibrium of a specialized economy with production\label{SpecializedProduction}}\\}
Assume \emph{a}, \emph{c} and \emph{d} for the consumers, and \emph{d} and \emph{g}--\emph{k} for the producers. For every specialization profile $\mathbf{k}$:
\begin{enumerate}
\item The quantity game $G_{\mathbf{k}}$, in which the prices are fixed at $p_{\mathbf{k}}$, the strategy sets are the sets $\mathbb{X}^i_{\mathbf{k}}$, the constraint correspondences are the budget sets \ref{eqn:75} of the consumers and the feasible sets \ref{eqn:71} of the producers, and the payoffs are $u^{i_d}$ and $\pi^{i_s}(p_{\mathbf{k}},\cdot)$, admits a Nash equilibrium $X^\star$.
\item For every Nash equilibrium $X^\star$ of $G_{\mathbf{k}}$, $(p_{\mathbf{k}},X^\star)$ is a transaction equilibrium of $\mathcal{E}_{\mathbf{k}}$.
\end{enumerate}
\end{proposition}

\vskip 0.5cm

\begin{proof}
\begin{enumerate}
\item We check the assumptions of Proposition \ref{Equilibrium in effective trade economy}. The strategy sets are non-empty, convex and compact. The feasible sets of the producers are continuous, with non-empty convex values, by Proposition \ref{ProducerFeasibleSet}. The budget sets of the consumers at the fixed prices $p_{\mathbf{k}}$ are non-empty, since they contain autarky, and convex, since their constraints are linear in $X^{i_d}$. They are continuous: upper semicontinuity follows from the closedness of their graph, the dividends being continuous; for lower semicontinuity, given a plan $X^{i_d}$ in the budget set and a converging sequence of strategies of the others, the plans $(1-\varepsilon_m)X^{i_d}$ satisfy the bilateral constraints, as in Section \ref{New way to consider the Consumer Problem}, and the budget constraint, since $(1-\varepsilon_m)\big(c^{i_d}-p^{i_d}\cdot S^{i_d}\big)\le(1-\varepsilon_m)D^{i_d}\le D^{i_d,m}$ when $D^{i_d}>0$, provided that $\varepsilon_m$ is also chosen such that $(1-\varepsilon_m)D^{i_d}\le D^{i_d,m}$, which is possible with $\varepsilon_m\to0$ since $D^{i_d,m}\to D^{i_d}$, and $(1-\varepsilon_m)\big(c^{i_d}-p^{i_d}\cdot S^{i_d}\big)\le0\le D^{i_d,m}$ when $D^{i_d}=0$. The payoffs are continuous; $u^{i_d}\circ T^{i_d}$ is quasi-concave in $X^{i_d}$ by assumption \emph{c}, $T^{i_d}$ being affine, and $\pi^{i_s}(p_{\mathbf{k}},\cdot)$ is linear in $X^{i_s}$. Hence $G_{\mathbf{k}}$ admits a Nash equilibrium.

\item Let $(p^i,X^i)$ be a strategy of agent $i$ in $\mathcal{E}_{\mathbf{k}}$. By Proposition \ref{DominantPrice}, its value does not exceed the value of $(e_{k_i},X^i)$ whenever $X^i$ is feasible at $p^i$, and, since $X^\star$ is a Nash equilibrium of $G_{\mathbf{k}}$, the value of $(e_{k_i},X^i)$ does not exceed that of $(e_{k_i},X^{i\star})$. Hence $(e_{k_i},X^{i\star})$ is a best response of agent $i$, for every $i$.
\end{enumerate}
\end{proof}

\vskip 0.5cm

The structure of Section \ref{New way to consider the Consumer Problem} extends to the economy with production. By Theorem \ref{Specialization}, every consumer sells at most one good at a transaction equilibrium. For a producer, the deviation used in the proof of Theorem \ref{Specialization}, which removes the sales of a good other than the one sold in maximal quantity and puts all the weight of the price on the latter, does not decrease the profit, and remains feasible by free disposal: specialization holds in the weak sense that an optimal plan can always be chosen specialized. As in part 2 of Proposition \ref{Compactness of equilibria}, an equilibrium of $\mathcal{E}_{\mathbf{k}}$ is an equilibrium of $\mathcal{E}$ when no agent can gain by changing its specialization.

\vskip 0.5cm

\begin{remark}{\textbf{Terms of trade in specialized economies\label{OneForOne}}\\}
Proposition \ref{SpecializedProduction} establishes the existence of an equilibrium in every specialized economy by a genuine fixed point argument, but the equilibrium may still be autarky. It also reveals a property of the model. At the prices $p_{\mathbf{k}}$, every traded unit has a value equal to one, whatever the good, so that the binding budget constraint of a consumer states that the number of units it buys equals the number of units it sells, plus its dividend, and the profit of a producer is the number of units it sells minus the number of units it buys. The terms of trade are one for one in physical units, as in all the equilibria of Table \ref{tab:2}. They are thus fixed by the individual normalization \ref{eqn:1}, which bounds the price of the good sold, rather than by supply and demand, and they depend on the units in which goods are measured. A normalization by a reference bundle $\omega\in\R^L_{++}$, $p^i\cdot\omega=1$, would make the model independent of the units, but the terms of trade would then be fixed by the ratios of the components of $\omega$. The modification below removes this property.
\end{remark}

\vskip 0.5cm

\subsection{Restoring concavity}

\hskip 0.5cm Following Section \ref{Topological Influences on Consumer Choice and Market Coordination}, the consumers have the utility \ref{eqn:64}, and the producers maximize
\begin{align}
\label{eqn:76}
\tilde\pi^{i_s}(p,X^{i_s})=\pi^{i_s}(p,X^{i_s})-\gamma^{i_s}(X^{i_s})+\varrho^{i_s}(p^{i_s},p^{-i_s}),
\end{align}
where $\varrho^{i_s}$ is given by \ref{eqn:65}, and $\gamma^{i_s}\ge0$ is a continuous and strictly convex cost of operating transactions, with $\gamma^{i_s}(0)=0$, borne by the producer, such as a cost of organization or of logistics. The distributed profit remains $\pi^{i_s}$. As in Section \ref{Market topology}, the prices are restricted to $P_\pi$, so that condition (C) holds everywhere.

\vskip 0.5cm

\begin{proposition}{\textbf{Equilibrium with production and strictly concave objectives\label{ProductionConcave}}\\}
Assume that the consumers satisfy the assumptions of Proposition \ref{ConcaveEquilibrium}, and that, for every producer $i_s$, the indirect profit
\begin{align*}
\hat\pi^{i_s}(p^{i_s},p^{-i_s},X^{-i_s})=\max_{X^{i_s}\in\mathcal{P}^{i_s}(X^{-i_s})}\Big(\pi^{i_s}(p,X^{i_s})-\gamma^{i_s}(X^{i_s})\Big)
\end{align*}
is continuously differentiable in $p^{i_s}$, with a gradient that is Lipschitz continuous with constant $\ell^{i_s}$, uniformly in $(p^{-i_s},X^{-i_s})$. If $\kappa(n-1)>\ell^i$ for every agent $i$:
\begin{enumerate}
\item Every agent has a unique best response, which depends continuously on the strategies of the other agents.
\item A transaction equilibrium of the economy with production exists, and the set of transaction equilibria is compact.
\end{enumerate}
\end{proposition}

\vskip 0.5cm

\begin{proof}
\begin{enumerate}
\item For the consumers, the argument is that of Proposition \ref{ConcaveEquilibrium}; the continuity of their budget sets extends to the dividends as in the proof of Proposition \ref{SpecializedProduction}, the dividends being continuous and non-negative. For a producer, the objective $\pi^{i_s}(p,\cdot)-\gamma^{i_s}$ of the inner problem is strictly concave, as the difference of a linear and a strictly convex function, on the non-empty convex compact set $\mathcal{P}^{i_s}(X^{-i_s})$, so that the inner problem has a unique solution, which is continuous by Proposition \ref{ProducerFeasibleSet} and part 2 of Theorem \ref{Berge's Maximum Theorem}. As in the proof of Proposition \ref{ConcaveEquilibrium}, $\hat\pi^{i_s}+\varrho^{i_s}$ is strictly concave in $p^{i_s}$ on $P_\pi$, so that the optimal price is unique and continuous.

\item By part 1, the best-response map is a continuous function from the non-empty, convex and compact set $P_\pi^n\times\prod_i\mathbb{X}^i$ into itself, which has a fixed point by Brouwer's fixed point theorem. The set of fixed points of a continuous map is closed, hence compact.
\end{enumerate}
\end{proof}

\vskip 0.5cm

With these objectives, a seller no longer puts all the weight of its price on the good it sells: moving away from the prices of the market has a cost, so that the price of the good sold is no longer pushed to the bound of the normalization, and the terms of trade are no longer fixed one for one, as in the equilibria of Table \ref{tab:4}. Specialization is no longer imposed (Remark \ref{NoSpecialization}). As in the exchange economy, autarky remains an equilibrium, and the uniqueness of the best responses does not imply the uniqueness of the equilibrium (Remark \ref{Uniqueness}).

\vskip 0.5cm
\noindent\textbf{Combining the two approaches.} The two approaches can be combined, the producers being specialized and the consumers having the augmented utility \ref{eqn:64}. This corresponds to producers that maximize their profit without relational preferences, and to consumers that value their trading relationships and the conformity of their prices with those of the market. Let $\mathbf{k}=(k_{i_s})_{1\le i_s\le n_s}$ assign a good to each producer, and let every producer $i_s$ be restricted to the set $\mathbb{X}^{i_s}_{\mathbf{k}}$ of plans that sell only good $k_{i_s}$, as in \ref{eqn:20}, while the consumers are not restricted.

\vskip 0.5cm

\begin{proposition}{\textbf{Specialized producers and augmented consumers\label{MixedProduction}}\\}
Assume \emph{d} and \emph{g}--\emph{k} for the producers, and that the consumers have the utility \ref{eqn:64}, choose their prices in $P_\pi$, and satisfy the assumptions of Proposition \ref{ConcaveEquilibrium}, with $\kappa(n-1)>\ell^{i_d}$ for every consumer $i_d$. Then the economy in which every producer $i_s$ is restricted to $\mathbb{X}^{i_s}_{\mathbf{k}}$ admits a transaction equilibrium, in which every producer sets the price $e_{k_{i_s}}$.
\end{proposition}

\vskip 0.5cm

\begin{proof}
By the argument of Proposition \ref{DominantPrice}, the price $e_{k_{i_s}}$ is optimal for every producer, whatever the strategies of the other agents, and we fix it. At this price, the profit of producer $i_s$ is linear in $X^{i_s}$, and its feasible set $\mathcal{P}^{i_s}(X^{-i_s})\cap\mathbb{X}^{i_s}_{\mathbf{k}}$ is non-empty, convex, compact and continuous, by the proof of Proposition \ref{ProducerFeasibleSet}, since the plans $(1-\varepsilon_m)X^{i_s}$ still sell only good $k_{i_s}$. By part 2 of Theorem \ref{Berge's Maximum Theorem}, its best response is non-empty, convex-valued and upper semicontinuous. A consumer buys from a producer at the price $1$ and from the other consumers at prices in $P_\pi$, so that every traded good has a positive price and, as in Proposition \ref{ProductionConcave}, the best response of every consumer is unique and continuous; the prices $e_{k_{i_s}}$ of the producers, being fixed, only add affine terms to $\varrho^{i_d}$. The product of these best responses is a non-empty, convex-valued and upper semicontinuous correspondence from the convex compact set $P_\pi^{n_d}\times\prod_{i_d}\mathbb{X}^{i_d}\times\prod_{i_s}\mathbb{X}^{i_s}_{\mathbf{k}}$ into itself. By Kakutani's fixed point theorem, it has a fixed point, which, together with the prices $e_{k_{i_s}}$, is a transaction equilibrium.
\end{proof}

\vskip 0.5cm

Neither the cost $\gamma^{i_s}$ nor the price term $\varrho^{i_s}$ is needed for the producers: the convexity of their best responses, which is all that Kakutani's theorem requires, follows from the linearity of the profit at fixed prices. In such an equilibrium, specialized and non-specialized sellers coexist (Remark \ref{PartialSpecialization}): the producers sell a single good, one for one, while the consumers may sell several goods at interior prices that depend on the market. As in Section \ref{Specialized economies with production}, the equilibrium is an equilibrium of the unrestricted economy when no producer can gain by changing the good it sells, and it may be autarky.

\vskip 0.5cm

\section{Exchange with money\label{Money}}

\hskip 0.5cm Let us introduce money. We consider a monetary economy in which money functions both as a medium of exchange and a store of value. The economy consists of $L+1$ markets, corresponding to $L$ goods plus money, with each market containing $n$ consumption units (no production).

\vskip 0.5cm

\subsection{Equilibrium with money}

\hskip 0.5cm Recall that the effective transaction of goods from consumer $i$ to consumer $j$ is defined as \mbox{$q_{ij} = \min\left(x^i_{ij}, x^j_{ij}\right)$}; it generates an equivalent monetary flow from consumer $j$ to consumer $i$: $m_{ji} = p^i \cdot q_{ij}$. Conversely, the effective demand of consumer $i$ from consumer $j$ is given by $q_{ji} = \min\left(x^i_{ji}, x^j_{ji}\right)$, with an equivalent monetary flow in the opposite direction: $m_{ij} = p^j \cdot q_{ji}$.\\

As before, following the alternative definition \ref{eqn:9}--\ref{eqn:10} of the consumer problem, the bilateral proposals coincide on the feasible set, and we define the final holdings of goods and money of consumer $i$ after trade as
\begin{align}
\label{eqn:77}
x^i = w^i + \sum_{j\neq i} x^i_{ji} - \sum_{j\neq i} x^i_{ij},\qquad m^i = \underline{m}^i - \sum_{j\neq i} p^j\cdot x^i_{ji} + \sum_{j\neq i} p^i\cdot x^i_{ij},
\end{align}
where $w^i$ and $\underline{m}^i$ represent consumer $i$'s initial holdings of goods and money, respectively.\\

\noindent\textbf{Prices.} Since money functions as a universally accepted medium of exchange, it must have the same price for all consumers, and this common price provides the common normalization of the economy, which enters every budget constraint through the monetary flows. Each consumer $i$ chooses a price vector $\tilde p^i=(\tilde p^i_1,\dots,\tilde p^i_L,\tilde p^i_{L+1})$ in the simplex of $\R^{L+1}_+$, subject to
\begin{align}
\label{eqn:78}
\tilde{p}^1_{L+1} = \tilde{p}^2_{L+1} = \dots = \tilde{p}^n_{L+1} = \tilde{p}_{L+1} > 0 \, .
\end{align}
Expressing the prices of goods in money, $p^i_k=\tilde p^i_k/\tilde p_{L+1}$, every price vector belongs to the set
\begin{align}
\label{eqn:79}
P_c=\Big\{p\in\R^L_+ \ \mid \ \sum_{k=1}^L p_k=c\Big\},\qquad c=\frac{1-\tilde p_{L+1}}{\tilde p_{L+1}}>0,
\end{align}
the same for all consumers. The constant $c$ is fixed by the common normalization, and no consumer can change it unilaterally; it measures the largest money price at which a good can be offered. In the rest of this section, prices are money prices $p^i\in P_c$, and the price of money is one.

\vskip 0.5cm

\noindent\textbf{The money constraint.} The realistic nature of transactions makes the Clower cash-in-advance hypothesis important \cite{Clower1967}, since each individual only demands goods that they can afford, and every transaction is equivalent to a monetary flow. Since every trade is paid, the value balance between purchases and sales, once the monetary flows are included, holds identically, and the only constraint is the one that bears on money holdings. We adopt the following form of the cash-in-advance constraint: the final money holding of consumer $i$ must lie in a closed interval $\mathscr{M}^i=[\underline{\mu}^i,\overline{\mu}^i]\subset\R_+$, with $\underline{\mu}^i<\underline{m}^i<\overline{\mu}^i$. Its lower bound reads
\begin{align}
\label{eqn:80}
\sum_{j\neq i} p^j \cdot x_{ji}^i \ \leq\ \underline{m}^i-\underline{\mu}^i + \sum_{j\neq i} p^i \cdot x_{ij}^i \, :
\end{align}
purchases are financed by the initial money holding, net of a reserve $\underline\mu^i\ge0$, and by the receipts of the period. When $\underline{m}^i=\underline\mu^i=0$, \ref{eqn:80} reduces to the transaction balance condition \ref{eqn:6} of the barter economy. A stricter version of the hypothesis, in which receipts can only be spent in the next period, would require $\sum_{j\neq i}p^j\cdot x^i_{ji}\le\underline m^i$; we do not adopt it here.\\

The consumer monetary budget set is
\begin{align}
\label{eqn:81}
\mathcal{M}^i(p,X^{-i}) = \Big\{ X^i\in \mathbb{X}^i \ \Big| \ x^i\in \mathscr{X}^i \ , \ m^i\in \mathscr{M}^i \ , \ x_{ij}^i \leq x_{ij}^j \ , \ x_{ji}^i \leq x_{ji}^j \ , \ j \neq i \Big\} \, ,
\end{align}
where $x^i$ and $m^i$ are given by \ref{eqn:77}, and the consumer's problem is
\begin{align}
\label{eqn:82}
\max_{p^i \in P_c}\ \max_{X^i\in \mathcal{M}^i(p,X^{-i})} \tilde u^i(p,X^i),\qquad \tilde u^i(p,X^i)=u^i\Big(x^i,\frac{m^i}{\mathbb{P}(p)}\Big)+\vartheta^i(X^i)+\varrho^i(p^i,p^{-i}),
\end{align}
where $\mathbb{P}$ is a continuous and positive price index, $\vartheta^i$ is the relational term of Section \ref{Market topology}, and $\varrho^i$ is the price term \ref{eqn:65}, written in units of $c$, $\varrho^i(p^i,p^{-i})=-\frac{\kappa}{2c^2}\sum_{j\neq i}\|p^j-p^i\|^2$, so that it does not depend on the unit of account. The function $u^i$ retains its previously specified properties, and is non-decreasing in the real cash balance.

\vskip 0.5cm

\begin{proposition}{\textbf{Monetary budget set\label{MonetaryBudget}}\\}
Under assumption \emph{d}, the correspondence $\mathcal{M}^i$ has non-empty, convex and compact values, and it is continuous at every $(p,X^{-i})\in P_c^n\times\prod_{j\neq i}\mathbb{X}^j$.
\end{proposition}

\vskip 0.5cm

\begin{proof}
The set $\mathcal{M}^i(p,X^{-i})$ contains autarky, since $w^i\in\mathscr{X}^i$ and $\underline m^i\in\mathscr{M}^i$; for fixed prices, all its constraints are linear in $X^i$, so that it is convex, and it is compact. Its graph is closed, since all its constraints are continuous in $(p,X)$, so that it is upper semicontinuous. For lower semicontinuity, let $X^i\in\mathcal{M}^i(p,X^{-i})$, let $(p^m,X^{-i,m})\to(p,X^{-i})$, and let $\varepsilon\in(0,1]$. The plan $(1-\varepsilon)X^i$ yields the final money holding $\varepsilon\,\underline m^i+(1-\varepsilon)m^i$ at the prices $p$, which lies in the interior of $\mathscr{M}^i$, since $\underline m^i$ does and $m^i\in\mathscr{M}^i$. By continuity of the money holding in the prices, it remains in $\mathscr{M}^i$ at the prices $p^m$ for $m$ large enough. Choosing $\varepsilon_m\downarrow0$ as in Section \ref{New way to consider the Consumer Problem} for the bilateral constraints, and slowly enough for the money constraint, the plans $(1-\varepsilon_m)X^i$ belong to $\mathcal{M}^i(p^m,X^{-i,m})$ and converge to $X^i$.
\end{proof}

\vskip 0.5cm

Money thus simplifies the continuity analysis of Section \ref{New way to consider the Consumer Problem}: the initial money holding provides a strict slack in the budget, so that neither condition (C) nor a price floor is needed.

\vskip 0.5cm

\begin{proposition}{\textbf{Equilibrium with money\label{MonetaryEquilibrium}}\\}
\begin{enumerate}
\item The autarky profile is a transaction equilibrium of the monetary economy, for every common price vector $p^1=\dots=p^n\in P_c$.
\item Assume, in addition to assumption \emph{d}, that $u^i$ is continuous and concave, that $\vartheta^i$ is continuous and strictly concave in $X^i$, and that the indirect utility of the inner problem in \ref{eqn:82} is continuously differentiable in $p^i$, with a gradient that is Lipschitz continuous with constant $\ell^i$, uniformly in the strategies of the others. If $\kappa(n-1)>c^2\ell^i$ for every $i$, every consumer has a unique best response, which depends continuously on the strategies of the others, a transaction equilibrium exists, and the set of transaction equilibria is compact.
\end{enumerate}
\end{proposition}

\vskip 0.5cm

\begin{proof}
\begin{enumerate}
\item If the other consumers propose no trade, the bilateral constraints force $X^i=0$, so that $x^i=w^i$ and $m^i=\underline m^i$ whatever the price of $i$; at a common price vector, $\varrho^i$ attains its maximum.

\item The proof is that of Proposition \ref{ConcaveEquilibrium}, with $\mathcal{M}^i$ in place of $\mathcal{B}^i\cap\mathfrak{T}^i$ and $P_c$ in place of $P_\pi$. The objective of the inner problem is strictly concave in $X^i$, since $x^i$ and $m^i$ are affine in $X^i$ for fixed prices, and $\mathcal{M}^i$ is continuous by Proposition \ref{MonetaryBudget}. The term $\varrho^i$ equals $-\frac{\kappa(n-1)}{2c^2}\|p^i\|^2$ plus terms that are affine in $p^i$, so that the price problem is strictly concave on the convex compact set $P_c$ when $\kappa(n-1)/c^2>\ell^i$. Brouwer's fixed point theorem applies to the continuous best-response map on $P_c^n\times\prod_i\mathbb{X}^i$.
\end{enumerate}
\end{proof}

\vskip 0.5cm

As in Remarks \ref{NoSpecialization} and \ref{Uniqueness}, the equilibrium of part 2 may be autarkic, by part 1, and it is not unique. The cycle construction of Proposition \ref{TradeEquilibrium} does not carry over directly either: besides the reasons given in Remark \ref{NoSpecialization}, money opens new deviations, since a member of the cycle may sell its good and keep part of its receipts as money instead of buying, so that the gains from trade must now be compared with the utility of real balances. The existence of an equilibrium with trade in the monetary economy thus remains an open question.

\vskip 0.5cm

\begin{remark}{\textbf{Selling prices and the price level\label{MoneyPriceLevel}}\\}
Without the price term, $\kappa=0$, the analysis of Section \ref{Production} applies: in a specialized economy, every seller puts all the weight of its price on the good it sells, so that every unit sold costs $c$ units of money, whatever the good. The selling prices are then fixed by the common normalization, the terms of trade are one for one, and the allocation is determined by the bilateral quantities, as in fixed-price equilibria; an increase in the money supply cannot raise the prices, and is absorbed by the velocity of money or by the volume of trade. With $\kappa>0$, moving away from the prices of the market is costly: the selling prices are interior, $p^i_{k}<c$, and depend on the conditions of the market, so that the price level and the relative prices are no longer fixed by the normalization.
\end{remark}

\vskip 0.5cm

The common price of money must be positive. If $\tilde{p}_{L+1}=0$, money has no value, so that a good can only be exchanged for nothing, at a zero price; as in Proposition \ref{PositivePrices}, sellers would then withdraw their goods from sale, and only autarky would remain. Money would play no role.\\

By treating money as the $(L+1)$th good, the monetary economy can be compared with the barter economy of Section \ref{New way to consider the Consumer Problem} with $L+1$ goods, in which money is an ordinary good that can be exchanged freely. In the monetary economy, every exchange of goods is paired with a monetary flow of the same value, so that the monetary budget set corresponds to a subset of the budget set of this barter economy. This leads to the following result.

\vskip 0.5cm

\begin{proposition}{\textbf{Optimality -- Third Comparison\label{Third Comparison}}\\}
For given strategies of the others, in which the proposals of money of the partners match the monetary flows, with the price of money equal to one for all consumers and the same utility function, the value of the monetary consumer problem cannot exceed the value of the consumer problem in the barter economy with $L+1$ goods, in which money is an ordinary good.
\end{proposition}

\vskip 0.5cm

\begin{proof}
Every plan of the monetary economy defines a plan of the barter economy with $L+1$ goods, in which each purchase of goods is paired with a sale of money of the same value, and each sale of goods with a purchase of money; this plan satisfies the transaction balance condition \ref{eqn:6} and yields the same final holdings. The feasible set of the monetary problem is therefore contained in that of the barter problem.
\end{proof}

\vskip 0.5cm

A final remark concerns the following accounting identities. By considering the net flows, we get the \textbf{local net quantity of money equation}
\begin{align}
\label{eqn:83}
\sum_{j\neq i} \left( p^j \cdot x_{ji}^i - p^i \cdot x_{ij}^i \right) &= \sum_{j\neq i} \left( m_{ij} - m_{ji} \right) = \underline m^i-m^i \, .
\end{align}

If we sum the identity $ m^i = \underline{m}^i + \sum_{j\neq i} \left( m_{ji} - m_{ij} \right) $ over $i$, the flows cancel, and we get the money market equilibrium condition (demand equals supply)
\begin{align}
\label{eqn:84}
M = \sum_{i=1}^n m^i = \sum_{i=1}^n \underline{m}^i=\underline{M} \, .
\end{align}

And if we consider one direction flows and sum over $i$, we get the \textbf{quantity of money equation}
\begin{align}
\label{eqn:85}
\sum_{i=1}^n \sum_{j\neq i} p^j \cdot x_{ji}^i = \sum_{i=1}^n \sum_{j\neq i} m_{ij} = \underline{M} \, v \, ,
\end{align}

\noindent where $v$ is the money velocity $v=\displaystyle \frac{\sum_{i=1}^n \sum_{j\neq i} m_{ij}}{\underline{M}}$, and the left-hand side is the value of all transactions. Relations \ref{eqn:83}--\ref{eqn:85} are identities, which hold for every feasible profile: as in Fisher's equation \cite{Fisher1922}, the content of the quantity theory lies in the behavior of its terms. Under the cash-in-advance constraint \ref{eqn:80}, receipts can be spent within the period, so that the velocity is not bounded; under the stricter version, the total value of payments could not exceed $\underline M$, and the velocity would satisfy $v\le1$.

\vskip 0.5cm

\subsection{Marginal effects \label{Marginal effects}}

\hspace{0.5cm} In this section, we discuss the possible effects of variations in prices and money on the economy; the following mechanisms are suggested by the structure of the consumer problem, and their occurrence depends on the configuration of the economy. Given a price matrix $p$, what happens when $p^j_k$ increases for some consumer $j$ and some real good $k$? From the stationarity conditions \ref{eqn:24}--\ref{eqn:25}, extended to the money constraint, one could expect one (or a combination) of the following reactions of a consumer $i$ that buys good $k$ from $j$:

\begin{enumerate}
\item \textbf{Money disbursement:} Decrease of $m^i$ causing $u^i$ to decrease.

\item \textbf{Income effect:} Decrease of $x^i_{ji,k}$ causing $u^i$ to decrease.

\item \textbf{Substitute seller:} Decrease of $x^i_{ji,k}$ and increase of $x^i_{li,k}$ for $l \neq j$.

\item \textbf{Substitute product:} Decrease of $x^i_{ji,k}$ and increase of $x^i_{hi,\ell}$ for some good $\ell \neq k$ and some seller $h$.

\item \textbf{Money hoarding:} Decrease of $x^i_{ji,k}$ and increase of $m^i$.

\item \textbf{Substitute other products (sell):} $x^i_{ji,k}$ is kept and $x^i_{ij,\ell}$ increases for some good $\ell\neq k$.

\item \textbf{Inflation effect:} $x^i_{ji,k}$ is kept and $p^i$ increases; with the price term $\varrho^i$, the increase of $p^j$ also raises the prices toward which the other consumers adjust.

\item \textbf{Substitute other products (buy):} $x^i_{ji,k}$ is kept and $x^i_{hi,\ell}$ decreases for some good $\ell\neq k$.

\end{enumerate}

Now, consider an increase $\underline{m}^i + \delta_m > \underline{m}^i$ in the money endowment of an individual $i$. With the consumer becoming wealthier (distribution disruption), the money constraint \ref{eqn:80} is relaxed, and three effects are conceivable, depending on its utility:

\begin{enumerate}
\item \textbf{Hoarding:} $m^i$ absorbs the increase in money, and nothing changes in the economy, the velocity of money decreasing.

\item \textbf{Quantity adjustment:} The consumer decides either to reduce $x_{ij}^i$ or to increase $x_{ji}^i$ (if other agents are interested in holding money), leading to an effect on trade.

\item \textbf{Price adjustment:} The prices adjust: the consumer may change its own price, and the sellers facing a larger demand may raise theirs, an increase that propagates to the other prices through the price term (inflation effect). By Remark \ref{MoneyPriceLevel}, this effect requires $\kappa>0$: without the price term, the selling prices are already at their upper bound $c$.

\end{enumerate}

One should also highlight the redistribution effect of this policy in creating inequalities.

\vskip 0.5cm

\subsection{Time, uncertainty, and open economy\label{Time, uncertainty, and open economy}}

\hskip 0.5cm The static framework developed above admits a natural extension to a dynamic environment under uncertainty. In such a setting, agents form expectations about future states of the economy and the actions of other agents, while production and consumption schedules need not be synchronized. Letting $\{\psi_t\}_{t\geq t_0}$ denote the stochastic evolution of states, individual decisions can be interpreted as solutions to intertemporal optimization problems in which agents maximize expected utility subject to feasibility, financial, and exchange constraints. The stricter cash-in-advance constraint, in which receipts can only be spent in the next period, provides a natural link between successive periods.\\

Expectations play a central role in coordinating decentralized decisions. Agents anticipate future prices, exchange opportunities, and counterpart behavior through subjective probability measures, which may or may not be mutually consistent. As a result, realized allocations need not coincide with anticipated ones. This discrepancy generates an endogenous adjustment process: revisions of beliefs, reallocation of resources, and updates in exchange patterns progressively reduce the mismatch between expectations and realizations. In this sense, equilibrium can be interpreted as a limiting configuration of a sequence of temporary equilibria, rather than as an instantaneous fixed point, in the spirit of the equilibrium of plans, prices and price expectations of \cite{Radner1972}.\\

\hskip 0.5cm The structure of the model also allows for a direct extension to an open economy framework. Introducing multiple currencies and cross-border exchange opportunities endogenizes exchange rates and international financial flows within the same decentralized interaction structure. In such an environment, expectations not only concern domestic variables but also foreign prices, exchange rates, and external liquidity conditions. This adds an additional layer of interdependence, as belief revisions in one economy propagate through trade, financial linkages, and currency markets, thereby affecting the global adjustment dynamics. These extensions are left for future research.

\vskip 0.5cm

\section{Conclusion}

\hskip 0.5cm In this paper, we introduced the effective trade model as a framework for decentralized exchange in which prices and bilateral trade proposals are jointly determined by individual agents, while actual transactions arise from the compatibility of the proposed trades. We formulated the consumer problem as a generalized game, established the existence of transaction equilibria and described their structure, characterized equilibrium behavior through Karush--Kuhn--Tucker conditions and a sensitivity analysis, analyzed their welfare properties, and studied adjustment processes. We examined market imperfections arising from indivisibilities and restrictions on the topology of exchange, and extended the framework to preferences over trading partners and prices, to production, and to monetary exchange. Numerical examples illustrate the theoretical results.\\

Several fundamental conclusions emerge from our analysis:

\begin{itemize}
\item \textbf{Effective exchange is determined by transaction feasibility rather than by desired supply and demand alone.} An agent's intention to buy or sell does not by itself generate a transaction; exchange requires bilateral compatibility between the proposed trades of the parties. Realized exchange is therefore an endogenous outcome of decentralized individual plans, and the bilateral constraints act as quantity signals, in the spirit of non-Walrasian equilibrium theory.

\item \textbf{Equilibria always exist, and equilibria with trade exist along trade cycles with gains from trade.} Autarky is always a transaction equilibrium: when no partner proposes to trade, no trade can take place. This coordination failure is intrinsic to bilateral exchange. When the economy contains a trade cycle with gains from trade, an equilibrium with trade can be constructed explicitly, each member of the cycle selling one good to its successor at the smallest of the preferred trade levels. Equilibria can also be obtained by fixed point arguments in specialized economies, in which prices are fixed by dominance, and in the augmented model, in which best responses are unique and continuous; in the absence of a trade cycle with gains from trade, and in the augmented and monetary models, the existence of an equilibrium with trade remains an open question.

\item \textbf{In the basic model, sellers specialize and prices are fixed by the normalization.} At a transaction equilibrium, every seller sells at most one good, at a positive price, and puts all the weight of its price on that good. Taking the proposals of its buyers as given, a seller has no reason to lower its price, which reaches the bound set by the normalization: the terms of trade are then one for one, and the allocation is determined by the bilateral quantities, as in a fixed-price equilibrium.

\item \textbf{Preferences can extend beyond consumption to the structure of exchange and to prices.} Agents may care not only about the goods they consume, but also about whom they trade with and how their prices compare with those of the market. These terms restore concavity: best responses become unique and continuous, specialization is no longer imposed, and selling prices become interior and responsive to market conditions. They also capture buyer--seller affinities, loyalty, market positioning and other network-based considerations.

\item \textbf{Transaction equilibria need not be unique or Pareto optimal.} Autarky can constitute a transaction equilibrium even when mutually beneficial exchanges are possible, so that the equilibrium is never unique as soon as an equilibrium with trade exists. The uniqueness of best responses does not imply the uniqueness of the equilibrium, and uniqueness must be sought for a selection of equilibria.

\item \textbf{The classical welfare correspondence between equilibrium and efficiency does not survive.} Transaction equilibria need not be Pareto optimal, and Pareto-optimal allocations need not be reachable through trade from the initial endowments. A planner who chooses both trades and prices provides a benchmark: its prices are determined by the trades, the incomes forming the stationary distribution of the chain of sales, and trade must be circular.

\item \textbf{Market imperfections can arise from the structure of exchange itself.} Indivisibilities can prevent mutually beneficial transactions from being realized, and restrictions on trading relationships constrain the set of feasible exchanges. In particular, market topology generates a middleman imperfection: for given strategies of the others, no agent can do better than in the unrestricted economy, and the numerical examples show that the topology changes the set of equilibria and their welfare.

\item \textbf{Adjustment processes reveal the structure of equilibria.} The planner's process converges to a social optimum at given prices, and the price adjustment of each seller converges to prices that put their weight on the goods sold in maximal quantity, a dynamic counterpart of specialization. The rest points of the joint dynamics are equilibria of specialized economies; the convergence of the joint dynamics remains open.

\item \textbf{Production and monetary exchange can be incorporated into the effective-trade framework.} Producers enter the generalized game as additional players. With money, the initial money holdings provide a slack that simplifies the analysis, and the quantity of money equation, an accounting identity, connects monetary circulation to the volume and structure of effective transactions; whether an increase in money raises prices or the volume of trade depends on whether prices respond to market conditions.

\end{itemize}

Taken together, these results suggest that the organization of transactions is itself a fundamental economic object: equilibrium depends not only on aggregate supply and demand, but also on the compatibility of decentralized plans, prices, bilateral transactions, and trading relationships. The effective trade model therefore provides a broader conception of exchange in which agents may care about both \emph{what} they consume and \emph{with whom} and \emph{how} they trade, while the structure and feasibility of the exchange network can shape both equilibrium outcomes and welfare. Several questions remain open: the existence of equilibria with trade beyond trade cycles with gains from trade, and in the augmented and monetary models; the selection among multiple equilibria; the convergence of the joint dynamics; and the formation of prices when sellers anticipate the reactions of their buyers. Future research may also extend the framework to dynamic and stochastic environments, endogenous network formation, and learning.

\vskip 0.5cm

\section*{Appendix}

\vskip 0.5cm

\subsection{Berge's Maximum Theorem}

\begin{theorem}{\textbf{Berge's Maximum Theorem\label{Berge's Maximum Theorem}} \cite[Ch.~VI, §3]{Berge1963}, \cite[Lemmas~17.29--17.30 and Theorem~17.31]{Aliprantis2006}}\\
Let $H$ and $\Theta$ be metric spaces, $f: H \times \Theta \to \R$ a continuous function, and $\mathcal{F} : \Theta \to 2^H$ an upper semicontinuous correspondence such that $\mathcal{F}(\theta)\subset H$ is non-empty and compact for each $\theta\in\Theta$. Define the value function $v: \Theta \to \R$ and the set of maximizers by
\begin{align*}
v(\theta) = \max_{x \in \mathcal{F}(\theta)} f(x, \theta), \qquad S(\theta) = \arg \max_{x \in \mathcal{F}(\theta)} f(x, \theta).
\end{align*}
Then:
\begin{enumerate}
\item $S(\theta)$ is non-empty and compact for each $\theta\in\Theta$, and $v$ is upper semicontinuous on $\Theta$;
\item if $\mathcal{F}$ is, moreover, lower semicontinuous at $\theta_0\in\Theta$, then $v$ is continuous at $\theta_0$ and $S$ is upper semicontinuous at $\theta_0$;
\item in particular, if $\mathcal{F}$ is continuous on $\Theta$, then $v$ is continuous on $\Theta$ and $S$ is upper semicontinuous on $\Theta$.
\end{enumerate}
\end{theorem}

\vskip 0.5cm

\begin{corollary}{\textbf{\cite{Sundaram1996}}\label{BergeUniqueness}\\}
Let $f: H \times \Theta \to \R$ be continuous, and $\mathcal{F} : \Theta \to 2^H$ be continuous and compact valued. Define $v$ and $S$ as in Theorem \ref{Berge's Maximum Theorem}.
\begin{enumerate}
\item Suppose $f(\cdot,\theta)$ is quasi-concave in $x$ for each $\theta$, and $\mathcal{F}$ is convex-valued on $\Theta$. Then $S$ is a convex-valued upper-semicontinuous correspondence.

\item If "quasi-concave" is replaced with "strictly quasi-concave", $S$ is single valued everywhere on $\Theta$, and hence defines a continuous function.

\item If f is quasi-concave on $H \times \Theta$, and $\mathcal{F}$ has a convex graph, then $v$ is a quasi-concave function, and $S$ is a convex-valued upper semicontinuous correspondence. If "quasi-concave" is replaced with "strictly quasi-concave", $S$ is single valued everywhere on $\Theta$, and hence defines a continuous function.
\end{enumerate}
\end{corollary}

\vskip 0.5cm

\begin{theorem}{\textbf{\label{Compactness of Nash equilibria} \cite{Laraki2019}}}\\
If a game $G=\{A^i,u^i\}_{1\leq i \leq n}$ is compact, continuous and quasi-concave, then its set of Nash equilibria is a non-empty and compact subset of $\prod_{i=1}^n A^i$.
\end{theorem}

\vskip 0.5cm

\begin{theorem}{\textbf{Kakutani fixed point theorem\label{Kakutani fixed point theorem} \cite{Starr2012}}} \\
Let $F : C \subseteq X \rightharpoonup X$ be a correspondence from a compact and convex subset $C$ of a topological space $X$. If $F$ is upper semi-continuous and $F(x)$ is a convex set for all $x\in C$, then $F$ has a fixed point $x^\star \in F(x^\star)$.
\end{theorem}

\vskip 0.5cm

\clearpage

\section*{Declarations}

\subsection*{Conflict of interest}

The author declares no conflict of interest.

\subsection*{Author's contribution}

The author is the sole contributor to this work, including its conception, analysis and writing.

\subsection*{Funding}

No funding was received for conducting this study.

\subsection*{Availability of data and materials}

No external data were used. The numerical examples are generated from the parameters given in Section \ref{Numerical simulation}; the computations were carried out with Wolfram Mathematica, and the code is available from the author upon request.

\subsection*{Declaration of generative AI and AI-assisted technologies in the writing process}

During the preparation of this work, the author used Claude (Anthropic) to improve the language of the manuscript, to check the consistency of proofs and computations, to suggest revisions of some arguments and passages. After using these tools, the author reviewed, verified and edited all the content, and takes full responsibility for the content of the publication.

\bibliographystyle{apalike}
\bibliography{BibliographieEco}

\end{document}